\documentclass[a4,12pt]{article}
\usepackage{jheppub}
\usepackage{amsmath,bm}
\usepackage{amssymb}
\usepackage{slashed}
\usepackage{graphicx}
\usepackage{color}
\usepackage[autostyle]{csquotes}
\usepackage{lscape}
\usepackage{subfig}
\usepackage{setspace}
\usepackage{bbm}
\usepackage{breqn}
\usepackage{wrapfig}
\usepackage{hyperref}
\usepackage{mathtools}
\usepackage{physics}
\usepackage{simpler-wick}
\newcommand{\be}{\begin{equation}}
\newcommand{\ee}{\end{equation}}
\newcommand{\bea}{\begin{eqnarray}}
\newcommand{\eea}{\end{eqnarray}}
\newcommand{\beas}{\begin{eqnarray*}}
\newcommand{\eeas}{\end{eqnarray*}}
\newcommand{\nn}{\vec n}
\newcommand{\ca}{\vec c_a}
\newcommand{\Xa}{\vec X_a}
\newcommand{\Ord}{\mathcal O}

\begin{document}

\title{All Order Gravitational Wave Tails from Classical Soft Theorem for Subleading Post-Newtonian Expansion}
\author[a]{Debanjan Karan,}
\author[a]{Babli Khatun}
\affiliation[a]{International Centre for Theoretical Sciences (ICTS-TIFR) \\ Tata Institute of Fundamental Research, Shivakote, Hesaraghatta, Bangalore 560089, India.}

\emailAdd{debanjan.karan@icts.res.in, babli.khatun@icts.res.in}

\abstract 
{\begin{abstract}
WThe low-frequency gravitational waveform emitted in a classical
scattering process contains an infinite tower of late- and early-time
tail terms of the form $u^{-r}(\ln|u|)^{r-1}$, $r \geq 1$. In the
Newtonian limit, the coefficients of this entire tower have recently
been shown to be universal, i.e.\ fixed by the asymptotic data of the
scattering objects alone. At the retained order in the velocity expansion, these leading-logarithmic tails are determined to all orders in Newton’s constant \(G\). We extend this result to the first subleading
order in the post-Newtonian expansion of the asymptotic states. We
consider scattering of an arbitrary number of incoming and outgoing
nonrelativistic objects interacting only through gravity at long
range, and we derive the complete tower of tail terms at this order. The correction arises from the first subleading term in the retardation expansion and is determined by the asymptotic Newtonian trajectories of the massive particles. Its coefficients are again
determined entirely by the asymptotic trajectories and are independent
of the details of the interaction. For $2\to2$ scattering, the
leading-logarithmic series can be summed in closed form. Our results
agree with a relativistic conjecture proposed in earlier work \cite{Karan:2026jzs}.
\end{abstract}}

\maketitle

%%%%%%%%%%%%%%%%%%%%%%%%%%%%%%%%%%%%%%%%%%%%%%%%%%%%%%%%%%%%%%%%%%%%%%%%%%%%%%%%%%%%%%%%%%%%%%%%%%%%%%%%%%%%%%%%%%%
\section{Introduction}

Gravitational radiation is generically produced in classical scattering
processes involving compact or extended objects spanning a wide range of
length scales and velocities. A natural question is whether one can obtain
an analytic expression for the resulting radiative gravitational field.
In general, the waveform depends sensitively on the details of the
interaction and on the trajectories followed by the scattering objects
during the collision, and a closed-form result is therefore out of reach
except in special limits. The classical soft graviton theorem provides a notable exception. It states that the low-frequency part of the gravitational waveform emitted in such a process can be computed without knowledge of the details of the scattering \cite{1804.09193,1806.01872,1808.03288,AtulBhatkar:2018kfi,1912.06413,2008.04376,2105.08739,2106.10741,2308.16807,2408.08851}. More precisely, a certain set of terms in the low-frequency expansion of the waveform are fixed entirely by the asymptotic data of the incoming and outgoing objects, namely their masses and momenta \cite{1912.06413}. Here ``soft'' refers to the regime in which the frequency $\omega$ of the emitted radiation is small compared to the inverse of the characteristic time scale of the scattering process. In this sense the classical soft theorem is a perturbative statement in which $\omega$, or equivalently the frequency at which the detector is sensitive, plays the role of the expansion parameter. The central feature of the classical soft graviton theorem is its universality: within its domain of validity, the waveform is determined solely by the asymptotic data of the scattering objects and is insensitive to the nature of the interactions that act during the collision. The predictions of the soft graviton theorem have also been tested in different frequency regimes and limiting cases \cite{Damour:1985cm,fBlanchet:1992br,0501041,1409.4555,1901.10986,2007.02077,DiVecchia:2023frv,2309.14925,2402.06361,2402.06604,2402.06533,Akhtar:2026pkm,Akhtar:2024lkk,Fucito:2024wlg}. In parallel,the link between soft factors and classical waveforms has also guided efforts to extract gravitational radiation directly from scattering amplitudes \cite{Kosower:2018adc,Bautista:2019tdr,Akhtar:2025fil,Paul:2026hyf}. Soft theorems have also been related to asymptotic symmetries \cite{1903.09133,1912.10229,Compere:2025tzr,2205.11477,2309.11220,2403.13053,2407.07978,DeAngelis:2025vlf,Compere:2026jmk,2603.09844,2603.22681,2512.21239,Banerjee:2026keq,2601.05951,2508.17919,Compere:2026vvi}.

In the low-frequency regime, the waveform contains more than the $\omega^{-1}$ pole associated with the gravitational memory effect \cite{mem1,mem2,mem3,mem4,Ilkhchi:2025gbd,Raclariu:2026mfs}. It also exhibits a tower of logarithmic terms of the form $\omega^{r-1} (\ln \omega)^r$, $r \geq 1$\footnote{In the time domain, such terms are denoted as $u^{-r}(\ln \vert u \vert )^{r-1}$ where the waveform is detected at a fixed retarded time $u$.} which arise from the long-range nature of the gravitational interaction and are commonly referred to as \emph{tails} in the literature \cite{1808.03288}. The coefficients of the first two members of this tower, $r=1$ and $r=2$, were previously shown to be universal, in the sense that they are fixed entirely by the asymptotic data of the scattering objects \cite{1912.06413,2008.04376}. In \cite{2501.07328}, it was also shown that for a long range electromagnetic force the universality exists even for an infinite tower of the tail terms in 4-dimension. The same has been proved in higher dimension \cite{Sahoo:2026ulx}. It was shown only recently \cite{Karan:2026jzs} that, in the Newtonian limit, this universality extends to the entire tower: for every $r \geq 1$, the coefficient of $\omega^{r-1}(\ln\omega)^{r}$ is determined solely by the asymptotic data of the incoming and outgoing objects. Over the past few decades, considerable progress has been made in computing analytic waveforms for generic scattering processes using perturbative schemes, most notably the post-Minkowskian (PM) \cite{Jakobsen:2021smu} and post-Newtonian (PN) approximations \cite{1310.1528,Bini:2024ijq}. Comparisons of amplitude-based scattering waveforms with results
from the multipolar post-Minkowskian formalism have demonstrated
agreement in the soft limit and, more generally, in the
post-Newtonian expansion, after accounting for asymptotic-frame
choices and infrared effects \cite{Georgoudis:2023eke,Bini:2024rsy}. 

In \cite{Karan:2026jzs} it was shown that, at leading order in the small-velocity expansion of the incoming and outgoing states, all universal terms in the soft expansion can be written in closed form. This result holds for an arbitrary number of massive scattering objects. In this paper we address the following question: can explicit expressions be obtained for all universal terms in the soft expansion when the asymptotic states are also expanded to
first subleading order in the PN expansion, i.e. at next-to-Newtonian order? To make the expansion explicit, we parametrize the asymptotic 3-velocities of the scattering objects as $\alpha\,\vec v_a$, where $\alpha$ is a small parameter that tracks the PN order. In this counting, all terms computed in \cite{Karan:2026jzs} arise at $\mathcal{O}(\alpha^{2})$. In this paper we compute the corrections at $\mathcal{O}(\alpha^{3})$. The velocity expansion is performed at fixed Newton’s constant \(G\), with no independent truncation in powers of \(G\). In particular, at fixed asymptotic masses and velocities, the coefficient of \(u^{-r}(\ln|u|)^{r-1}\) carries a factor \(G^{r+1}\). Retaining the entire tower for \(r\geq1\) therefore determines the leading-logarithmic waveform to all orders in \(G\) at the velocity order considered.

At this order, the logarithmic terms in the soft radiation probe properties of the source beyond those captured at $\mathcal{O}(\alpha^{2})$. The terms of \cite{Karan:2026jzs} are fixed entirely by the mass quadrupole moment of the initial and final scattering states. The $\mathcal{O}(\alpha^{3})$ correction, by contrast, also depends on the mass octupole moment and on the angular-momentum density \cite{RevModPhys.52.299, Goldberger:2009qd}. It is therefore natural to ask whether the leading-logarithmic contributions at this order require new information about the encounter, or whether they are still determined entirely by the asymptotic data. These results also provide a further check of the arbitrary-velocity conjecture proposed in section~5 of \cite{Karan:2026jzs}, beyond its leading Newtonian limit. Such agreement tests the conjecture at an additional velocity order without constituting a proof at general
velocities.

A related issue concerns the dynamics required for this calculation. A correction to the Newtonian waveform need not arise solely from a
correction to the Newtonian force law. The source is evaluated on the
past light cone of the detector, which gives the retarded-time argument
$u+\vec{n}\cdot \vec{x}$ at leading order in $1/R$. Expanding this
argument introduces one additional power of velocity into the waveform,
producing the $O(\alpha^3)$ correction considered here. Conservative
corrections to Newtonian dynamics are instead quadratic in velocity
and affect the waveform at higher order. Consequently, the Newtonian trajectory
coefficients obtained in \cite{Karan:2026jzs} suffice to determine the source moments needed here. Conservation of energy and momentum must nevertheless be imposed on the total source, including the gravitational stresses. Approximating the energy and momentum densities by their leading matter contributions is justified only after this conservation law has been used. In particular, this approximation does not remove the gravitational
interaction from the calculation.

We now state the scope of the calculation. We work at radiative order
in the distance $R$ to the detector, keeping only the $1/R$ part of the
field, and we compute the contribution sourced by the asymptotic
massive objects. The nonrelativistic assumption applies only to the
asymptotic motion. Nothing is assumed about the velocities or field
strengths reached during the encounter itself, since the universal
terms do not depend on them.\\
One caveat concerns the memory terms. Energy carried off by the
emitted gravitational radiation also sources a memory contribution,
the so-called nonlinear or null memory \cite{Tolish:2014bka}, which is not included in the
massive-particle terms computed here \cite{Tolish:2014oda}. Our memory terms are therefore
complete through $\mathcal{O}(\alpha^{3})$ only if this radiative
contribution is negligible at that order, or if it is added separately. With this scope in place, we turn to the result and to the
approximation under which its coefficients are fixed.

\textbf{Summary of the results:}
    The leading-logarithmic tail terms including the memory at order $\mathcal{O}(\alpha^3)$ are summarized here. 
{\allowdisplaybreaks
\begin{align}\label{eij 1 multiparticle w intro}
    e^{ij}_{(1)}(w)= & \frac{2G\alpha^3}{R}\sum_a m_a(\nn\cdot\vec{v}_a)v_a^iv_a^j + \frac{G\alpha^3}{R} \sum_{r=1}^{\infty}\frac{(\ln w)^{r-1}}{w^r} \sum_a m_a  \Big[2\delta_{r1}(\nn\cdot\ca)v_a^iv_a^j +2\delta_{r2}\Big\{ (\nn\cdot\vec{v}_a)c_a^ic_a^j \notag 
    \\
    & -(\nn\cdot\ca)(v_a^ic_a^j+v_a^jc_a^i)\Big\} +2\delta_{r3}(\nn\cdot\ca)c_a^ic_a^j +(r-1)(r-2)\Big\{ (r+1)(\nn\cdot\vec d_{r-1,a})v_a^iv_a^j \notag \\
    & -(r-1)(\nn\cdot\vec{v}_a) \left(v_a^id_{r-1,a}^j+v_a^jd_{r-1,a}^i +c_a^id_{r-2,a}^j+c_a^jd_{r-2,a}^i\right) +(r-1)(\nn\cdot\vec d_{r-2,a}) (v_a^ic_a^j+v_a^jc_a^i) \notag \\
   &  -(r-3)(\nn\cdot\ca) \left(v_a^id_{r-2,a}^j+v_a^jd_{r-2,a}^i +c_a^id_{r-3,a}^j+c_a^jd_{r-3,a}^i\right)  +(r-3)(\nn\cdot\vec d_{r-3,a})c_a^ic_a^j \Big\}\Big].
\end{align}}
Here $G$ is Newton's constant, $i,j$ are spatial indices, and the sum runs over the outgoing particles. Here $\vec{n}$ is the direction of the wave-detector placed at a distance $R$. Let $\vec{X}_a(u)$ denote the position of the $a$th particle and define
$\widetilde{\vec{X}}_a(w)\equiv\vec{X}_a(w/\alpha^3)$, where
$w=\alpha^3u$. At the order retained here, $\vec{c}_a$ and
$\vec{d}_{r,a}$ are the coefficients of $\ln w$ and
$(\ln w/w)^r$, respectively, in the asymptotic expansion of
$\alpha^2\widetilde{\vec{X}}_a(w)$. These coefficients are the same
as those obtained in~\cite{Karan:2026jzs} in the Newtonian limit and are determined
by the asymptotic masses and velocities of the scattering objects:
\be \label{c_a}
\vec c_a = G\, \sum_{b\ne a}  m_b\, {\vec v_a -\vec v_b\over |\vec v_a-\vec v_b|^3}\, ,
\ee
\begin{equation}
\begin{split}
 \vec d_{r,a}=&-\frac{G}{2^r r(r+1)}\sum_{b\ne a} \frac{m_b}{|\vec{v}_a-\vec v_b|^3}\Bigg[
  (\vec{v}_a-\vec v_b)\sum_{k=[(r+1)/2]}^r \frac{(-1)^k(2k+1)!}{k!(2k-r)!(r-k)!}
  \\
 &\hspace{10mm}\times \frac{[(\vec{v}_a-\vec v_b)\cdot(\ca-\vec c_b)]^{2k-r} |\ca-\vec c_b|^{2(r-k)}}{|\vec{v}_a-\vec v_b|^{2k}}
 \\
 &+2(\ca-\vec c_b)\sum_{k=[r/2]}^{r-1} \frac{(-1)^k(2k+1)!}{k!(2k+1-r)!(r-k-1)!}
 \\
 &\hspace{10mm}\times \frac{[(\vec{v}_a-\vec v_b)\cdot(\ca-\vec c_b)]^{2k+1-r} |\ca-\vec c_b|^{2(r-k-1)}}{|\vec{v}_a-\vec v_b|^{2k}}
 \Bigg],\qquad r\geq1.
 \label{r2:d intro}
\end{split}
\end{equation}
where $[s]$ denotes the largest integer not larger than $s$. The sums use the lower limit appropriate to the parity of $r$ and run in integer steps. Since $\vec{d}_{r,a}$ is determined in terms of the coefficients $\vec{c}_a$, with the expression of $\vec{c}_a$ given in terms of the asymptotic data, it determines the logarithmic trajectory required for the leading-logarithmic waveform. Together with energy-momentum conservation, the asymptotic Newtonian trajectories\footnote{We will justify later that even to determine order $\alpha^3$ waveform, the Newtonian trajectory is sufficient.} determine eq.\eqref{eij 1 multiparticle w intro} at the retained order, including both matter and gravitational contributions.\\
We compare the waveform at this order with the waveform obtained in the Newtonian limit in \cite{Karan:2026jzs}:
\begin{align}
e^{ij}(w)
=& \frac{2G\alpha^{2}}{R}\sum_a m_a v_a^i v_a^j +\frac{G\alpha^{2}}{R} \sum_{r=1}^{\infty}\frac{(\ln w)^{r-1}}{w^{r}} \sum_a m_a  \Bigg[ \delta_{r1}(v_a^i c_a^j+v_a^j c_a^i) -2\delta_{r2}c_a^i c_a^j \notag \\
 &\hspace{25mm} +(r-1)(r-2) \left( v_a^i d_{r-1,a}^{j} +v_a^j d_{r-1,a}^{i} +c_a^i d_{r-2,a}^{j} +c_a^j d_{r-2,a}^{i}\right)\Bigg].
\end{align}
Comparing the waveform at $\mathcal{O}(\alpha^2)$ with the waveform at $\mathcal{O}(\alpha^3)$ it is manifest that in the Newtonian limit the waveform does not depend on the direction of the detector. Also in the PN expansion to first sub-leading order expansion the memory\footnote{In \cite{Karan:2026jzs} there is a discussion related to the memory term.} has an extra factor of $\vec{n}\cdot \vec{v}_a$ while the tail terms are comparatively complicated as compared to the Newtonian case.\\
Specializing the multiparticle coefficients $\vec{c}_a$, $\vec{d}_{r,a}$ to
two outgoing particles and substituting them into \eqref{eij 1 multiparticle w intro}
gives a geometric series that can be summed explicitly. With
$\vec v=\vec v_1-\vec v_2$, $v=|\vec v|>0$,
$\gamma=G(m_1+m_2)/v^3$ and $z=\ln w/w$, the retained late time wave-form is
\begin{equation}
\begin{split}
  e^{ij}_{(1)}(w) =\frac{2G\alpha^3}{R} & \sum_{a=1}^2m_a(\vec n\cdot\vec v_a)v_a^iv_a^j + \frac{2G m_1m_2 \alpha^3}{R(m_1+m_2)} \sum_{r=1}^\infty (-1)^{r-1} \gamma^r \frac{(\ln w)^{r-1}}{w^r}
  \\
  & \times \Bigg\{\frac{\vec n\cdot\vec v}{2} \Big[(v_1^i+v_2^i)v^j+(v_1^j+v_2^j)v^i\Big] - (r-1) \frac{\vec n\cdot(m_1\vec v_1+m_2\vec v_2)}{m_1+m_2}v^iv^j  \Bigg\}
 \label{r3:delta-closed late intro}
\end{split}
\end{equation}
For two incoming particles, define
$\vec v'=\vec v'_1-\vec v'_2$, $v'=|\vec v'|>0$,
$\gamma'=-G(m'_1+m'_2)/v^{\prime 3}$ and $z'=\ln|w|/w$.
The same replacement rules give the early-time wave-form
\begin{equation}
\begin{split}
  e^{ij}_{(1)}(w)=\frac{2G\alpha^3}{R} & \sum_{a=1}^2 m_a'(\vec n\cdot\vec v_a') {v_a'}^i {v_a'}^j + \frac{2G m_1' m_2' \alpha^3}{R(m_1' + m_2')} \sum_{r=1}^\infty (-1)^{r-1} (\gamma')^r \frac{(\ln \vert w \vert)^{r-1}}{w^r}
  \\
  & \times \Bigg\{\frac{\vec n\cdot\vec v'}{2} \Big[({v_1'}^i+{v_2'}^i){v'}^j+({v_1'}^j+{v_2'}^j) {v'}^i \Big] - (r-1) \frac{\vec n\cdot(m_1'\vec v_1' +m_2' \vec v_2')}{m_1'+m_2'} {v'}^i {v'}^j  \Bigg\}
 \label{r3:delta-closed early intro}
\end{split}
\end{equation}

The rest of the paper is organized as follows. Section~\ref{s2} derives the
asymptotic trajectory coefficients and the corresponding early- and
late-time multiparticle wave-forms in PN expansion of asymptotic scattering states to first sub-leading order expansion. Section~\ref{s3} specializes the calculation to two particles in each asymptotic state. Section~\ref{s4} explicitly shows the agreement of waveform at the first subleading order in the PN expansion with the all-order conjecture proposed in \cite{Karan:2026jzs}. Appendix~\ref{Err} estimates the corrections omitted at various stages while deriving the waveform. In Appendix~\ref{app B} we have reproduced the subsubleading waveform both at the Newtonian and the subleading order in PN expansion from the conjecture. Appendix \ref{app C} computes the quadratic graviton vertex that leads to the gravitational energy momentum contribution to the waveform. In Appendix \ref{App D} the circular average is worked out in detail.
%%%%%%%%%%%%%%%%%%%%%%%%%%%%%%%%%%%%%%%%%%%%%%%%%%%%%%%%%%%%%%%%%%%%%%%%%%%%%%%%%%%%%%%%%%%%%%%%%%%%%%%%%%%%%%%%%%%
\section{General formula for the waveform}\label{s2}
%%%%%%%%%%%%%%%%%%%%%%%%%%%%%%%%%%%%%%%%%%%%%%%%%%%%%%%%%%%%%%%%%%%%%%%%%%%%%%%%%%%%%%%%%%%%%%%%%%%%%%%%%%%%%%%%%%%
The calculation in this section isolates the first velocity correction to the all order gravitational waveform in the Newtonian limit derived in \cite{Karan:2026jzs}. We use the conventions and the trajectory expansion of \cite{Karan:2026jzs}, following especially its section-2. The additional source information required here is the first spatial moment of the total energy-momentum tensor. Reducing that moment by energy--momentum conservation will express the correction in terms of the same trajectories that determine the leading waveform in the Newtonian limit.

Consider massive incoming and outgoing particles with asymptotically small velocities. Their masses and velocities are denoted by $\lbrace m_a', \vec{v}_a' \rbrace$ and $\lbrace m_a, \vec{v}_a \rbrace$, respectively. The finite interaction region need not admit a Newtonian description; the expansion concerns the asymptotic portions of the motion when the particles are almost free, although there is long-range weak gravitational force. We use $c=1$ and the Minkowski metric with signature $\eta_{\mu\nu} = (-,+,+,+)$. The formulas below determine the massive-particle contribution. Identifying its constant term with the complete memory also requires any omitted null-memory contribution to be negligible at the retained order; otherwise that contribution must be added.\\
Let $T^{\mu\nu}(t,\vec x)$ denote the total energy-momentum tensor (sum of the matter and gravitational energy-momentum tensor) that depends on the space-time coordinates $(t,\vec x)$. We define the trace-reversed metric fluctuation by $e^{\mu\nu}(t', \vec x')$ which is produced at a space-time point $(t',\vec x')$ by the energy momentum tensor. We normalize $e^{\mu\nu}$ via:
\begin{equation}
 g_{\mu\nu}=\eta_{\mu\nu}+2h_{\mu\nu},\qquad
 e_{\mu\nu}=h_{\mu\nu}-\tfrac12\eta_{\mu\nu}\eta^{\rho\sigma}h_{\rho\sigma},
 \qquad
 h_{\mu\nu}=e_{\mu\nu}-\tfrac12\eta_{\mu\nu}\eta^{\rho\sigma}e_{\rho\sigma}.
 \label{r2:metric}
\end{equation}
All indices are raised and lowered with the Minkowski metric $\eta$. With retarded boundary conditions in de Donder gauge, the Green function is
\begin{equation}
 G(x,x')=-\frac{1}{4\pi|\vec x-\vec x'|} \delta(t-t'-|\vec x-\vec x'|).
\end{equation}
Using the retarded Green's function one can further invert $ \Box e^{\mu\nu} = - 8 \pi G T^{\mu\nu}$ to obtain the field at some distant detector located at a distance $R$
\begin{equation}
 e^{\mu\nu}(t',\vec x')=2G\int d^3x\,\frac{T^{\mu\nu}(t(\vec x),\vec x)}{|\vec x'-\vec x|},
\label{r2:green}
\end{equation}
where the retarded Green's function imposes the constraint 
\begin{equation}
 (t'-t(\vec x))^2=(\vec x'-\vec x)^2,\qquad t'>t(\vec x).
\end{equation}
For $\vec x'=R\nn$, $\nn^2=1$, and $t'=R+u$, this becomes
\begin{equation}
 (R+u-t(\vec x))^2=R^2-2R\nn\cdot\vec x+|\vec x|^2.
\end{equation}
Thus $t(\vec x)=u+\nn\cdot\vec x+\Ord(R^{-1})$. Keeping only the leading $1/R$ term gives
\begin{equation}
 e^{\mu\nu}(t',\vec{x}')=\frac{2G}{R}\int d^3x\, T^{\mu\nu}(u+\nn\cdot\vec x,\vec x).
 \label{r2:retarded}
\end{equation}
Here and below $T^{\mu\nu}$ is the symmetric conserved total source, including both matter and gravitational contributions. In \cite{Karan:2026jzs}, this waveform is computed at Newtonian limit. In the following we will focus on deriving the subleading correction to the Newtonian waveform. At the order considered here, the entire $O(\alpha^3)$ correction
arises from the first-order expansion in $\vec{n}\cdot\vec{x}$
of the retarded-time argument of $T^{\mu \nu}$. Corrections to the
Newtonian source and trajectories contribute at higher order. Thus, we write
\begin{align}
 e^{ij}(t', \vec{x}')&=\frac{2G}{R}\int d^3x\,T^{ij}(u,\vec x)+ \frac{2G}{R}n_k \frac{d}{d u} \int d^3x\,x^k T^{ij}(u,\vec x) + \mathcal{E}_1^{ij},
 \label{r2:E1split}
\end{align}
where $\mathcal{E}_1^{ij}$ captures the higher order terms due to Taylor expanding $T^{ij}(u+\vec{n}\cdot \vec{x})$. In appendix \ref{Error1} we argue that the error $\mathcal{E}_1^{ij}$ contributes to the correction of the Newtonian waveform at higher order than what we intend to derive. In \cite{Karan:2026jzs}, it is shown that the first term in eq.\eqref{r2:E1split} produces the waveform in the Newtonian limit. So we will only focus on the next term that is believed to produce the waveform at the subleading order correcting the Newtonian result 
\begin{equation}\label{eij 1}
    e^{ij}_{(1)}(t', \vec{x}') = \frac{2G}{R}n_k\frac{d}{d u} \mathcal{S}^{kij},~~ \mathcal{S}^{kij} = \int d^3x\,x^kT^{ij}(u,\vec x).
\end{equation}
$\mathcal{S}^{kij}$ is defined to be the stress current and is symmetric in its last two indices, while the symmetry is not manifest in all three indices. Therefore one could decompose it in a way that there is a part which is fully symmetric in all three of its indices and the rest of it does not obey the symmetry 
\begin{equation}\label{stress current}
    \mathcal{S}^{kij} = \frac{1}{3} \left( \mathcal{S}^{kij} + \mathcal{S}^{ikj} + \mathcal{S}^{jki} \right) + \frac{1}{3} \Big[ \left( \mathcal{S}^{kij} - \mathcal{S}^{ikj} \right) + \left( \mathcal{S}^{kij} - \mathcal{S}^{jki} \right)  \Big].
\end{equation}
In the following we show that the fully symmetric part of $\mathcal{S}^{kij}$ can be written in terms of the so-called mass octupole, $\mathcal{I}^{kij}$. And the remaining part can be written in terms of the first spatial moment of the angular-momentum density, $\mathcal{J}^{ki\vert j}$ defined as \footnote{These integrals are actually divergent and hence one needs to understand them by introducing a large spherical cutoff. See the discussion below eq.\eqref{A.8} for details.}
\begin{equation}\label{moments}
 \mathcal{I}^{kij} = \int d^3x\,x^k x^i x^j T^{00}(u, \vec{x}),~~~~\mathcal{J}^{ki\vert j} = \int\dd^3x\,x^j\left( x^kT^{0i} (u, \vec{x})-x^iT^{0k}(u, \vec{x})\right).
\end{equation}
The moments in eq. \eqref{moments} need not converge individually. We introduce a large-distance cutoff and take the required time derivatives before removing it. Appendix \ref{Error1} shows that this prescription gives finite differentiated moments and justifies the integrations by parts used below.\\
From the conservation of the total energy-momentum tensor\footnote{In particular, $\partial_t T^{00} = - \partial_i T^{i0}$ and $\partial_t T^{i0} = -\partial_j T^{ji}$. Differentiating the first relation and using the second yields $\partial_t^2 T^{00} = \partial_i \partial_j T^{ij}$.}, namely $\partial_\mu T^{\mu \nu}=0$ we can write the mass octupole moment as follows
\begin{equation}\label{relation I with S}
\begin{split}
     \partial_u^2 \mathcal{I}^{kij} = \int d^3x\, x^k x^i x^j \partial_l \partial_m T^{lm} = & 2 \int d^3x\, \left( x^j T^{ik} + x^i T^{jk} + x^k T^{ij} \right)
    \\
    = & 2 \left( \mathcal{S}^{jik} + \mathcal{S}^{ijk} + \mathcal{S}^{kij}  \right).
\end{split}
\end{equation}
It shows that the fully symmetric part of the stress current $\mathcal{S}^{kij}$ in eq.\eqref{stress current} can be written in terms of the mass octupole. Note that to write the second equality we have used integration by parts under the assumption that the boundary term vanishes at infinity for large $\vert \vec{x} \vert$. In principle, one could show that $\partial_i \partial_j T^{ij}(t, \vec{x})$ falls off as $1/r^6$ and hence vanish.\\
To obtain the remaining part of $\mathcal{S}^{kij}$ we use the conservation law $\partial_t T^{i0} = -\partial_j T^{ji}$ and compute its moment as follows
\begin{equation}\label{Ti0 moment}
    \partial_u \int d^3x\, x^j x^k T^{i0} = - \int d^3x\, x^j x^k \partial_m T^{mi} = \int d^3 x\, \left( x^k T^{ji} + x^j T^{ki} \right).
\end{equation}
To write the second equality we have used integration by parts and dropping the boundary terms. Now interchanging the indices $i \leftrightarrow k$ and then subtracting the resulting equation from eq.\eqref{Ti0 moment} we obtain
\begin{equation}\label{del u of Ti0 moment}
\begin{split}
    \partial_u \int d^3x\, x^j \left( x^k T^{i0} - x^i T^{k0} \right) = \mathcal{S}^{kji} - \mathcal{S}^{ijk} \implies \partial_u \mathcal{J}^{ki \vert j} = \mathcal{S}^{kji} - \mathcal{S}^{ijk}.
\end{split}
\end{equation}
By interchanging $i \leftrightarrow j$ in eq.\eqref{del u of Ti0 moment} and adding the resulting expression with it yields 
\begin{equation}\label{relation J with S}
    \partial_u \left( \mathcal{J}^{ki \vert j} +  \mathcal{J}^{kj \vert i} \right) = \left( \mathcal{S}^{kji} - \mathcal{S}^{ijk} \right) + \left( \mathcal{S}^{kji} - \mathcal{S}^{jik} \right).
\end{equation}
This relates the remaining part of the stress current in eq.\eqref{stress current} with the time derivative of the spatial moment of the angular momentum density. \\
Now using the eq.\eqref{relation I with S} and \eqref{relation J with S} in eq.\eqref{stress current} we obtain 
\begin{equation}
    \mathcal{S}^{kij} = \frac{1}{6} \partial_u^2 \mathcal{I}^{kij} + \frac{1}{3} \partial_u \left( \mathcal{J}^{ki \vert j} +  \mathcal{J}^{kj \vert i} \right).
\end{equation}
Therefore the leading correction to the Newtonian waveform can be written as
\begin{equation}\label{eij 1 trajectories}
    e^{ij}_{(1)} = \frac{G}{R} \frac{d^2}{du^2} \left[ \frac{1}{3} \frac{d}{du} \left( n_k \mathcal{I}^{kij} \right) + \frac{2}{3} n_k \left( \mathcal{J}^{ki \vert j} +  \mathcal{J}^{kj \vert i} \right) \right].
\end{equation}
This expression reflects the fact that the waveform in the PN expansion of asymptotic scattering states to first sub-leading order is determined by the octupole and the spatial moment of the angular momentum density. To evaluate it, let $\vec{X}_a(t)$ be the trajectory of the $a$-th particle. With this, one can find $\mathcal{I}^{kij}$ and $\mathcal{J}^{ki \vert j}$ in terms of the mass and trajectories of the particles\footnote{From $T^{00}$ and $T^{0i}$ we only keep the terms that contribute to the order of $\alpha$ we are dealing with. In particular, in the expression of $\mathcal{I}^{kij}$ and $\mathcal{J}^{kj\vert i}$ we replace 
\begin{equation}\label{T00 and T0i}
    T^{00}_{\text{matter}} \simeq \sum_a m_a \delta^3 \left( \vec{x} - \vec{X}_a \right),~~T^{0i}_{\text{matter}} \simeq \sum_a m_a \dot{X}_a^i(u) \delta^3 \left( \vec{x} - \vec{X}_a \right).
\end{equation}
Of course, there will be corrections due to the velocity of the particles as well as the gravitational potential energy that we will ignore for now. Later we will argue that these corrections lead to waveform at higher order than what we are considering here.}
{\allowdisplaybreaks
\begin{align}
    & \frac{d}{du} \left( n_k \mathcal{I}^{kij} \right) = \sum_a m_a \left[ \left( \vec{n} \cdot \dot{\vec{X}}_a (u) \right)  X_a^i(u) X_a^j(u) + \left( \vec{n}\cdot \vec{X}_a(u) \right) \dot{X}_a^i(u) X_a^j(u) + \left( \vec{n}\cdot \vec{X}_a(u) \right) X_a^i(u) \dot{X}_a^j(u) \right] \notag
    \\
    & n_k \mathcal{J}^{ki \vert j} = \sum_a m_a \left[ \left( \vec{n}\cdot \vec{X}_a(u) \right) \dot{X}_a^i(u) X_a^j(u) -  \left( \vec{n} \cdot \dot{\vec{X}}_a(u) \right) X_a^i(u) X_a^j(u) \right]
\end{align}}
where the overdot denotes derivative with respect to $u$. \\
Substituting these expressions in eq.\eqref{eij 1 trajectories} we get 
\begin{equation}\label{eij 1 du2}
    e^{ij}_{(1)} = \frac{G}{R} \frac{d^2}{du^2} \sum_a m_a \left[ \left( \vec{n} \cdot \vec{X}_a(u) \right) \left( \dot{X}_a^i(u) X_a^j(u) + X_a^i(u) \dot{X}_a^j (u) \right) - \left( \vec{n} \cdot \dot{\vec{X}}_a (u)\right) X_a^i (u) X_a^j(u) \right] + \mathcal{E}^{ij}_2.
\end{equation}
In the appendix \eqref{Error2} we will argue that the error $\mathcal{E}^{ij}_2$, appearing due to the replacement given by eq.\eqref{T00 and T0i} to account for kinetic energies of the objects and the gravitational energy density, can be ignored at this order. For now we proceed by dropping the term $\mathcal{E}^{ij}_2$.\\

In \cite{Karan:2026jzs}, Newtonian approximation corresponds to the assumption that the asymptotic velocities of the incoming and the outgoing particles are small. Here also we use the same notion except that we explore the subleading correction to the waveform. We shall implement the non-relativistic limit by labelling these velocities as $\alpha v_a$ for a small parameter $\alpha$. We also define rescaled time variable $w$
\begin{equation}
    w = \alpha^3 u ,\qquad z = \frac{\ln w}{w}.
\end{equation}
Eq.\eqref{eij 1 du2} now can be written as follows
{\allowdisplaybreaks
\begin{align}\label{eij 1 all X(w)}
   e^{ij}_{(1)}= \frac{G\alpha^9}{R} & \sum_a m_a \Bigg[ 2 \left( \vec{n} \cdot \frac{d \vec{X}_a(w)}{dw} \right) \frac{d X_a^i(w)}{dw} \frac{d X_a^j(w)}{dw} + \left( \vec{n} \cdot \frac{d \vec{X}_a(w)}{dw} \right) \Bigg( X_a^i(w) \frac{d^2 X_a^j (w)}{dw^2}  \notag \\
   & + X_a^j (w) \frac{d^2 X_a^i(w)}{dw^2} \Bigg)- \left( \vec{n} \cdot \frac{d^2 \vec{X}_a(w)}{dw^2} \right) \left(X_a^i(w) \frac{ d X_a^j(w)}{d w} +X_a^j(w) \frac{ d X_a^i(w)}{dw} \right) \notag \\
   & +3 \left( \vec{n} \cdot \vec{X}_a(w) \right) \left(\frac{d X_a^i(w)}{dw} \frac{d^2 X_a^j(w)}{dw^2} + \frac{d X_a^j(w)}{dw} \frac{d^2 X_a^i(w)}{dw^2} \right) \notag \\
   &+ \left(\vec{n} \cdot \vec{X}_a(w) \right)\left( X_a^i (w) \frac{d^3 X_a^j(w)}{dw^3} + X_a^j(w) \frac{d^3 X_a^i(w)}{dw^3} \right) -\left( \vec{n} \cdot \frac{d^3 \vec{X}_a(w)}{dw^3} \right) X_a^i (w) X_a^j (w) \Bigg].
\end{align}}

We shall show in the appendix \eqref{Error1} and \eqref{Error2} that in this approximation $\mathcal{E}_1^{ij}$ and $\mathcal{E}^{ij}_2$ can be ignored and we will only focus on $e^{ij}_{(1)}$.\\
We begin with the ansatz that the relevant part of the trajectory\footnote{For a full relativistic trajectory, see \cite{Nadai:2026ruw}.} of the $a$-th particle takes the form \cite{Karan:2026jzs}:
\begin{equation}
 \Xa=\alpha^{-2}\left[ \Big{\lbrace} \vec{v}_a w +\ca wz+\vec b_a+\sum_{r\geq1}\vec d_{r,a}z^r +\Ord(w^{-1}) \Big{\rbrace} + \mathcal{O}(\alpha) \right],
 \qquad v_a,c_a,d_{r,a} \sim \mathcal{O}(1).
 \label{r2:ansatz}
\end{equation}
It can be shown that this kind of solution can be obtained from the equations of motion. Then we have the first three derivatives of the trajectory 
\begin{align}
 \frac{d \vec{X}_a}{d w}
  &=\alpha^{-2}\left[\vec{v}_a + \ca w^{-1} - w^{-1}\sum_{r\geq1}r\vec d_{r,a}z^r+\Ord(w^{-2})\right] + \mathcal{O}\left(\alpha^{-1}\right),
 \label{r2:first-derivative}\\
 \frac{d^2\Xa}{d w^2}
  &=\alpha^{-2}w^{-2}\left[-\ca +\sum_{r\geq1}r(r+1)\vec d_{r,a}z^r+\Ord(w^{-1})\right] + \mathcal{O}\left(\alpha^{-1}\right),
 \label{r2:second-derivative}\\
 \frac{d^3\Xa}{d w^3}
  &=\alpha^{-2}w^{-3}\left[2\ca -\sum_{r\geq1}r(r+1)(r+2)\vec d_{r,a}z^r+\Ord(w^{-1})\right] + \mathcal{O}\left(\alpha^{-1}\right).
 \label{r2:third-derivative}
\end{align}
We now substitute these expressions in eq.\eqref{eij 1 all X(w)} to obtain the waveform at order $\alpha^3$. Note that according to the strategy mentioned above, after substituting all the expressions we will collect terms at order $w^0, w^{-1}$ and $w^{-1}z^r$ where $r \geq 1$. Therefore we get the following expression 
{\allowdisplaybreaks
\begin{align}\label{eij 1 after subs X}
  e^{ij}_{(1)}(w)= & \frac{G \alpha^3}{R} \sum_a m_a\Bigg\{ 2(\nn \cdot \vec{v}_a)v_a^iv_a^j +\frac2w\nn\cdot\left(\ca-\sum_{r\geq1}r\vec d_{r,a}z^r\right)v_a^iv_a^j + \frac{2\nn\cdot\vec{v}_a}{w}\Bigg[ v_a^i\left(c_a^j-\sum_{r\geq1}r d_{r,a}^jz^r\right) \notag
  \\
  & + \left( i \leftrightarrow j \right) \Bigg] +\frac{\nn\cdot\vec{v}_a}{w}\Bigg[ (v_a^i+c_a^iz)\left(-c_a^j+\sum_{r\geq1}r(r+1)d_{r,a}^jz^r\right) + \left( i \leftrightarrow j \right) \Bigg] \notag
  \\
  & -\frac1w\nn\cdot\left(-\ca+\sum_{r\geq1}r(r+1)\vec d_{r,a}z^r\right) \Big[ (v_a^i+c_a^iz)v_a^j+ \left( i \leftrightarrow j \right) \Big] \notag
 \\
  & +\frac{3\nn\cdot(\vec{v}_a+\ca z)}w\Bigg[ v_a^i\left(-c_a^j+\sum_{r\geq1}r(r+1)d_{r,a}^jz^r\right) +  \left( i \leftrightarrow j \right) \Bigg] \notag
  \\
  & +\frac{\nn\cdot(\vec{v}_a+\ca z)}w\Bigg[ (v_a^i+c_a^iz)\left(2c_a^j-\sum_{r\geq1}r(r+1)(r+2)d_{r,a}^jz^r\right) +  \left( i \leftrightarrow j \right) \Bigg] \notag
  \\
  & -\frac1w\nn\cdot\left(2\ca-\sum_{r\geq1}r(r+1)(r+2)\vec d_{r,a}z^r\right) (v_a^i+c_a^iz)(v_a^j+c_a^jz)\Bigg\}. 
\end{align}}
By substituting $z = \ln w/w$ we can put the expression in the following form where the first term represents the constant value of the waveform and the remaining term produces the coefficients of $(\ln w)^{r-1}/w^r$ for $r \geq 1$
{\allowdisplaybreaks
\begin{align}\label{eij 1 multiparticle w}
    e^{ij}_{(1)}(w)= & \frac{2G\alpha^3}{R}\sum_a m_a(\nn\cdot\vec{v}_a)v_a^iv_a^j + \frac{G\alpha^3}{R} \sum_{r=1}^{\infty}\frac{(\ln w)^{r-1}}{w^r} \sum_a m_a \Bigg[2\delta_{r1}(\nn\cdot\ca)v_a^iv_a^j +2\delta_{r2}\Big\{ (\nn\cdot\vec{v}_a)c_a^ic_a^j \notag 
    \\
    & -(\nn\cdot\ca)(v_a^ic_a^j+v_a^jc_a^i)\Big\} +2\delta_{r3}(\nn\cdot\ca)c_a^ic_a^j +(r-1)(r-2)\Bigg\{ (r+1)(\nn\cdot\vec d_{r-1,a})v_a^iv_a^j \notag \\
    & -(r-1)(\nn\cdot\vec{v}_a) \Big(v_a^id_{r-1,a}^j+v_a^jd_{r-1,a}^i +c_a^id_{r-2,a}^j+c_a^jd_{r-2,a}^i\Big) +(r-1)(\nn\cdot\vec d_{r-2,a}) (v_a^ic_a^j+v_a^jc_a^i) \notag \\
   &  -(r-3)(\nn\cdot\ca) \Big(v_a^id_{r-2,a}^j+v_a^jd_{r-2,a}^i +c_a^id_{r-3,a}^j+c_a^jd_{r-3,a}^i\Big)  +(r-3)(\nn\cdot\vec d_{r-3,a})c_a^ic_a^j \Bigg\}\Bigg].
\end{align}}
All quantities inside the square bracket are independent of $w$. The three Kronecker-delta terms account for the low-order contributions involving only $\vec{v}_a$ and $\ca$; the remaining terms involve the contributions from the trajectory coefficients $\vec d_{r-1,a}$, $\vec d_{r-2,a}$ and $\vec d_{r-3,a}$.\\
We now solve for the expressions of the trajectory coefficients $\vec{c}_a$ and $\vec{d}_{r,a}$ using the rescaled force equation
\begin{equation}
 m_a\frac{d^2\Xa}{d w^2} =\alpha^{-6}m_a\frac{d^2\Xa}{d u^2}  =-\alpha^{-6}G\sum_{b\ne a}m_am_b \frac{\Xa-\vec X_b}{|\Xa-\vec X_b|^3}+\mathcal{E}_3.
 \label{r2:E3}
\end{equation}
where $\mathcal{E}_3$ is an error term that takes into account the effect of additional forces other than just the Newtonian gravitational force. We show in appendix \eqref{Error3} that $\mathcal{E}_3$ does not contribute to the desired order. \\
One needs to now substitute eq.\eqref{r2:ansatz} in RHS and \eqref{r2:second-derivative} the LHS of the force equation. From eq.\eqref{r2:second-derivative} we see that for fixed $z$ every tail enters at the same order $w^{-2}$ as $\vec c_a$. In the separation $\vec X_a - \vec X_b$, by contrast, $\vec b_a$ and the tails are a relative $\mathcal{O}(w^{-1})$ correction, so they affect the force only at $\mathcal{O}(\alpha^{-2} w^{-3})$. Therefore dropping this contribution and $\mathcal{E}_3$ gives 
\begin{align}
 \alpha^{-2}w^{-2}
  \left[-\ca+\sum_{r\geq1}r(r+1)\vec d_{r,a}z^r\right] =-\alpha^{-2}G w^{-2}\sum_{b\ne a}m_b
   \frac{\vec{v}_a-\vec v_b+(\ca-\vec c_b)z}  {|\vec{v}_a-\vec v_b+(\ca-\vec c_b)z|^3}
       +\Ord(\alpha^{-2}w^{-3}).
 \label{r2:force-sub}
\end{align}
By taking a Taylor expansion of the denominator\footnote{First one could take a Taylor series expansion of the denominator using 
\begin{equation}
 \frac{1}{|\vec v_{ab}+\vec c_{ab}z|^3}
=\frac{1}{|\vec v_{ab}|^3}\sum_{k\ge0}\sum_{j=0}^{k}
\frac{(-1)^k(2k+1)!}{2^{k+j}\,k!\,j!\,(k-j)!}\,
\frac{(\vec v_{ab}\cdot\vec c_{ab})^{k-j}\,|\vec c_{ab}|^{2j}}{|\vec v_{ab}|^{2k}}\;z^{k+j}.
\end{equation}
Multiplying the double sum by the numerator $\vec v_{ab}+\vec c_{ab}z$ of eq \eqref{r2:force-sub}, the $z^r$ coefficient receives two contributions: from
$\vec v_{ab}$, the terms with $k+j=r$, and from $\vec c_{ab}z$, the terms
with $k+j=r-1$. In the first we set $j=r-k$, and the condition
$0\le j\le k$ restricts the sum to $[(r+1)/2]\le k\le r$. In the second we
set $j=r-1-k$, which gives $[r/2]\le k\le r-1$.} in the RHS and comparing the coefficients of $z^0$ and $z^r$ we obtain the following trajectory coefficients \cite{Karan:2026jzs}
\begin{align}
 \ca=G\sum_{b\ne a}m_b\frac{\vec{v}_a-\vec v_b}{|\vec{v}_a-\vec v_b|^3}.
 \label{r2:c}
\end{align}
\allowdisplaybreaks{
\begin{align}
 \vec d_{r,a}=&-\frac{G}{2^r r(r+1)}\sum_{b\ne a} \frac{m_b}{|\vec{v}_a-\vec v_b|^3}\Bigg[
  (\vec{v}_a-\vec v_b)\sum_{k=[(r+1)/2]}^r \frac{(-1)^k(2k+1)!}{k!(2k-r)!(r-k)!} \notag
  \\
 &\hspace{10mm}\times \frac{[(\vec{v}_a-\vec v_b)\cdot(\ca-\vec c_b)]^{2k-r} |\ca-\vec c_b|^{2(r-k)}}{|\vec{v}_a-\vec v_b|^{2k}} \notag
 \\
 &+2(\ca-\vec c_b)\sum_{k=[r/2]}^{r-1} \frac{(-1)^k(2k+1)!}{k!(2k+1-r)!(r-k-1)!} \notag
 \\
 &\hspace{10mm}\times \frac{[(\vec{v}_a-\vec v_b)\cdot(\ca-\vec c_b)]^{2k+1-r} |\ca-\vec c_b|^{2(r-k-1)}}{|\vec{v}_a-\vec v_b|^{2k}}
 \Bigg],\qquad r\geq1.
 \label{r2:d}
\end{align}}
where $[s]$ denotes the largest integer not larger than $s$. \\
We now display the constant, $w^{-1}$ and $w^{-2}\ln w$ terms of the waveform that will be helpful in the next section when we will compare them with the known/conjectured result \cite{2407.04128}. Expanding the position-space expression eq.\eqref{eij 1 multiparticle w} yields
\begin{align}
 e^{ij}_{(1)}(w)=\frac{2G\alpha^3}{R}\sum_a m_a\Bigg[&  (\nn\cdot\vec{v}_a)v_a^iv_a^j +\frac{\nn\cdot\ca}{w}v_a^iv_a^j +\frac{\ln w}{w^2}\Big{\lbrace} (\nn\cdot\vec{v}_a)c_a^ic_a^j -(\nn\cdot\ca)(v_a^ic_a^j+v_a^jc_a^i)\Big{\rbrace}\Bigg].
 \label{r2:first-three}
\end{align}
A similar expression gives the waveform at large negative $w$, with $m_a$ and $\vec{v}_a$ replaced by $m'_a$ and $\vec{v}'_a$. One further difference arises, however. When $w$ is negative, the denominator of eq.~\eqref{r2:force-sub} gives $|w|^3 = -w^3$ instead of $w^3$. This sign can be absorbed by flipping the signs of all the masses in the computation of the trajectories. Consequently, in the
trajectories $m_a$ must be replaced by $-m'_a$ rather than by $m'_a$; for instance, eq.~\eqref{r2:c} picks up an additional minus sign for the incoming particles.

Several results in the scattering events of relativistic particles are known in terms of the Fourier transformation of the waveform, so we derive these as well. We define the Fourier transform of $e^{ij}$ with respect to time by \cite{Karan:2026jzs}
\begin{equation}
 \widetilde e^{ij}(\nu)=\alpha^{-3}\int_{-\infty}^{\infty} d w\, e^{i\nu w}e^{ij}(w),\qquad
 e^{ij}(w)=\alpha^3\int_{-\infty}^{\infty}\frac{d\nu}{2\pi} e^{-i\nu w}\widetilde e^{ij}(\nu)
 \label{r2:Fourier}
\end{equation}
Note that due to the rescaling of the retarded time $w=\alpha^3 u$, $\nu$ is related to the variable $\omega$ conjugate to $u$ via $\nu = \omega/\alpha^3$.\\

Using Fourier transformation identities \cite{Karan:2026jzs} we can write the waveform both for large positive $w$ (from eq.\eqref{r2:first-three}) and large negative $w$ as follows 
{\allowdisplaybreaks
\begin{align}\label{eij 1 multiparticle w Fr}
    & \tilde{e}^{ij}_{(1)}(\nu)= \frac{2Gi}{R}\sum_a \frac{m_a(\nn\cdot\vec{v}_a)v_a^iv_a^j}{\nu + i \epsilon} + \frac{G}{R} \sum_{r=1}^{\infty} \frac{(-1)^r}{r!}i^{r-1} \nu^{r-1} [\ln ( \nu +i \epsilon)]^r \sum_a m_a \Bigg[2\delta_{r1}(\nn\cdot\ca)v_a^iv_a^j  \notag 
    \\
    & +2\delta_{r2}\Big\{ (\nn\cdot\vec{v}_a)c_a^ic_a^j -(\nn\cdot\ca)(v_a^ic_a^j+v_a^jc_a^i)\Big\} +2\delta_{r3}(\nn\cdot\ca)c_a^ic_a^j +(r-1)(r-2)\Bigg\{ (r+1)(\nn\cdot\vec d_{r-1,a})v_a^iv_a^j \notag \\
    & -(r-1)(\nn\cdot\vec{v}_a) \Big(v_a^id_{r-1,a}^j+v_a^jd_{r-1,a}^i +c_a^id_{r-2,a}^j+c_a^jd_{r-2,a}^i\Big) +(r-1)(\nn\cdot\vec d_{r-2,a}) (v_a^ic_a^j+v_a^jc_a^i) \notag \\
   &  -(r-3)(\nn\cdot\ca) \Big(v_a^id_{r-2,a}^j+v_a^jd_{r-2,a}^i +c_a^id_{r-3,a}^j+c_a^jd_{r-3,a}^i\Big)  +(r-3)(\nn\cdot\vec d_{r-3,a})c_a^ic_a^j \Bigg\}\Bigg].
\end{align}}
And similarly following the replacement discussed below eq.\eqref{r2:first-three} one can write the waveform in the frequency space at early time, $w \to - \infty$.\\
Before concluding the section we will add a few remarks:  \\
At this PN order, we have shown that mass octupole and the angular momentum density solely contribute to the waveform. However it is worth mentioning that the error terms neglected here are quite important in the next PN order. In the appendices we have argued why the error terms $\mathcal{E}^{ij}_1, \mathcal{E}^{ij}_2$ and $\mathcal{E}_3$ can be ignored at this order.\\
Another important point to mention is that although the angular momentum density contributes to the waveform in the PN expansion at this order, there will be no contribution due to the intrinsic spin of the objects/particles which are scattering. In \cite{Kidder:1995zr}, the intrinsic spin term is claimed to have the form $X_a \mathbf{S}$ where the total spin $\mathbf{S}$ scales as $\mathcal{O}(1)$ and $X_a \sim \mathcal{O}(\alpha^{-2})$. Therefore, since the waveform contains two derivatives with respect to $u$, it also scales as $\mathcal{O}(\alpha^6)$. So the double time derivative acting on the intrinsic spin term contributes to $\mathcal{O}(\alpha^4)$ in the waveform. As a result we can ignore this contribution at this PN order. 

%%%%%%%%%%%%%%%%%%%%%%%%%%%%%%%%%%%%%%%%%%%%%%%%%%%%%%%%%%%%%%%%%%%%%%%%%%%%%%%%%%%%%%%%%%%%%%%%%%%%%%%%%%%%%%%%%%%
\section{Waveform for $2 \to 2$ scattering}\label{s3}
%%%%%%%%%%%%%%%%%%%%%%%%%%%%%%%%%%%%%%%%%%%%%%%%%%%%%%%%%%%%%%%%%%%%%%%%%%%%%%%%%%%%%%%%%%%%%%%%%%%%%%%%%%%%%%%%%%%
We specialize to the case when both the initial and final states contain two particles. To solve for the trajectory coefficients of two final state particles with asymptotic velocities $\alpha \vec{v}_1, \alpha \vec{v}_2$ and masses $m_1, m_2$ respectively, we substitute the following ansatz in the force equation \eqref{r2:force-sub}
\begin{equation}
 \vec v=\vec v_1-\vec v_2,\qquad v=|\vec v|>0,\qquad \vec c_a=c_a\vec v,\qquad \vec d_{r,a}=d_{r,a}\vec v, \qquad a=1,2.
 \label{r3:ansatz}
\end{equation}
The masses and rescaled asymptotic velocities are held fixed in the $\alpha$ expansion. As mentioned in section~\ref{s2} and also argued in Appendix \ref{Error3}, the Newtonian recursion in \eqref{r2:force-sub} suffices through order $\alpha^3$ in the waveform. Substitution of \eqref{r3:ansatz} into the force equation determines the trajectory coefficients as follows \cite{Karan:2026jzs}
\begin{equation}\label{c d for 2}
\begin{split}
    & c_1=\frac{m_2}{m_1+m_2}\gamma,\qquad c_2=-\frac{m_1}{m_1+m_2}\gamma, \qquad\gamma\equiv\frac{G(m_1+m_2)}{v^3} 
    \\
    & d_{r,a}=\frac{(-1)^{r+1}}{r}c_a\gamma^{r},\qquad r\geq1,\quad a=1,2.
\end{split}
\end{equation}
We need to substitute eq.\eqref{r3:ansatz} and \eqref{c d for 2} in eq.\eqref{eij 1 after subs X} to obtain the waveform for the two outgoing particles. We find the following expressions useful to derive the two-particle waveform for $w \to \infty$
\begin{align}
 \vec c_a-\sum_{r\geq1}r\vec d_{r,a}z^r &=\frac{c_a \vec v}{1+\gamma z},
 \label{r3:velocity-sum}\\
 -\vec c_a+\sum_{r\geq1}r(r+1)\vec d_{r,a}z^r  &=-\frac{c_a \vec v}{(1 + \gamma z)^2},
 \label{r3:acceleration-sum}\\
 2\vec c_a-\sum_{r\geq1}r(r+1)(r+2)\vec d_{r,a}z^r &=\frac{2c_a \vec v}{(1+\gamma z)^3}.
 \label{r3:direct-replacements}
\end{align}
Therefore using eq.\eqref{r3:velocity-sum}-\eqref{r3:direct-replacements} in eq.\eqref{eij 1 after subs X} we obtain 
\begin{align}
  e^{ij}_{(1)}(w)=\frac{2G\alpha^3}{R}\sum_{a=1}^2m_a\Bigg\{& (\vec n\cdot\vec v_a)v_a^iv_a^j +\frac{c_a(\vec n\cdot\vec v)}{w (1 +\gamma z)}v_a^iv_a^j + \frac{z}{w (1+\gamma z)^3}\Bigg[
 \frac{c_a(\vec n\cdot\vec v)}\gamma (\gamma v_a^i-c_a v^i)(\gamma v_a^j-c_a v^j)\notag
  \\
  &\qquad+\left(\vec n\cdot\vec v_a -\frac{c_a}{\gamma}\vec n\cdot\vec v\right)
 \Big\{\gamma^2z c_a(v_a^iv^j+v_a^jv^i) +c_a^2(1-\gamma z)v^iv^j\Big\}\Bigg]\Bigg\}.
 \label{r3:direct-regrouped}
\end{align}
Performing the summation over the two particles yields
\begin{align}
 e^{ij}_{(1)}(w)=\frac{2G\alpha^3}{R}\Bigg[& \sum_{a=1}^2m_a(\vec n\cdot\vec v_a)v_a^iv_a^j + \frac{m_1m_2\gamma}{(m_1+m_2)w} \Bigg\{\frac{\vec n\cdot\vec v}{2(1+\gamma z)} \bigl[(v_1^i+v_2^i)v^j+(v_1^j+v_2^j)v^i\bigr]\notag\\
 &\hspace{60mm} +\frac{\gamma z}{(1+\gamma z)^2} \frac{\vec n\cdot(m_1\vec v_1+m_2\vec v_2)}{m_1+m_2}v^iv^j  \Bigg\}\Bigg].
 \label{r3:delta-closed}
\end{align}
This expression should be thought of as a resummed waveform for 2 outgoing particles.\\ 
We can now use the following series where the expansion is in power of $z=\ln w/w$ 
\begin{equation}
    \frac{\gamma}{1 + \gamma z} = \sum_{r=1}^\infty (-1)^{r-1} \gamma^r z^{r-1}, \qquad \frac{\gamma^2 z}{(1 + \gamma z)^2} = -\sum_{r=1}^\infty (-1)^{r-1} (r-1) \gamma^r z^{r-1}
\end{equation}
so as to obtain the waveform
\begin{equation}
\begin{split}
  e^{ij}_{(1)}(w) =\frac{2G\alpha^3}{R} & \sum_{a=1}^2m_a(\vec n\cdot\vec v_a)v_a^iv_a^j + \frac{2G m_1m_2 \alpha^3}{R(m_1+m_2)} \sum_{r=1}^\infty (-1)^{r-1} \gamma^r \frac{(\ln w)^{r-1}}{w^r}
  \\
  & \times \Bigg\{\frac{\vec n\cdot\vec v}{2} \Big[(v_1^i+v_2^i)v^j+(v_1^j+v_2^j)v^i\Big] - (r-1) \frac{\vec n\cdot(m_1\vec v_1+m_2\vec v_2)}{m_1+m_2}v^iv^j  \Bigg\}
 \label{r3:delta-closed late}
\end{split}
\end{equation}
This is the late time waveform in the position space when the final state contains exactly two outgoing particles. The first term in the expression is $w$-independent and hence contributes to the gravitational memory while the remaining series over $r$ determines the leading logarithmic terms.\\

The expression reduces to a more simpler form in the center of mass frame. To analyze this explicitly, substitute the velocity of both the particles in this frame: $\vec{v}_1 = m_2 \vec{v}/(m_1+m_2)$ and $\vec{v}_2 = - m_1 \vec{v} / (m_1 + m_2)$ in the above expression. As a result the memory term becomes
\begin{equation}
    \sum_{a=1}^2m_a(\vec n\cdot\vec v_a)v_a^iv_a^j  = \frac{m_1 m_2 \delta m}{(m_1 +m_2)^2} (\vec{n} \cdot \vec{v}) v^i v^j ,~~~\delta m = m_2 - m_1.
\end{equation}
For this frame choice, the second term in the tail coefficient vanishes directly because of the spatial momentum conservation. And the  first term can be written in the following form
\begin{equation}
\begin{split}
    & \sum_{r=1}^\infty (-1)^{r-1} \gamma^r \frac{(\ln w)^{r-1}}{w^r} \times \Bigg\{\frac{\vec n\cdot\vec v}{2} \Big[(v_1^i+v_2^i)v^j+(v_1^j+v_2^j)v^i\Big] \Bigg\}
    \\
    = & \frac{\delta m (\vec{n} \cdot \vec{v})  v^iv^j}{m_1+m_2} \sum_{r=1}^\infty (-1)^{r-1} \gamma^r \frac{(\ln w)^{r-1}}{w^r}.
\end{split}
\end{equation}
So the waveform in the PN expansion to first sub-leading order order will be given by the following expression \footnote{We thank Alok Laddha for suggesting that we include the center of mass frame discussion in its present form.} 
\begin{equation}\label{2to2n eij CM}
    e^{ij}_{(1)}(w) \Big{\vert}_{\text{CM}} = \frac{2G \alpha^3}{R} \frac{m_1 m_2 \delta m}{(m_1 + m_2)^2} (\vec{n} \cdot \vec{v} ) v^i v^j \left[ 1 + \sum_{r=1}^\infty (-1)^{r-1} \gamma^r \frac{(\ln w)^{r-1}}{w^r} \right]
\end{equation}
Therefore, the waveform at this order of PN expansion is proportional to the mass difference which was also reported in \cite{Blanchet:1996pi}\footnote{See eq.3b and 4b where $0.5$ PN expansion of the waveform $h_+$ and $h_\times$ were derived for gravitational wave.}.

If the incoming state also consists of two particles with asymptotic velocities $\alpha \vec{v}_1', \alpha \vec{v}_2'$ and masses $m_1,' m_2'$ respectively, the waveform at early time (described by large negative $w$) can be given by the following expression
\begin{equation}
\begin{split}
  e^{ij}_{(1)}(w)=\frac{2G\alpha^3}{R} & \sum_{a=1}^2 m_a'(\vec n\cdot\vec v_a') {v_a'}^i {v_a'}^j + \frac{2G m_1' m_2' \alpha^3}{R(m_1' + m_2')} \sum_{r=1}^\infty (-1)^{r-1} (\gamma')^r \frac{(\ln \vert w \vert)^{r-1}}{w^r}
  \\
  & \times \Bigg\{\frac{\vec n\cdot\vec v'}{2} \Big[({v_1'}^i+{v_2'}^i){v'}^j+({v_1'}^j+{v_2'}^j) {v'}^i \Big] - (r-1) \frac{\vec n\cdot(m_1'\vec v_1' +m_2' \vec v_2')}{m_1'+m_2'} {v'}^i {v'}^j  \Bigg\}
 \label{r3:delta-closed early}
\end{split}
\end{equation}
where we define $\gamma'$ as
\begin{equation}
    \gamma' = - \frac{G(m_1'+m_2')}{{v'}^3}.
\end{equation}
The appearance of the extra minus sign in this formula can be understood following the same logic described below the eq.\eqref{r2:first-three}. \\

The waveforms both at early and late times can be written in the frequency space using the Fourier transformation identities described in \cite{Karan:2026jzs}
{\allowdisplaybreaks
\begin{align}\label{2 to 2 freq}
    & \tilde{e}^{ij}_{(1)}(\nu) = \frac{2iG}{R} \left(  \frac{1}{\nu + i \epsilon} \sum_{a=1}^2m_a(\vec n\cdot\vec v_a)v_a^iv_a^j - \frac{1}{\nu - i \epsilon} \sum_{a=1}^2 m_a'(\vec n\cdot\vec v_a') {v_a'}^i {v_a'}^j \right) \notag
    \\
    & \qquad - \frac{2G m_1m_2}{R(m_1+m_2)} \sum_{r=1}^\infty i^{r-1} \frac{\gamma^r}{r!} \nu^{r-1} \left[ \ln (\nu + i \epsilon) \right]^{r} \Bigg\{\frac{\vec n\cdot\vec v}{2} \Big[(v_1^i+v_2^i)v^j+(v_1^j+v_2^j)v^i\Big] \notag
    \\
    & \hspace{9 cm} - (r-1) \frac{\vec n\cdot(m_1\vec v_1+m_2\vec v_2)}{m_1+m_2}v^iv^j  \Bigg\}
    \notag \\
    & \qquad + \frac{2G m_1' m_2'}{R(m_1' + m_2')} \sum_{r=1}^\infty i^{r-1} \frac{(\gamma')^r}{r!} \nu^{r-1} \left[ \ln (\nu - i\epsilon) \right]^{r} \Bigg\{\frac{\vec n\cdot\vec v'}{2} \Big[({v_1'}^i+{v_2'}^i){v'}^j+({v_1'}^j+{v_2'}^j) {v'}^i \Big] \notag
    \\
    & \hspace{9 cm} - (r-1) \frac{\vec n\cdot(m_1'\vec v_1' +m_2' \vec v_2')}{m_1'+m_2'} {v'}^i {v'}^j  \Bigg\}
\end{align}}
Note that the waveform at this order vanishes for equal mass as a consequence of the fact that it is proportional to the mass difference. This can also be argued just by looking at the eq.\eqref{eij 1} along with the trajectory eq.\eqref{r2:ansatz}. In the CM frame and equal mass, the trajectories are related by $\vec{X}_1 = - \vec{X}_2$. The source is therefore invariant under the reflection $\vec{x} \to -\vec{x}$. Since $e^{ij}_{(1)}$ is just an odd moment of the reflection symmetric source, it must vanish.

%%%%%%%%%%%%%%%%%%%%%%%%%%%%%%%%%%%%%%%%%%%%%%%%%%%%%%%%%%%%%%%%%%%%%%%%%%%%%%%%%%%%%%%%%%%%%%%%%%%%%%%%%%%%%%%%%%%%%%%%%%%%%%%%%
\section{Comparison with the all-orders relativistic conjecture}\label{s4}
%%%%%%%%%%%%%%%%%%%%%%%%%%%%%%%%%%%%%%%%%%%%%%%%%%%%%%%%%%%%%%%%%%%%%%%%%%%%%%%%%%%%%%%%%%%%%%%%%%%%%%%%%%%%%%%%%%%%%%%%%%%%%%%%%
In \cite{Karan:2026jzs}, an all-orders expression for gravitational-wave tails was conjectured for relativistic particles. We test this conjecture by expanding its late-time waveform in the small-velocity parameter \(\alpha\), retaining the Newtonian contribution at order \(\alpha^2\) and the first subleading contribution at order \(\alpha^3\) in a PN expansion. We compare the resulting leading-logarithmic tails with those derived in Section 2. Our starting point is 
\begin{equation}
\begin{aligned}
\widetilde e^{\mu\nu}(\omega)
={}&-\exp\left\{
-2iG\omega\ln(\omega+i\epsilon)
\sum_a p_a\cdot n
\right\}
\sum_{k\geq0}\frac{1}{k!}
A_{(k)}^{\mu\nu}(-i\omega)^{k-1}
\bigl[\ln(\omega+i\epsilon)\bigr]^k
\end{aligned}
\label{eq:frequency-waveform}
\end{equation}
At the order of interest, the exponential factor does not contribute (see \cite{Karan:2026jzs} for details), so we drop it from the analysis. Performing the inverse Fourier transform, the waveform at fixed retarded time $u$ takes the form
\begin{equation}\label{eij in terms of u K}
\begin{split}
& e^{ij}\left(\frac{\ln u}{u} \right) = - A^{ij}{(0)} + \sum_{k \geq 1} A^{ij}_{(k)} \frac{(\ln u)^{k-1}}{u^k} = - A^{ij}{(0)} + \frac{1}{R u}\mathcal{K}^{ij}\left(\frac{\ln u}{u}\right)
\end{split}
\end{equation}
which can be written in terms of the scaled retarded time $w = \alpha^3 u$ upto leading-log accuracy
\begin{equation}
    e^{ij}\left(\alpha^3 z \right) =  - A^{ij}{(0)} + \frac{\alpha^3}{R w}\mathcal{K}^{ij}\left(\alpha^3 z \right)~~~~~\text{where}~~z = \frac{\ln w}{w}
\end{equation}
where $A^{ij}_{(0)}$ is the well-known memory term. The second equality in eq.\eqref{eij in terms of u K} follows from the conjecture of \cite{Karan:2026jzs}, in which
\begin{equation}
\mathcal{K}^{ij}(\tilde{z}) = \mathcal{K}^{ij}_X(\tilde{z}) + \mathcal{K}^{ij}_H(\tilde{z}).
\end{equation}
Here $\mathcal{K}^{ij}_X$ and $\mathcal{K}^{ij}_H$ are the contributions from the matter energy-momentum tensor and the gravitational energy-momentum tensor respectively. To state their expressions we first mention the common covariant inputs  
\allowdisplaybreaks{
\begin{align}
& D_{ab} =\big[(p_a\cdot p_b)^2-p_a^2p_b^2\big]^{1/2}, \qquad \gamma_{ab} =\frac{(p_a\cdot p_b) \big[\frac32p_a^2p_b^2-(p_a\cdot p_b)^2\big]} {D_{ab}^3} =\frac{\sigma_{ab}(\sigma_{ab}^2-\frac32)}{(\sigma_{ab}^2-1)^{3/2}}, ~~~\sigma_{ab}=-\frac{p_a\cdot p_b}{m_am_b}, \notag \\
& C_a^\mu  =\sum_{b\ne a}\Big[ 2G\gamma_{ab}p_b^\mu -\frac{Gp_a^2(p_b^2)^2}{D_{ab}^3}p_a^\mu \Big], \qquad \mathcal{D}_b(q) =\big[(p_b\cdot q)^2-p_b^2q^2\big]^{1/2} \notag \\
& A_{b,\mu\nu} =p_{b\mu}p_{b\nu}-\tfrac12p_b^2\eta_{\mu\nu}, \qquad (A_b\cdot p_a)^\mu
=(p_a\cdot p_b)p_b^\mu-\tfrac12p_b^2p_a^\mu, \qquad p_a\cdot A_b\cdot p_a
=(p_a\cdot p_b)^2-\tfrac12p_b^2p_a^2.
\end{align}}
For the matter trajectories introduce the formal argument $\check z$ and the functions
\begin{equation}\label{F U}
\begin{aligned}
& ~~~~~~~~~~~~~~~~~~~~~~~~~~~~~~~~~~ q_{ab}^\mu(\check z)=p_a^\mu+\check z(C_a^\mu-C_b^\mu),\\
& \mathcal{F}_{ab}^\mu(\check z)
=\frac{2G}{\mathcal{D}_b(q_{ab})^3}\Big\{ 2(A_b\cdot p_a)^\mu
\big[(p_b\cdot q_{ab})(p_a\cdot p_b) -p_b^2(p_a\cdot q_{ab})\big] -(p_a\cdot A_b\cdot p_a)
\big[(p_b\cdot q_{ab})p_b^\mu-p_b^2q_{ab}^\mu\big]
\Big\}.\\
& ~~~~~~~~~~~~~~~~~~~~~~~~~~~~ \mathcal{F}_a^\mu(\check z)
=\sum_{b\ne a} \mathcal{F}_{ab}^\mu(\check z) =\sum_{r\ge0} \mathcal{F}_{a,r}^\mu\check z^r, \qquad \mathcal{F}_a^\mu(0)=-C_a^\mu \\
& ~~~~~~~~~~~~~~~~~~~~~~~~~~~~ \mathcal{U}_a^\mu(\check z) =-\int_0^1 \mathcal{F}_a^\mu(s\check z)\,ds
=-\sum_{r\ge0}\frac{\mathcal{F}_{a,r}^\mu}{r+1}\check z^r.
\end{aligned}
\end{equation}
The matter contribution is then
\begin{equation}\label{KX}
\begin{aligned}
\mathcal{K}_X^{ij}(\tilde z)
&=2G\sum_a\frac{1}{1+\tilde z\,n\cdot C_a} \left[ p_a^i \mathcal{U}_a^j(\zeta_a(\tilde{z}))
+p_a^j \mathcal{U}_a^i(\zeta_a(\tilde{z})) +\frac{p_a^ip_a^j}{d_a}\,n\cdot \mathcal{U}_a(\zeta_a(\tilde{z})) \right], \\
d_a &=-n\cdot p_a, \qquad \zeta_a(\tilde z) =\frac{\tilde z\,d_a} {1+\tilde z\,n\cdot C_a}.
\end{aligned}
\end{equation}
Since each power of the formal variable $\tilde{z}$ generates an extra factor of $\ln u/u$ in a tail, with the rescaled retarded time $w=\alpha^3 u$, we should evaluate $\mathcal{K}_X^{ij}(\tilde z)$ at $\tilde{z}=\alpha^3 z$ where $z= \ln w/w$. This is useful because the derived waveform is written in terms of the scaled retarded time $w$.
Also one should be careful with the notation $d_a$ since this is different from the trajectory coefficients $\vec{d}_{r,a}$.

We now state the gravitational contribution to the waveform. Let $e_1$ and $e_2$ be any fixed orthonormal spacelike basis
of the two-dimensional plane orthogonal to both momenta
$p_a$ and $p_b$ of the outgoing particles.
\begin{equation}\label{proj e}
\begin{gathered}
e_A\cdot e_B=\delta_{AB},\qquad
p_a\cdot e_A=p_b\cdot e_A=0,\qquad A,B=1,2,\\
\Pi_{ab}^{\mu\nu}=e_1^\mu e_1^\nu+e_2^\mu e_2^\nu,\qquad
e_\phi^\mu=e_1^\mu\cos\phi+e_2^\mu\sin\phi, \qquad 0 \leq \phi \leq 2\pi,
\end{gathered}
\end{equation}
where $\Pi^{\mu\nu}_{ab}$ is the projector onto that plane and $e^\mu_\phi$ is a unit vector.\\
Some further useful scalars and Lorentz vectors are defined as 
\begin{equation}\label{scalars and vectors of KH}
\begin{aligned}
& \varpi_{ab}^\mu =\frac{p_b\cdot n}{D_{ab}^2} \big[(p_a\cdot p_b)p_a^\mu-p_a^2p_b^\mu\big], \qquad p_a\cdot\varpi_{ab}=0,\qquad p_b\cdot\varpi_{ab}=p_b\cdot n, \\
& k_a^\mu(\chi)=e_\phi^\mu+\chi\varpi_{ab}^\mu,\qquad
k_b^\mu(\chi)=-e_\phi^\mu+\chi(n^\mu-\varpi_{ab}^\mu),\qquad k_a^\mu +k_b^\mu =\chi n^\mu, \\
& \beta_{ab}=n\cdot C_b+\varpi_{ab}\cdot(C_a-C_b),\qquad
b_\phi=e_\phi\cdot(C_a-C_b),\qquad \chi_\phi=-\frac{\tilde z\,b_\phi}{1+\tilde z\beta_{ab}}.
\end{aligned}
\end{equation}
We then define the gravitational contribution to the waveform as follows
\begin{equation}\label{G2}
\begin{aligned}
& \mathcal{K}_H^{\mu\nu}(\widetilde z) =-\sum_{a<b}\frac{16G^2}{D_{ab}}
 \left\langle \frac{\Phi_{ab}^{\mu\nu}(\chi_\phi;\phi)}{1+\widetilde z\beta_{ab}}
 \right\rangle,\qquad \langle f\rangle=\frac{1}{2\pi}\int_0^{2\pi}f(\phi)\,d\phi \\
& \Phi_{ab}^{\mu\nu}(\chi;\phi) =\frac{\mathcal{V}^{\mu\nu}[A_a,k_a(\chi);A_b,k_b(\chi)]}
 {[k_a(\chi)\cdot k_a(\chi)][k_b(\chi)\cdot k_b(\chi)]},
\end{aligned}
\end{equation}
where $\mathcal{V}^{\mu\nu}[A_a,k_a(\chi);A_b,k_b(\chi)]$ is defined as follows
\begin{equation}\label{G2 gr}
\begin{aligned}
\mathcal{G}_{(2)}^{\mu\nu}[H]\Big{\vert}_{\varepsilon_a\varepsilon_b}
&=\mathcal{V}^{\mu\nu}[A_a,k_a;A_b,k_b]e^{i(k_a+k_b)\cdot x}, \qquad H_{\mu\nu} =\varepsilon_a A_{a,\mu\nu}e^{ik_a\cdot x}
 +\varepsilon_b A_{b,\mu\nu}e^{ik_b\cdot x},
\end{aligned}
\end{equation}
where $H_{\mu\nu}$ is a superposition of two plane waves. $\mathcal{G}^{\mu\nu}_{(2)}$ is defined as follows
\begin{equation}\label{G2 R H}
\begin{aligned}
\mathcal{G}_{(2)}^{\mu\nu}[H]
={}&R_{(2)}^{\mu\nu}
 -H^{\mu\rho}R_{(1)\rho}{}^\nu
 -H^{\nu\rho}R_{(1)\rho}{}^\mu -\tfrac12\eta^{\mu\nu}
 \bigl(\operatorname{tr}R_{(2)}
       -H^{\rho\sigma}R_{(1)\rho\sigma}\bigr)
 +\tfrac12H^{\mu\nu}\operatorname{tr}R_{(1)}\\
&~~~~~~~~~~~~~~ +\tfrac12\operatorname{tr}H
 \bigl(R_{(1)}^{\mu\nu}
       -\tfrac12\eta^{\mu\nu}\operatorname{tr}R_{(1)}\bigr)
\end{aligned}
\end{equation}
where the Christoffel connections and Ricci tensors are defined as follows: 
\begin{equation}
\begin{split}
\Gamma_{(1)\sigma\mu\nu}
&=\tfrac12\bigl(
  \partial_\mu H_{\sigma\nu}
 +\partial_\nu H_{\sigma\mu}
 -\partial_\sigma H_{\mu\nu}\bigr),\\
\Gamma_{(1)\mu\nu}^{\rho}
&=\eta^{\rho\sigma}\Gamma_{(1)\sigma\mu\nu},\qquad
\Gamma_{(2)\mu\nu}^{\rho}
=-H^{\rho\sigma}\Gamma_{(1)\sigma\mu\nu},\\
R_{(1)\mu\nu}
&=\partial_\rho\Gamma_{(1)\mu\nu}^{\rho}
 -\partial_\nu\Gamma_{(1)\mu\rho}^{\rho},\\
R_{(2)\mu\nu}
&=\partial_\rho\Gamma_{(2)\mu\nu}^{\rho}
 -\partial_\nu\Gamma_{(2)\mu\rho}^{\rho}
 +\Gamma_{(1)\rho\lambda}^{\rho}\Gamma_{(1)\mu\nu}^{\lambda}
 -\Gamma_{(1)\nu\lambda}^{\rho}\Gamma_{(1)\mu\rho}^{\lambda}.
\end{split}
\end{equation}
where all traces are taken with the flat metric $\eta^{\mu\nu}$.

\subsection{Matter contribution to the all-order waveform}

We expand the matter contribution in eq.\eqref{KX} in the non-relativistic limit. The required components of \(F_{ab}^{\mu}(\check z)\) are
\begin{equation}\label{Fab i and 0}
\begin{aligned}
\mathcal{F}_{ab}^i(\check z)
&=-\frac{Gm_b}{\alpha^2} \frac{v_{ab}^i+\frac{\check z}{m_a\alpha^3}c_{ab}^i} {\left|\vec v_{ab}
+\frac{\check z}{m_a\alpha^3}\vec c_{ab}\right|^3} + \mathcal{O}(1),
\\
\mathcal{F}_{ab}^0(\check z) &=-\frac{Gm_b}{\alpha} \frac{(2\vec v_a-\vec v_b)\cdot
\left(\vec v_{ab} +\frac{\check z}{m_a\alpha^3}\vec c_{ab}\right)} {\left|\vec v_{ab}
+\frac{\check z}{m_a\alpha^3}\vec c_{ab}\right|^3} + \mathcal{O} (\alpha).
\end{aligned}
\end{equation}
One can now use these expressions to compute $\mathcal{U}_a^i (\check z)$ following the definition quoted in eq.\eqref{F U}
\begin{equation}\label{Ua as a function of check z}
\begin{aligned}
\mathcal{U}_a^i(\check z)
=-\int_0^1 \mathcal{F}_a^i(s\check z)\,ds & =\frac{G}{\alpha^2}\sum_{b\ne a}m_b
\int_0^1 \frac{v_{ab}^i+\frac{s\check z}{m_a\alpha^3}c_{ab}^i} {\left|\vec v_{ab}
+\frac{s\check z}{m_a\alpha^3}\vec c_{ab}\right|^3}\,ds+ \mathcal{O}(1)
\\
&=\frac{1}{\alpha^2}\int_0^1 \left[ c_a^i-\sum_{r\ge1}r(r+1)d_{r,a}^i
\left(\frac{s\check z}{m_a\alpha^3}\right)^r \right]ds+ \mathcal{O}(1)
\\
&=\frac{1}{\alpha^2} \left[ c_a^i-\sum_{r\ge1}r\,d_{r,a}^i \left(\frac{\check z}{m_a\alpha^3}\right)^r \right]+O(1).
\end{aligned}
\end{equation}
The second line follows from the equation of motion \eqref{r2:force-sub}; integrating over \(s\) gives the third.\\
Since in $\mathcal{K}_X^{ij}(\tilde z)$, $\mathcal{U}_a$ has an argument $\zeta_a (\tilde{z})$ we now compute the following quantity 
\begin{equation}\label{zeta a in alpha3z}
\begin{split}
\zeta_a (\alpha^3 z)  = \frac{\alpha^3 z\,d_a} {1+\alpha^3 z\,n\cdot C_a} = & m_a\alpha^3z \left[1-\alpha\vec n\cdot\vec v_a+ \mathcal{O}(\alpha^2)\right]
\left[1-\alpha z\,\vec n\cdot\vec c_a+ \mathcal{O}(\alpha^2)\right]
\\
= & m_a\alpha^3z \left[1-\alpha\vec n\cdot(\vec v_a+z\vec c_a)+ \mathcal{O}(\alpha^2) \right],
\end{split}
\end{equation}
where we have used the following expressions that go into the definition of $\zeta_a$ as given in eq.\eqref{KX}
\begin{equation}\label{3 fac}
\begin{aligned}
& d_a = p_a^0-\vec n\cdot\vec p_a =m_a\left[1-\alpha\vec n\cdot\vec v_a+ \mathcal{O}(\alpha^2)\right],
\\
& n\cdot C_a =-C_a^0+\vec n\cdot\vec C_a =\alpha^{-2}\vec n\cdot\vec c_a+ \mathcal{O}(\alpha^{-1}),
\\
& \frac{1}{1+\alpha^3z\,n\cdot C_a} =1-\alpha z\,\vec n\cdot\vec c_a+ \mathcal{O}(\alpha^2).
\end{aligned}
\end{equation}
Substituting eq.\eqref{zeta a in alpha3z} in eq.\eqref{Ua as a function of check z} we obtain 
{\allowdisplaybreaks
\begin{align} \label{U both pieces}
\mathcal{U}_a^i(\zeta_a(\alpha^3z))
&=\frac{1}{\alpha^2}
\left[
c_a^i-\sum_{r\ge1}r\,d_{r,a}^i
\left(\frac{\zeta_a(\alpha^3z)}{m_a\alpha^3}\right)^r
\right]+ \mathcal{O}(1) \notag
\\
&=\frac{1}{\alpha^2}
\Big[
c_a^i-\sum_{r\ge1}r\,d_{r,a}^iz^r
\left\{1-r\alpha\vec n\cdot(\vec v_a+z\vec c_a)\right\} \Big]+ \mathcal{O}(1) \notag
\\
&=\frac{1}{\alpha^2}
\Big[c_a^i-\sum_{r\ge1}r\,d_{r,a}^iz^r\Big] +\frac{1}{\alpha} \left[\vec n\cdot(\vec v_a+z\vec c_a)\right] \sum_{r\ge1}r^2d_{r,a}^iz^r+ \mathcal{O}(1).
\end{align}}
Note that to obtain the second line we have substituted the expression for $\zeta_a(\alpha^3 z)$ and taken a binomial expansion for small $\alpha$.\\
Substituting eq.\eqref{zeta a in alpha3z} and eq.\eqref{3 fac} into the first two terms of eq.\eqref{KX} gives
\begin{equation}\label{pU sym}
\begin{aligned}
&\frac{p_a^i \mathcal{U}_a^j(\zeta_a (\alpha^3 z))+p_a^j \mathcal{U}_a^i(\zeta_a(\alpha^3 z))}
{1+\alpha^3z\,n\cdot C_a}
\\
=&\frac{m_a}{\alpha} \Big[ v_a^i \Big(c_a^j-\sum_{r\ge1}r\,d_{r,a}^jz^r\Big)
+(i\leftrightarrow j) \Big] + m_a\left[\vec n\cdot(\vec v_a+z\vec c_a)\right]
\Big[ v_a^i\sum_{r\ge1}r^2d_{r,a}^jz^r +(i\leftrightarrow j) \Big]
\\
&\qquad \qquad ~~~~~~~~~~~~~~~~~ -m_a z(\vec n\cdot\vec c_a)
\Big[ v_a^i\Big(c_a^j-\sum_{r\ge1}r\,d_{r,a}^jz^r\Big)
+(i\leftrightarrow j) \Big]+ \mathcal{O}(\alpha).
\end{aligned}
\end{equation}
To manipulate the rightmost term in the second line of the above equation we use the following relation 
\begin{equation}
\sum_{r\ge1}r^2d_{r,a}^jz^r
= \Big(c_a^j-\sum_{r\ge1}r\,d_{r,a}^jz^r\Big) - \Big(c_a^j-\sum_{r\ge1}r(r+1)d_{r,a}^jz^r\Big).
\end{equation}
Using this expression we can see that there will be cancellation between a part of the rightmost term in the second line and the term in the third line of eq.\eqref{pU sym}. Then we get the following expression 
\begin{equation}\label{KX first term}
\begin{aligned}
&\frac{p_a^i \mathcal{U}_a^j(\zeta_a (\alpha^3 z))+p_a^j \mathcal{U}_a^i(\zeta_a(\alpha^3 z))}
{1+\alpha^3z\,n\cdot C_a}
\\
&=\frac{m_a}{\alpha}
\Big[ v_a^i\Big(c_a^j-\sum_{r\ge1}r\,d_{r,a}^jz^r\Big) +(i\leftrightarrow j)
\Big] + m_a(\vec n\cdot\vec v_a) \Big[ v_a^i\Big(c_a^j-\sum_{r\ge1}r\,d_{r,a}^jz^r\Big) +(i\leftrightarrow j) \Big]
\\
&\qquad ~~~~~~ -m_a\left[\vec n\cdot(\vec v_a+z\vec c_a)\right]
\Big[ v_a^i\Big(c_a^j-\sum_{r\ge1}r(r+1)d_{r,a}^jz^r\Big) +(i\leftrightarrow j) \Big]+ \mathcal{O}(\alpha).
\end{aligned}
\end{equation}
We now evaluate the third term in the expression for
$\mathcal K_X^{ij}$ in eq.\eqref{KX}. We need to derive $\mathcal{U}_a^0 (\zeta_a(\tilde{z}))$ using eq.\eqref{Fab i and 0} and plugging it in the definition \eqref{F U} we obtain 
{\allowdisplaybreaks
\begin{align}\label{Ua0 as a function of check z}
\mathcal{U}_a^0(\check z)
=-\int_0^1 \mathcal{F}_a^0(s\check z)\,ds & =\frac{G}{\alpha}\sum_{b\ne a}m_b\int_0^1
\frac{(2\vec v_a-\vec v_b)\cdot
\left(\vec v_{ab}
+\frac{s\check z}{m_a\alpha^3}\vec c_{ab}\right)}
{\left|\vec v_{ab}
+\frac{s\check z}{m_a\alpha^3}\vec c_{ab}\right|^3}
\,ds+ \mathcal{O}(\alpha) \notag
\\
&=\alpha\,\vec v_a\cdot \vec{\mathcal{U}}_a(\check z)
+\frac{G}{\alpha}\sum_{b\ne a}m_b\int_0^1
\frac{\vec v_{ab}\cdot
\left(\vec v_{ab}
+\frac{s\check z}{m_a\alpha^3}\vec c_{ab}\right)}
{\left|\vec v_{ab}
+\frac{s\check z}{m_a\alpha^3}\vec c_{ab}\right|^3}
\,ds+ \mathcal{O}(\alpha).
\end{align}}
The remaining integrand is a total derivative, so the integral depends on the boundary values
\begin{equation}
    \int_0^1
\frac{\vec{v}_{ab} \cdot \left(\vec v_{ab} +\frac{s\check z}{m_a\alpha^3}\vec c_{ab}\right)} {\left|\vec v_{ab} +\frac{s\check z}{m_a\alpha^3}\vec c_{ab}\right|^3}
\,ds  = \int_0^1 \frac{d}{ds} \left[  \frac{s}{\left|\vec v_{ab} +\frac{s\check z}{m_a\alpha^3}\vec c_{ab}\right|} \right]\,ds = \frac{1}{\left|\vec v_{ab} +\frac{\check z}{m_a\alpha^3}\vec c_{ab}\right|} 
\end{equation}
Substituting the integral in eq.\eqref{Ua0 as a function of check z} and using the expression for $\mathcal{U}_a^i(\check z)$ from eq.\eqref{Ua as a function of check z} yield the following expression for $\mathcal{U}_a^0(\check z)$
\begin{equation}
\mathcal{U}_a^0(\check z)
=\frac{1}{\alpha}
\left\{
\vec v_a\cdot
\left[
\vec c_a-\sum_{r\ge1}r\,\vec d_{r,a}
\left(\frac{\check z}{m_a\alpha^3}\right)^r
\right]
+G\sum_{b\ne a}
\frac{m_b}
{\left|\vec v_{ab}
+\frac{\check z}{m_a\alpha^3}\vec c_{ab}\right|}
\right\}
+ \mathcal{O}(\alpha).
\end{equation}
Taking the argument $\check z = \zeta_a(\alpha^3 z)$, keeping the leading order term in $\alpha$ we find the following expression for $\mathcal{U}_a^0(\zeta_a(\alpha^3 z))$ 
\begin{align} \label{Ua0 alpha}
& \mathcal{U}_a^0(\zeta_a(\alpha^3 z)) =\frac{1}{\alpha} \left\{ \vec v_a\cdot
\Big(\vec c_a-\sum_{r\ge1}r\,\vec d_{r,a}z^r\Big)
+G\sum_{b\ne a} \frac{m_b}{|\vec v_{ab}+z\vec c_{ab}|}
\right\}+ \mathcal{O}(1)
\end{align}
By comparing eq.\eqref{U both pieces} with eq.\eqref{Ua0 alpha} one can see that in the small $\alpha$ limit $\mathcal{U}_a^i$ dominates over $\mathcal{U}_a^0$. This observation will be useful in evaluating the following quantities where it is sufficient to keep the spatial term while computing $n\cdot \mathcal{U}_a$.\\
To find the third term in the definition of $\mathcal{K}_X^{ij}$ we therefore compute 
\begin{align}
    & n\cdot \mathcal{U}_a(\zeta_a(\alpha^3 z)) =- \mathcal{U}_a^0(\zeta_a(\alpha^3 z))+\vec n\cdot\vec{\mathcal{U}}_a(\zeta_a(\alpha^3 z))=\frac{1}{\alpha^2}\,\vec n\cdot
\Big(\vec c_a-\sum_{r\ge1}r\,\vec d_{r,a}z^r\Big) + \mathcal{O}(\alpha^{-1}),
\\
& ~~~~~~~~~~~~~~~~~~~~~~~~~~~~~~~~~ \frac{p_a^ip_a^j}{d_a} =\alpha^2m_av_a^iv_a^j+ \mathcal{O}(\alpha^3).
\end{align}
Then combining all the factors together we get 
\begin{equation}\label{KX second term}
\frac{1}{1+\alpha^3z\,n\cdot C_a}
\frac{p_a^ip_a^j}{d_a}\,n\cdot \mathcal{U}_a(\zeta_a(\alpha^3 z))
= m_a\,\vec n\cdot
\Big( \vec c_a-\sum_{r\ge1}r\,\vec d_{r,a}z^r \Big)
v_a^iv_a^j+ \mathcal{O}(\alpha).
\end{equation}
Summing up eq.\eqref{KX first term} and \eqref{KX second term} we substitute in eq.\eqref{KX} that yields  
\begin{equation}
\begin{aligned}
\mathcal{K}_X^{ij}(\alpha^3z)
={}&\frac{2G}{\alpha}\sum_a m_a \Big[
v_a^i\Big(c_a^j-\sum_{r\ge1}r\,d_{r,a}^jz^r\Big) +(i\leftrightarrow j) \Big]
\\
&+2G\sum_a m_a\Bigg\{ \vec n\cdot \Big( \vec c_a-\sum_{r\ge1}r\,\vec d_{r,a}z^r \Big)
v_a^iv_a^j +(\vec n\cdot\vec v_a) \Big[ v_a^i\Big(c_a^j-\sum_{r\ge1}r\,d_{r,a}^jz^r\Big) +(i\leftrightarrow j) \Big]
\\
&\qquad ~~~~~ -\left[\vec n\cdot(\vec v_a+z\vec c_a)\right]
\Big[ v_a^i\Big(c_a^j-\sum_{r\ge1}r(r+1)d_{r,a}^jz^r\Big) +(i\leftrightarrow j) \Big]
\Bigg\}+ \mathcal{O}(\alpha).
\end{aligned}
\end{equation}
Using this expression in eq.\eqref{eij in terms of u K} we can write down the waveform due to the matter contribution 
\begin{equation}\label{X waveform}
\begin{aligned}
e_X^{ij}(\alpha^3z)
={}&\frac{2G \alpha^2}{wR}\sum_a m_a \Big[
v_a^i\Big(c_a^j-\sum_{r\ge1}r\,d_{r,a}^jz^r\Big) +(i\leftrightarrow j) \Big]
\\
&+ \frac{2G\alpha^3}{wR} \sum_a m_a\Bigg\{ \vec n\cdot \Big( \vec c_a-\sum_{r\ge1}r\,\vec d_{r,a}z^r \Big)
v_a^iv_a^j +(\vec n\cdot\vec v_a) \Big[ v_a^i\Big(c_a^j-\sum_{r\ge1}r\,d_{r,a}^jz^r\Big) +(i\leftrightarrow j) \Big]
\\
&\qquad ~~~~~~~~~~~~~~~~~~~~ -\left[\vec n\cdot(\vec v_a+z\vec c_a)\right]
\Big[ v_a^i\Big(c_a^j-\sum_{r\ge1}r(r+1)d_{r,a}^jz^r\Big) +(i\leftrightarrow j) \Big]
\Bigg\}.
\end{aligned}
\end{equation}

\subsection{Gravity contribution to the all-order waveform}

We now expand the gravitational contribution \(K_H^{ij}\) to obtain the waveform through order \(\alpha^3\). We begin with the numerator \(\mathcal{V}^{ij}\) of \(\Phi_{ab}^{ij}\) which can be computed by following the definitions given in eq.\eqref{G2} and eq.\eqref{G2 gr}:
\begin{align}
\mathcal{V}^{ij}[A_a,k_a;A_b,k_b] = \mathcal{G}_{(2)}^{ij}[H]\Big{\vert}_{\varepsilon_a\varepsilon_b} e^{-i(k_a+k_b)\cdot x}.
\end{align}
One could directly use the formula for $\mathcal{G}_{(2)}^{ij}[H]$ following eq.\eqref{G2 R H} to obtain the expression for $\mathcal{V}^{ij}$. However, it would be easier to simplify the components of $H_{\mu\nu}$ just by power counting with respect to $\alpha$ so that we can already ignore various contributions to the Christoffel connections and the curvature tensors at higher order in $\alpha$.  To do that note the definition $A_{a,\mu\nu}=p_{a\mu}p_{a\nu}-\tfrac12p_a^2\eta_{\mu\nu}$,
together with $p_a^2=-m_a^2$ and $\vec p_a=\alpha p_a^0\vec v_a$, gives
\begin{equation}
\begin{aligned}
A_{a,00}
&=(p_a^0)^2-\tfrac12m_a^2
 =\tfrac12m_a^2+ \mathcal{O}(\alpha^2),\\
A_{a,ij}
&=p_a^ip_a^j+\tfrac12m_a^2\delta_{ij}
 =\tfrac12m_a^2\delta_{ij}+ \mathcal{O}(\alpha^2),\\
A_{a,0i}
&=-p_a^0p_a^i
 =-\alpha m_a^2v_a^i+ \mathcal{O}(\alpha^3).
\end{aligned}
\label{khv:amplitudes}
\end{equation}
The same formulas hold for $b$. A spatial component of \eqref{G2 R H} has no free temporal
index, and every contracted temporal index occurs twice. Consequently,
each term containing an off-diagonal amplitude $A_{0i}$ or a temporal
wave vector $k^0$ \footnote{\label{ka0 kb0} From eq.\eqref{proj e} and \eqref{scalars and vectors of KH} note that $p_a\cdot k_a=0$ and $p_b \cdot k_b=0$ that further lead to \begin{align}
k_a^0
&=\alpha\vec v_a\cdot\vec e_\phi
 +\chi_\phi\varpi_{ab}^0
 = \mathcal{O}(\alpha), \notag \\
k_b^0
&=-\alpha\vec v_a\cdot\vec e_\phi
 +\chi_\phi(1-\varpi_{ab}^0)
 = \mathcal{O}(\alpha) \notag.
\end{align}} contains an even total number of these factors.
Such terms therefore start at $A_{0i}A_{0j}$, $A_{0i}k^0$, or
$k^0k^0$, all of order $\alpha^2$. The diagonal-amplitude corrections
also start at $\alpha^2$. We may thus evaluate the numerator through
order $\alpha$ by making the substitutions
\begin{equation}
\begin{gathered}
H_{ij}=\delta_{ij}H_{00},\qquad
H_{0i}=0,\qquad
\partial_0H_{\mu\nu}=0,\qquad H_{00}
=\varepsilon_a\frac{m_a^2}{2}e^{i\vec k_a\cdot\vec x}
+\varepsilon_b\frac{m_b^2}{2}e^{i\vec k_b\cdot\vec x}.
\end{gathered}
\label{khv:field-substitution}
\end{equation}
Appendix~\ref{app:vertex} derives the spatial quadratic tensor
\eqref{khv:G2-spatial} from the connections and curvature tensors
of the field $H_{\mu\nu}$. We now substitute the two plane waves in \eqref{khv:field-substitution}. After removing the common factor $\varepsilon_a\varepsilon_b e^{i(\vec k_a+\vec k_b)\cdot\vec x}$, each term in \eqref{khv:G2-spatial} produces the following 
\begin{equation}
\begin{aligned}
\left.H_{00}\partial_i\partial_jH_{00}
\right|_{\varepsilon_a\varepsilon_b}
e^{-i(\vec k_a+\vec k_b)\cdot\vec x}
&=-\frac{m_a^2m_b^2}{4}
  (k_a^ik_a^j+k_b^ik_b^j),\\
\left.(\partial_iH_{00})(\partial_jH_{00})
\right|_{\varepsilon_a\varepsilon_b}
e^{-i(\vec k_a+\vec k_b)\cdot\vec x}
&=-\frac{m_a^2m_b^2}{4}
  (k_a^ik_b^j+k_b^ik_a^j),\\
\left.H_{00}\nabla^2H_{00}
\right|_{\varepsilon_a\varepsilon_b}
e^{-i(\vec k_a+\vec k_b)\cdot\vec x}
&=-\frac{m_a^2m_b^2}{4}
  (|\vec k_a|^2+|\vec k_b|^2),\\
\left.(\partial_kH_{00})(\partial_kH_{00})
\right|_{\varepsilon_a\varepsilon_b}
e^{-i(\vec k_a+\vec k_b)\cdot\vec x}
&=-\frac{m_a^2m_b^2}{2}\vec k_a\cdot\vec k_b.
\end{aligned}
\label{khv:bilinears}
\end{equation}
Thus at leading order in $\alpha$ we obtain 
{\allowdisplaybreaks
\begin{align}\label{khv:vertex-general}
\mathcal{V}^{ij}[A_a,k_a;A_b,k_b]
=\frac{m_a^2m_b^2}{4}\Bigl[
&-k_a^ik_a^j-k_b^ik_b^j
-\tfrac12(k_a^ik_b^j+k_b^ik_a^j)+\delta^{ij}\bigl(
 |\vec k_a|^2+|\vec k_b|^2
 +\tfrac32\vec k_a\cdot\vec k_b\bigr)\Bigr]
+ \mathcal{O}(\alpha^2).
\end{align}}
Using \(\vec k_a+\vec k_b=\chi_\phi\vec n\), we rewrite eq.\eqref{khv:vertex-general} in terms of the sum and difference of the wave vectors:
\begin{equation}
\begin{aligned}
& k_a^ik_a^j+k_b^ik_b^j =\tfrac12(k_a+k_b)^i(k_a+k_b)^j
 +\tfrac12(k_a-k_b)^i(k_a-k_b)^j,\\
& k_a^ik_b^j+k_b^ik_a^j =\tfrac12(k_a+k_b)^i(k_a+k_b)^j
 -\tfrac12(k_a-k_b)^i(k_a-k_b)^j,\\
& |\vec k_a|^2+|\vec k_b|^2+\tfrac32\vec k_a\cdot\vec k_b
=\tfrac78|\vec k_a+\vec k_b|^2
 +\tfrac18|\vec k_a-\vec k_b|^2 \simeq \tfrac18|\vec k_a-\vec k_b|^2 + \mathcal{O}(\alpha^2)
\end{aligned}
\label{khv:sum-difference}
\end{equation}
Note that in the last equation we have discarded the term $\vert \vec{k}_a +\vec{k}_b \vert^2$ because it is of order $\alpha^2$.\footnote{From eq.\eqref{scalars and vectors of KH} we know $k_a^\mu + k_b^\mu = \chi n^\mu$. At $\chi = \chi_\phi$ by taking non-relativistic limit we get 
\begin{align}
b_\phi
&=-e_\phi^0(C_a^0-C_b^0)
  +\vec e_\phi\cdot(\vec C_a-\vec C_b)
 =\alpha^{-2}\vec e_\phi\cdot\vec c_{ab}+ \mathcal{O}(1), \notag \\
\chi_\phi
&=-\frac{\alpha^3z\,b_\phi}{1+\alpha^3z\beta_{ab}}
 =-\frac{\alpha z\,\vec e_\phi\cdot\vec c_{ab}}
         {1+\alpha^3z\beta_{ab}}
 + \mathcal{O}(\alpha^3) \notag.
\end{align}
which implies $\vec{k}_a + \vec{k}_b \sim \mathcal{O}(\alpha)$.}
Substituting these into \eqref{khv:vertex-general} yields
\allowdisplaybreaks{
\begin{align}
 \mathcal{V}^{ij}[A_a,k_a;A_b,k_b]
=\frac{m_a^2m_b^2}{32}
 \Bigl[
 \delta^{ij}|\vec k_a-\vec k_b|^2
 -2(k_a-k_b)^i(k_a-k_b)^j
 \Bigr]
+ \mathcal{O}(\alpha^2).
\label{khv:vertex-final}
\end{align}}
This expression gives the numerator of $\Phi^{ij}_{ab}(\chi_\phi; \phi)$ as defined in eq.\eqref{G2}.
The denominator of $\Phi^{\ij}_{ab}$ can be evaluated as follows
\begin{equation}
\begin{aligned}
k_a\cdot k_a
&=\frac14\bigl[(k_a-k_b)+\chi_\phi n\bigr]^2
 =\frac14(k_a-k_b)^2
 +\frac{\chi_\phi}{2}n\cdot(k_a-k_b),\\
k_b\cdot k_b
&=\frac14\bigl[-(k_a-k_b)+\chi_\phi n\bigr]^2
 =\frac14(k_a-k_b)^2
 -\frac{\chi_\phi}{2}n\cdot(k_a-k_b).
\end{aligned}
\label{kh:denominator-factors}
\end{equation}
By multiplying the two terms $k_a \cdot k_a$ with $k_b \cdot k_b$ one can see that there will be cancellation between the cross terms. And the remaining terms give the expression as follows 
\begin{equation}
\begin{aligned}
(k_a\cdot k_a)(k_b\cdot k_b) =\frac1{16}|\vec k_a-\vec k_b|^4
 -\frac18|\vec k_a-\vec k_b|^2(k_a^0-k_b^0)^2 +\frac1{16}(k_a^0-k_b^0)^4
 -\frac{\chi_\phi^2}{4}\bigl[n\cdot(k_a-k_b)\bigr]^2.
\end{aligned}
\label{kh:denominator-exact}
\end{equation}
By footnote \ref{ka0 kb0}, one could show that the last three terms have orders $\alpha^2,\alpha^4,\alpha^2$ and hence we ignore them. Consequently,
\begin{equation}
(k_a\cdot k_a)(k_b\cdot k_b)
=\frac1{16}|\vec k_a-\vec k_b|^4+ \mathcal{O}(\alpha^2).
\label{kh:denominator}
\end{equation}
Dividing \eqref{khv:vertex-final} by \eqref{kh:denominator} gives 
\begin{equation}
\Phi_{ab}^{ij}(\chi_\phi;\phi)
=\frac{m_a^2m_b^2}{2}
\left[
\frac{\delta^{ij}}{|\vec k_a-\vec k_b|^2}
-\frac{2(k_a-k_b)^i(k_a-k_b)^j}
      {|\vec k_a-\vec k_b|^4}
\right]
+ \mathcal{O}(\alpha^2).
\label{eq:paper-Phi}
\end{equation}
Both numerator and denominator have now been evaluated to the required
order. The remaining calculation is to take the angular average of this
tensor. In particular, we want to evaluate
\begin{equation}
\begin{aligned}
&\left\langle
\frac{\Phi_{ab}^{ij}(\chi_\phi;\phi)}
     {1+\alpha^3z\beta_{ab}}
\right\rangle = \frac{m_a^2m_b^2}{2(1+\alpha^3z\beta_{ab})}
\frac{1}{2\pi}\int_0^{2\pi}
d\phi
\left[ \frac{\delta^{ij}}{|\vec k_a-\vec k_b|^2}
-\frac{2 (k_a-k_b)^i(k_a-k_b)^j} {|\vec k_a-\vec k_b|^4} \right] + \mathcal{O}(\alpha^2).
\end{aligned}
\label{khavg:integration-task main}
\end{equation}
In the Appendix \ref{App D} we have shown the detailed calculation of the angular average. Eq.\eqref{khavg:final-average} gives
\begin{equation}
\begin{aligned}
 &\left\langle
 \frac{\Phi_{ab}^{ij}(\chi_\phi;\phi)}
      {1+\alpha^3z\beta_{ab}}
 \right\rangle
 =\frac{m_a^2m_b^2|\vec v_{ab}|}{8}
 \Bigg[
 \frac{(v_{ab}^i+zc_{ab}^i)(v_{ab}^j+zc_{ab}^j)}
      {|\vec v_{ab}+z\vec c_{ab}|^3}\\
 &\qquad~~~~~~~~~~~~~~~~~~~~~~~~ -\frac{\alpha z}{2}\frac{d}{dz}\left\{
 [\vec n\cdot(\vec v_a+\vec v_b+z(\vec c_a+\vec c_b))]
 \frac{(v_{ab}^i+zc_{ab}^i)(v_{ab}^j+zc_{ab}^j)}
      {|\vec v_{ab}+z\vec c_{ab}|^3}
 \right\}
 \Bigg]+ \mathcal{O}(\alpha^2).
\end{aligned}
\label{khavg:final-average main}
\end{equation}
Hence from eq.\eqref{G2} we get 
\begin{equation}
\begin{aligned}
 K_H^{ij}(\alpha^3z)
 ={}&-\frac{2G^2}{\alpha}\sum_{a<b}m_am_b
 \frac{(v_{ab}^i+zc_{ab}^i)(v_{ab}^j+zc_{ab}^j)}
      {|\vec v_{ab}+z\vec c_{ab}|^3}\\
 &+G^2z\sum_{a<b}m_am_b\frac{d}{dz}\left\{
 [\vec n\cdot(\vec v_a+\vec v_b+z(\vec c_a+\vec c_b))]
 \frac{(v_{ab}^i+zc_{ab}^i)(v_{ab}^j+zc_{ab}^j)}
      {|\vec v_{ab}+z\vec c_{ab}|^3}
 \right\}+ \mathcal{O}(\alpha).
\end{aligned}
\label{khavg:generator}
\end{equation}
As a result the waveform due to the gravitational energy-momentum tensor can be represented as follows
\begin{equation}\label{H waveform}
\begin{split}
    e^{ij}_H(\alpha^3 z)=& -\frac{2G^2 \alpha^2}{Rw} \sum_{a<b}m_am_b
 \frac{(v_{ab}^i+zc_{ab}^i)(v_{ab}^j+zc_{ab}^j)}
      {|\vec v_{ab}+z\vec c_{ab}|^3}\\
 &+ \frac{G^2z \alpha^3}{w R}\sum_{a<b}m_am_b\frac{d}{dz}\left\{
 [\vec n\cdot(\vec v_a+\vec v_b+z(\vec c_a+\vec c_b))]
 \frac{(v_{ab}^i+zc_{ab}^i)(v_{ab}^j+zc_{ab}^j)}
      {|\vec v_{ab}+z\vec c_{ab}|^3}
 \right\}.
\end{split}
\end{equation}

\subsection{Combined contribution and matching with the PN waveform}
Since in the PN expansion we are interested in the waveform at order $\alpha^3$, we combine the matter and gravity contribution from eqs\eqref{X waveform} and \eqref{H waveform} to obtain the following 
\begin{equation}\label{eij tot X H}
\begin{split}
    e^{ij}_{(1)}(\alpha^3 z) =& \frac{2G\alpha^3}{wR} \sum_a m_a \Bigg{\lbrace} \vec n\cdot \Big( \vec c_a-\sum_{r\ge1}r\,\vec d_{r,a}z^r \Big)
   v_a^iv_a^j +(\vec n\cdot\vec v_a) \Big[ v_a^i\Big(c_a^j-\sum_{r\ge1}r\,d_{r,a}^jz^r\Big) +(i\leftrightarrow j) \Big]
   \\
   &\qquad ~~~~~~~~~~~~~~~~~ -\left[\vec n\cdot(\vec v_a+z\vec c_a)\right]
   \Big[ v_a^i\Big(c_a^j-\sum_{r\ge1}r(r+1)d_{r,a}^jz^r\Big) +(i\leftrightarrow j) \Big] \Bigg{\rbrace}
   \\
   & + \frac{G^2z \alpha^3}{w R}\sum_{a<b}m_am_b\frac{d}{dz}\left\{ [\vec n\cdot(\vec v_a+\vec v_b+z(\vec c_a+\vec c_b))] \frac{(v_{ab}^i+zc_{ab}^i)(v_{ab}^j+zc_{ab}^j)}
      {|\vec v_{ab}+z\vec c_{ab}|^3} \right\}.
\end{split}
\end{equation}
The subscript `\((1)\)' denotes the first subleading contribution in the PN expansion; the dependence on \(z\) retains the leading-logarithmic series to all orders. The matter contribution is already expressed in terms of the trajectory coefficients. We now rewrite the gravitational contribution in the same form.\\
We can rewrite the last line of eq.\eqref{eij tot X H} by using the following algebraic relation\footnote{To check the validity of the relation, one can substitute $\vec{v}_{ab}=\vec{v}_a - \vec{v}_b$ and the same for $\vec{c}_{ab}$ and expand out all the terms.}
\begin{equation}\label{numerator identity}
\begin{aligned}
&\big[\vec n\cdot(\vec v_a+\vec v_b+z(\vec c_a+\vec c_b))\big]
(v_{ab}^i+zc_{ab}^i)(v_{ab}^j+zc_{ab}^j)
\\
={}&[\vec n\cdot(\vec v_a+z\vec c_a)]
\big[(v_a^i+zc_a^i)(v_{ab}^j+zc_{ab}^j)
+(i\leftrightarrow j)\big]
-[\vec n\cdot(\vec v_b+z\vec c_b)]
\big[(v_b^i+zc_b^i)(v_{ab}^j+zc_{ab}^j)
+(i\leftrightarrow j)\big]
\\
&~~~~~~~~~~~~~~~-[\vec n\cdot(\vec v_{ab}+z\vec c_{ab})]
\big[(v_a^i+zc_a^i)(v_a^j+zc_a^j)
-(v_b^i+zc_b^i)(v_b^j+zc_b^j)\big]
\end{aligned}
\end{equation}
Substituting the first two terms of the RHS of eq.\eqref{numerator identity} in eq.\eqref{eij tot X H} yields the following expression
\begin{equation}
\begin{aligned}
&G\sum_{a<b}\frac{m_am_b}{|\vec v_{ab}+z\vec c_{ab}|^3}
\Big\{
[\vec n\cdot(\vec v_a+z\vec c_a)]
\big[(v_a^i+zc_a^i)(v_{ab}^j+zc_{ab}^j)
+(i\leftrightarrow j)\big]
\\
&\hspace{32mm}
-[\vec n\cdot(\vec v_b+z\vec c_b)]
\big[(v_b^i+zc_b^i)(v_{ab}^j+zc_{ab}^j)
+(i\leftrightarrow j)\big]
\Big\}
\\
&=\sum_a m_a[\vec n\cdot(\vec v_a+z\vec c_a)]
\left[
(v_a^i+zc_a^i)
G\sum_{b\ne a}m_b
\frac{v_{ab}^j+zc_{ab}^j}
     {|\vec v_{ab}+z\vec c_{ab}|^3}
+(i\leftrightarrow j)
\right].
\end{aligned}
\label{khm:ordered-symmetric}
\end{equation}
The last equality follows by interchanging the dummy indices
$a\leftrightarrow b$ in the second term. Under this interchange,
the summation range becomes $a>b$, while
$\vec v_{ba}+z\vec c_{ba}=-(\vec v_{ab}+z\vec c_{ab})$
cancels the explicit minus sign and leaves the denominator unchanged.
The two sums then combine into $\sum_a\sum_{b\ne a}$, after which
factors independent of $b$ are taken outside the inner sum.\\
For the last term in \eqref{numerator identity}, the same reversal gives
\begin{equation}
\begin{aligned}
&-G\sum_{a<b}m_am_b
\frac{\vec n\cdot(\vec v_{ab}+z\vec c_{ab})}
     {|\vec v_{ab}+z\vec c_{ab}|^3}
\big[(v_a^i+zc_a^i)(v_a^j+zc_a^j)
-(v_b^i+zc_b^i)(v_b^j+zc_b^j)\big]
\\
&=-\sum_a m_a(v_a^i+zc_a^i)(v_a^j+zc_a^j)
\,\vec n\cdot\left(
G\sum_{b\ne a}m_b
\frac{\vec v_{ab}+z\vec c_{ab}}
     {|\vec v_{ab}+z\vec c_{ab}|^3}
\right).
\end{aligned}
\label{khm:ordered-diagonal}
\end{equation}
Following the equation of motion of a particle under Newtonian gravity (see e.g. eq.\eqref{r2:force-sub})
\begin{equation}
G\sum_{b\ne a}m_b
\frac{\vec v_{ab}+z\vec c_{ab}}
     {|\vec v_{ab}+z\vec c_{ab}|^3}
=\vec c_a-\sum_{r\ge1}r(r+1)\vec d_{r,a}z^r.
\label{khm:force}
\end{equation}
Substituting it into \eqref{khm:ordered-symmetric} and \eqref{khm:ordered-diagonal} yields
\allowdisplaybreaks{
\begin{align}\label{khm:pair-to-particle}
&G\sum_{a<b}m_am_b
[\vec n\cdot(\vec v_a+\vec v_b+z(\vec c_a+\vec c_b))]
\frac{(v_{ab}^i+zc_{ab}^i)(v_{ab}^j+zc_{ab}^j)}
     {|\vec v_{ab}+z\vec c_{ab}|^3} \notag
\\
&=\sum_a m_a\Bigg\{
[\vec n\cdot(\vec v_a+z\vec c_a)]
\Big[
(v_a^i+zc_a^i)
\Big(c_a^j-\sum_{r\ge1}r(r+1)d_{r,a}^jz^r\Big)
+(i\leftrightarrow j)
\Big] \notag
\\
&\hspace{24mm}
-\vec n\cdot\Big(
\vec c_a-\sum_{r\ge1}r(r+1)\vec d_{r,a}z^r
\Big)
(v_a^i+zc_a^i)(v_a^j+zc_a^j)
\Bigg\}.
\end{align}}
Thus eq.\eqref{eij tot X H} becomes the following
\begin{align} \label{khm:particle-derivative}
e^{ij}_{(1)}(\alpha^3 z) =& \frac{2G\alpha^3}{wR} \sum_a m_a \Bigg{\lbrace} \vec n\cdot \Big( \vec c_a-\sum_{r\ge1}r\,\vec d_{r,a}z^r \Big)
   v_a^iv_a^j +(\vec n\cdot\vec v_a) \Big[ v_a^i\Big(c_a^j-\sum_{r\ge1}r\,d_{r,a}^jz^r\Big) +(i\leftrightarrow j) \Big] \notag
   \\
   &\qquad ~~~~~~~~~~~~~~~~~ -\left[\vec n\cdot(\vec v_a+z\vec c_a)\right]
   \Big[ v_a^i\Big(c_a^j-\sum_{r\ge1}r(r+1)d_{r,a}^jz^r\Big) +(i\leftrightarrow j) \Big] \Bigg{\rbrace} \notag
   \\
   & + \frac{G \alpha^3}{w R} \sum_a m_a z\frac{d}{dz}\Bigg\{
[\vec n\cdot(\vec v_a+z\vec c_a)]
\Big[
(v_a^i+zc_a^i)
\Big(c_a^j-\sum_{r\ge1}r(r+1)d_{r,a}^jz^r\Big)
+(i\leftrightarrow j)
\Big] \notag
\\ 
& ~~~~~~~~~~~~~~~~~~~~~~~~~~~~~~~~~~~~~~~~~~~~~~ -\vec n\cdot\Big(
\vec c_a-\sum_{r\ge1}r(r+1)\vec d_{r,a}z^r
\Big)
(v_a^i+zc_a^i)(v_a^j+zc_a^j)
\Bigg\}.
\end{align}
Now the derivatives to be substituted in eq.\eqref{khm:particle-derivative} are listed as follows
\begin{equation}
\begin{aligned}
z\frac{d}{dz}(v_a^i+zc_a^i)
&=zc_a^i=(v_a^i+zc_a^i)-v_a^i,
\\
z\frac{d}{dz}[\vec n\cdot(\vec v_a+z\vec c_a)]
&=z\vec n\cdot\vec c_a
=\vec n\cdot(\vec v_a+z\vec c_a)-\vec n\cdot\vec v_a,
\\
z\frac{d}{dz}\left(
\vec c_a-\sum_{r\ge1}r(r+1)\vec d_{r,a}z^r
\right)
&=-\sum_{r\ge1}r^2(r+1)\vec d_{r,a}z^r.
\end{aligned}
\label{khm:derivatives}
\end{equation}
With these expressions along with the following identity 
\begin{equation}
\begin{aligned}
-\sum_{r\ge1}r^2(r+1)\vec d_{r,a}z^r
={}&
\left(2\vec c_a
-\sum_{r\ge1}r(r+1)(r+2)\vec d_{r,a}z^r\right) -2\left(\vec c_a
-\sum_{r\ge1}r(r+1)\vec d_{r,a}z^r\right).
\end{aligned}
\label{khm:rearrangements}
\end{equation}
Eq.\eqref{khm:particle-derivative} be put in following form after some cancellations and addition of similar terms
\allowdisplaybreaks{
\begin{align}\label{khm:surviving-generator}
e^{ij}_{(1)}(\alpha^3 z) = &\frac{G\alpha^3}{wR} \sum_a m_a \Bigg{\lbrace} 2 \vec n\cdot \Big( \vec c_a-\sum_{r\ge1}r\,\vec d_{r,a}z^r \Big)
   v_a^iv_a^j + 2 (\vec n\cdot\vec v_a) \Big[ v_a^i\Big(c_a^j-\sum_{r\ge1}r\,d_{r,a}^jz^r\Big) +(i\leftrightarrow j) \Big] \notag
\\
&\quad + \frac{G \alpha^3}{wR} \sum_a m_a\Bigg\{ (\vec n\cdot\vec v_a) \Big[(v_a^i+zc_a^i)
\Big(-c_a^j+\sum_{r\ge1}r(r+1)d_{r,a}^jz^r\Big) +(i\leftrightarrow j) \Big] \notag
\\
&\quad ~~~~~~~~~~~~~~~~~~~~~~~~ -\vec n\cdot\Big( -\vec c_a+\sum_{r\ge1}r(r+1)\vec d_{r,a}z^r \Big) \Big[(v_a^i+zc_a^i)v_a^j+(i\leftrightarrow j)\Big] \notag
\\
&\quad ~~~~~~~~~~~~~~~~~~~~~~~~ - 3 [\vec n\cdot(\vec v_a+z\vec c_a)] \Big[ v_a^i \Big(c_a^j - \sum_{r\ge1}r(r+1)d_{r,a}^jz^r\Big) +(i\leftrightarrow j) \Big] \notag
\\
&\quad ~~~~~~~~+[\vec n\cdot(\vec v_a+z\vec c_a)] \Big[ (v_a^i+zc_a^i)
\Big(2c_a^j-\sum_{r\ge1}r(r+1)(r+2)d_{r,a}^jz^r\Big) +(i\leftrightarrow j)
\Big] \notag
\\
&\quad ~~~~~~~~~~~~~~~~~~~~~~~ -\vec n\cdot\Big( 2\vec c_a-\sum_{r\ge1}r(r+1)(r+2)\vec d_{r,a}z^r \Big) (v_a^i+zc_a^i)(v_a^j+zc_a^j) \Bigg\}.
\end{align}}
This expression agrees with the time-dependent, leading-logarithmic part of eq.\eqref{eij 1 after subs X} at order \(\alpha^3\).

%%%%%%%%%%%%%%%%%%%%%%%%%%%%%%%%%%%%%%%%%%%%%%%%%%%%%%%%%%%%%%%%%%%%%%%%%%%%%%%%%%%%%%%%%%%%%%%%%%%%%%%%%%%%%%%%%%%
\section{Conclusion}
%%%%%%%%%%%%%%%%%%%%%%%%%%%%%%%%%%%%%%%%%%%%%%%%%%%%%%%%%%%%%%%%%%%%%%%%%%%%%%%%%%%%%%%%%%%%%%%%%%%%%%%%%%%%%%%%%%%

Building on the Newtonian analysis of \cite{Karan:2026jzs}, we have determined the leading-logarithmic gravitational waveform tails at PN expansion of asymptotic scattering states to first sub-leading order in the post-Newtonian expansion. At order $\alpha^3$, we determine the coefficients of the leading-logarithmic terms $u^{-r}(\ln|u|)^{r-1}$ in the gravitational waveform for every $r\geq1$, allowing arbitrary numbers of incoming and outgoing massive objects. These
coefficients depend on the asymptotic masses and velocities, without requiring a model of the dynamics in the finite interaction region, subject to the assumptions on omitted radiative contributions stated in the analysis.

The correction is governed by the first spatial moment of the stress tensor. Conservation of the total energy-momentum tensor reduces this moment to the mass octupole and the first spatial
moment of the angular-momentum density. A useful feature of the calculation is that both can be evaluated on Newtonian trajectories at the required accuracy. The logarithmic trajectory
coefficients $\vec c_a$ and $\vec d_{r,a}$ in eq.\eqref{r2:c} and \eqref{r2:d}, substituted into eq.\eqref{eij 1 multiparticle w}, therefore determine the entire leading-logarithmic waveform.

The constant, $u^{-1}$ and $u^{-2}\ln|u|$ terms agree with the corresponding classical soft-graviton results. For two particles in each asymptotic state, the leading-logarithmic series admits a closed-form sum. In the centre-of-mass frame, the order-$\alpha^3$ contribution is proportional to the mass difference and vanishes for equal masses. This cancellation reflects the odd spatial parity of the source moments that generate the correction: an equal-mass two-body configuration is invariant under spatial reflection, while these moments change sign. The cancellation concerns this velocity order, rather than the complete gravitational waveform.

The calculation also tests proposed relativistic expressions beyond their leading Newtonian limits. The multiparticle comparison is essential, since agreement in two-body scattering
does not determine the general dependence on the asymptotic data. The agreement with the corresponding expansion of the conjecture in \cite{Karan:2026jzs} provides a further check at order $\alpha^3$. At this order, the comparison also probes observation-direction dependence through factors such as $n\cdot p_a$ and contractions involving $n_\rho p_a^{[\mu}p_b^{\rho]}$, beyond the information accessible at leading Newtonian order. These checks support the conjecture but do not establish it at arbitrary velocities.

Among several questions that can now be addressed within the same framework, next velocity order presents a different problem. At order $\alpha^4$, further retardation terms must be combined with corrections to the trajectories and source densities. Their combined contribution would test whether the highest-logarithmic coefficients retain the same dependence
on asymptotic data once these effects enter together. The equal-mass two-body configuration offers a useful special case: its order-$\alpha^3$ term vanishes, so the next order isolates contributions that not constrained by the odd-parity cancellation. Higher-order multiparticle calculations would provide additional tests of the relativistic conjecture, while a direct derivation at arbitrary velocities remains the broader objective.

%%%%%%%%%%%%%%%%%%%%%%%%%%%%%%%%%%%%%%%%%%%%%%%%%%%%%%%%%%%%%%%%%%%%%%%%%%%%%%%%%%%%%%%%%%%%%%%%%%%%%%%%%%%%%%%%%%%
\section*{Acknowledgement}
We thank the authors of \cite{Karan:2026jzs} for insightful discussions. We are grateful to Alok Laddha, Biswajit Sahoo and Ashoke Sen for carefully reading the manuscript at various stages and for suggesting major improvements to earlier drafts. DK thanks CMI and IMSc for their warm hospitality during the final stages of this work. This work was supported by the Department of Atomic Energy, Government of India, under Project No. RTI4019.

\section*{Acknowledgement for AI uses} 
We acknowledge the use of ChatGPT-6 Astra (OpenAI) and Claude Opus 5.5 (Anthropic) for algebraic manipulations in Section \ref{s4} and the Appendix \ref{App D}.
Claude Opus 5.5 identified the agreement of eq.\eqref{2to2n eij CM} with ref.\cite{Blanchet:1996pi}. All results were checked independently by the authors who take full responsibility for its content, including any errors.
%%%%%%%%%%%%%%%%%%%%%%%%%%%%%%%%%%%%%%%%%%%%%%%%%%%%%%%%%%%%%%%%%%%%%%%%%%%%%%%%%%%%%%%%%%%%%%%%%%%%%%%%%%%%%%%%%%%

\appendix

%%%%%%%%%%%%%%%%%%%%%%%%%%%%%%%%%%%%%%%%%%%%%%%%%%%%%%%%%%%%%%%%%%%%%%%%%%%%%%%%%%%%%%%%%%%%%%%%%%%%%%%%%%%%%%%%%%%
\section{Analysis of the error terms}\label{Err}
%%%%%%%%%%%%%%%%%%%%%%%%%%%%%%%%%%%%%%%%%%%%%%%%%%%%%%%%%%%%%%%%%%%%%%%%%%%%%%%%%%%%%%%%%%%%%%%%%%%%%%%%%%%%%%%%%%%
In this appendix we show that the error terms $\mathcal{E}_1, \mathcal{E}_2$ and $\mathcal{E}_3$ can be ignored at the order we have derived the waveform.
%%%%%%%%%%%%%%%%%%%%%%%%%%%%%%%%%%%%%%%%%%%%%%%%%%%%%%%%%%%%%%%%%%%%%%%%%%%%%%%%%%%%%%%%%%%%%%%%%%%%%%%%%%%%%%%%%%%
\subsection{Analysis of the error term $\mathcal{E}_1^{ij}$} \label{Error1}
%%%%%%%%%%%%%%%%%%%%%%%%%%%%%%%%%%%%%%%%%%%%%%%%%%%%%%%%%%%%%%%%%%%%%%%%%%%%%%%%%%%%%%%%%%%%%%%%%%%%%%%%%%%%%%%%%%%
In this appendix we will argue that the error $\mathcal{E}^{ij}_1$ encountered in eq.\eqref{r2:E1split} contributes to the order higher $\alpha$ than what we are considering here. The expression of $\mathcal{E}_1^{ij}$ is the following 
\begin{equation}\label{A1}
\begin{split}
    & \mathcal{E}_1^{ij} = \frac GR n_kn_l\partial_u^2  \mathcal{S}^{kl\vert ij} +\frac{2G}{R}\sum_{q\geq3}\frac1{q!}\partial_u^q \int d^3x\,(\vec n\cdot\vec x)^qT^{ij}
    \\
    & \mathcal{S}^{kl\vert ij} = \int d^3x\,x^kx^lT^{ij}
\end{split}
\end{equation}
In attempting to rewrite the second moment of the stress tensor $\mathcal{S}^{kl\vert ij}$ from eq.\eqref{A1} using the conservation law we get 
\begin{equation}\label{A.3}
  n_k n_l \partial_u^2 \mathcal{I}^{klij} = 2 n_k n_l \mathcal{S}^{kl \vert ij} + 4 n_k n_l \left( \mathcal{S}^{ki\vert lj} + \mathcal{S}^{kj \vert li}  \right) + 2 n_k n_l \mathcal{S}^{ij \vert kl}
\end{equation}
where we have defined the mass hexadecapole moment as follows
\begin{equation}
    \mathcal{I}^{klij} = \int d^3x\,x^kx^l x^i x^j T^{00}
\end{equation}
Since we are interested in the expression for $\mathcal{S}^{kl\vert ij}$, following eq.\eqref{A.3} we therefore obtain 
\begin{equation}\label{A.5}
    n_k n_l \mathcal{S}^{kl \vert ij} = \frac{1}{2}  n_k n_l \partial_u^2 \mathcal{I}^{klij}  - 2 n_k n_l \left( \mathcal{S}^{ki\vert lj} + \mathcal{S}^{kj \vert li}  \right) -  n_k n_l \mathcal{S}^{ij \vert kl}
\end{equation}
To simplify the RHS, we consider the cubic moment of the momentum density and rewrite it in the following form using the conservation law and integration by parts while ignoring the boundary terms
\begin{equation}\label{A.6}
    \partial_u \int d^3x\, x^k x^i x^j T^{0l} = \mathcal{S}^{ij \vert kl} + \mathcal{S}^{kj \vert li} +\mathcal{S}^{ki \vert lj}
\end{equation}
Note that the second moment of the stress tensor $\mathcal{S}^{kl \vert ij}$ is symmetric in its last two indices what we have also used to get the above form. Then substituting \eqref{A.6} in \eqref{A.5} we obtain 
\begin{equation}
    n_k n_l \mathcal{S}^{kl \vert ij} = \frac{1}{2}  n_k n_l \partial_u^2 \mathcal{I}^{kl ij}  - 2 n_k n_l \partial_u \int d^3x\, x^k x^i x^j T^{0l} +  n_k n_l \mathcal{S}^{ij \vert kl}
\end{equation}
It shows that the conservation law fixes the first two terms but does not eliminate the last stress moment. Substituting this in eq.\eqref{A1} we therefore get 
\begin{equation}
\begin{split}
    \mathcal{E}_1^{ij} = \frac{G}{2R} n_kn_l\partial_u^4  \mathcal{I}^{kl ij} - \frac{2G}{R} n_k n_l \partial_u^3 \int d^3x\, x^k x^i x^j T^{0l} + \frac{G}{R} n_k n_l \partial_u^2 \mathcal{S}^{ij \vert kl} 
    \\
    +\frac{2G}{R}\sum_{q\geq3}\frac1{q!}\partial_u^q \int d^3x\,(\vec n\cdot\vec x)^qT^{ij} 
\end{split}
\end{equation}
This expression is sufficient to justify that $\mathcal{E}_1^{ij} \sim \mathcal{O}(\alpha^4)$ and higher order terms. Given the expressions of the mass and the momentum density (see eq.\eqref{T00 and T0i}) even at the lowest order in $\alpha$ (meaning that we only focus on the matter contribution to the components $T^{00}, T^{0i}$ and $T^{ij}$ at the Newtonian limit), one could summarize the scaling\footnote{Recall that due to the scaling $w = \alpha^3 u$, we have $\partial_u \sim \alpha^3 , ~~ X^i(u) \sim \alpha^{-2}, ~~ \dot{X}^i(u) \sim \alpha $},
\begin{equation}\label{A.8}
\begin{split}
    & \partial_u^4 \mathcal{I}^{klij} \simeq \alpha^{12} \sum_a m_a X^k_a X^l_a X^i_a X^j_a \sim \mathcal{O}(\alpha^4),\qquad \partial_u^3 \int d^3x\, x^k x^i x^j T^{0l} \simeq \alpha^9 \sum_a m_a X_a^k X_a^i X_a^j \dot{X}_a^l \sim \mathcal{O}(\alpha^4)
    \\
    & \partial_u^2 \mathcal{S}^{ij \vert kl}  \simeq \alpha^6 \sum_a m_a X_a^i X_a^j \dot{X}_a^k \dot{X}_a^l \sim \mathcal{O}(\alpha^4),\qquad \partial_u^q \int d^3x\, x^{k_1}\cdots x^{k_q} T^{ij} \simeq \alpha^{3q} \alpha^{-2q} \alpha^2 \simeq \mathcal{O}(\alpha^{q+2}) \geq \mathcal{O}(\alpha^5)
\end{split}
\end{equation}
An important subtle point about all the spatial integrals is that they are generically divergent and hence one should really think of them with a large spherical cutoff, say $\Lambda$. For instance, if we consider the integral in the first line of eq.\eqref{A1} with a spherical cutoff, the integrand schematically behaves as $r^{q-2}$ that manifestly diverges with increasing $q$ ($q \geq 3$), given the Newtonian gravitational stress falls as $r^{-4}$. For the second stress moment, the divergent monopole term is time independent and is annihilated by the two time derivatives as argued in \cite{Karan:2026jzs}. This does not hold for all higher moments: the monopole–quadrupole part of the exterior gravitational stress behaves as \(M Q_{kl}(u)/r^6\), so the fourth moment generally retains a divergence proportional to \(\Lambda M Q_{kl}^{(4)}(u)\) even after differentiation which is the case for $q=4$. However, at the retained order of waveform, the divergent part involve only the conserved multipole contribution, whose time derivative actually vanishes and do not contribute to the waveform.\\
This completes the proof that the error we have ignored while writing eq.\eqref{eij 1} from eq.\eqref{r2:E1split} contributes at $\mathcal{O}(\alpha^4)$ and hence can be neglected in our computation.\\

%%%%%%%%%%%%%%%%%%%%%%%%%%%%%%%%%%%%%%%%%%%%%%%%%%%%%%%%%%%%%%%%%%%%%%%%%%%%%%%%%%%%%%%%%%%%%%%%%%%%%%%%%%%%%%%%%%%
\subsection{Analysis of the error term $\mathcal{E}_2^{ij}$} \label{Error2}
%%%%%%%%%%%%%%%%%%%%%%%%%%%%%%%%%%%%%%%%%%%%%%%%%%%%%%%%%%%%%%%%%%%%%%%%%%%%%%%%%%%%%%%%%%%%%%%%%%%%%%%%%%%%%%%%%%%
In this appendix we will show that the correction term we ignored in eq.\eqref{eij 1 du2} capturing the contribution due to the kinetic energies of the particles and the gravitational potential energy contribute to the higher order in $\alpha$.

\subsubsection{Kinetic energy of matter particles}

Omitting gravitational-potential corrections for the moment, the first kinetic correction to the energy density becomes the following 
\begin{equation}
    \delta T^{00}_{\mathrm{kinetic}} \simeq \frac12\sum_a m_a|\dot{\vec{X}}_a|^2 \delta^{(3)}(\vec x-\Xa).
\end{equation}
It is already argued in \cite{Karan:2026jzs} that the contribution of this correction to the quadrupole moment appears at $\mathcal{O}(\alpha^4)$. In the following we show that the above kinetic correction to the energy density contributes to the mass octupole term at higher order in $\alpha$. The change in the mass octupole due to the kinetic energy correction
\begin{equation}
\begin{split}
    & \delta \mathcal{I}_{\mathrm{kinetic}}^{kij} =\int d^3x\,x^kx^ix^j\delta T^{00}_{\mathrm{kinetic}}=\frac12\sum_a m_a |\dot{\vec{X}}_a|^2 X_a^kX_a^iX_a^j \sim \Ord(\alpha^{2-6})=\Ord(\alpha^{-4})
\end{split}
\end{equation}
Consequently the correction acquired by the waveform becomes 
\begin{equation}
     \frac{G}{3R}n_k\partial_u^3\delta \mathcal{I}_{\mathrm{kinetic}}^{kij} =\frac{G}{6R}n_k\partial_u^3 \sum_a m_a|\dot{\vec{X}}_a|^2 X_a^kX_a^iX_a^j \sim \mathcal{O}(\alpha^{9-4})= \mathcal{O}(\alpha^5)
\end{equation}
It therefore proves that the kinetic correction to the energy density through the mass octupole term can be ignored at the order of the waveform we are considering here.\\

We now show the same conclusion remains valid for the correction through the first spatial moment of the angular momentum density. At leading order the correction to the momentum density becomes 
\begin{equation}
    \delta T^{0i}_{\text{kinetic}} \simeq \frac{1}{2} \sum_a m_a |\dot{\vec{X}}_a|^2 \dot{X}_a \delta^{(3)}(\vec x-\Xa)
\end{equation}
Due to this the change in $\mathcal{J}^{ki \vert j}$ becomes 
\begin{equation}
    \delta \mathcal{J}^{ki \vert j}_{\text{kinetic}} = \frac{1}{2} \sum_a m_a |\dot{\vec{X}}_a|^2 X_a^j \left( X_a^k \dot{X}_a^i - X_a^i \dot{X}_a^k \right) \sim \mathcal{O}(\alpha^{2-2-1}) = \mathcal{O}(\alpha^{-1})
\end{equation}
As a result the correction to the waveform becomes 
\begin{equation}
\begin{split}
    \frac{2G}{3R} n_k \partial_u^2  \delta \mathcal{J}^{ki \vert j}_{\text{kinetic}} = \frac{G}{3R} n_k \partial_u^2 \sum_a m_a |\dot{\vec{X}}_a|^2 X_a^j \left( X_a^k \dot{X}_a^i - X_a^i \dot{X}_a^k \right) \sim \mathcal{O}(\alpha^{6-1}) = \mathcal{O}(\alpha^5)
\end{split}
\end{equation}
Therefore it completes the proof that the kinetic correction to the matter and momentum energy density does not affect the waveform at order $\alpha^3$.

\subsubsection{Gravitational potential energy}

Now we estimate the order of the corrections to the mass octupole\footnote{In \cite{Karan:2026jzs} it was already shown that the gravitational energy correction to the mass quadrupole affects the waveform at $\mathcal{O}(\alpha^4)$.} due to the gravitational energy correction and consequently its effect on the waveform. \\
Since the magnitude of the mutual interaction energy is 
\begin{equation}
  \Phi = \frac{G m_a m_b}{|\Xa-\vec X_b|} =\frac{G m_a m_b\alpha^2} {|w|\,|\vec{v}_a-\vec v_b+(\ca-\vec c_b)z|} +\hbox{subleading terms in}~w
\end{equation}
which is relative $\alpha^2$ as compared with the rest mass energy. Due to the gravitational interaction the correction to the waveform through the mass octupole term becomes
\begin{equation}
    \frac{G}{3R}n_k\partial_u^3\delta \mathcal{I}_{\mathrm{potential}}^{kij} = \frac{G}{3R}n_k\partial_u^3\left[  X^kX^iX^jG\sum_{a<b}\frac{m_am_b}{|\Xa-\vec X_b|}\right] \simeq \mathcal{O}(\alpha^{9-6+2}) \sim \mathcal{O}(\alpha^5)
\end{equation}
which therefore does not contribute to the order $\alpha^3$ waveform. In $T^{00}$ there are other non-linear terms that involve spatial derivatives of the potential. Since from the trajectory expression we know that $\vec{X}_a \sim \alpha^{-2}$, each spatial derivative gives an extra factor of $\alpha^2$ and as a result those terms will also not contribute at this order. In particular $\partial_k \Phi \partial^k \Phi \sim \mathcal{O}(\alpha^8)$. As a result, the waveform will contribute at $\mathcal{O}(\alpha^4)$ through mass quadrupole and at $\mathcal{O}(\alpha^5)$ through the mass octupole. So we can ignore the effect.\\

Due to the gravitational interaction the leading order correction to the momentum density becomes
\begin{equation}
     \delta T^{0i}_{\text{potential}} \simeq \sum_a m_a \Phi_a \dot{X}_a \delta^{(3)}(\vec x-\Xa)
\end{equation}
which gives 
\begin{equation}
    \delta \mathcal{J}^{ki \vert j}_{\text{potential}} = \sum_a m_a \Phi_a X_a^j \left( X_a^k \dot{X}_a^i - X_a^i \dot{X}_a^k \right) \sim \mathcal{O}(\alpha^{2-2-1}) = \mathcal{O}(\alpha^{-1}).
\end{equation}
As a result the correction to the waveform becomes 
\begin{equation}
\begin{split}
    \frac{2G}{3R} n_k \partial_u^2  \delta \mathcal{J}^{ki \vert j}_{\text{potential}} = \frac{2G}{3R} n_k \partial_u^2 \sum_a m_a \Phi_a X_a^j \left( X_a^k \dot{X}_a^i - X_a^i \dot{X}_a^k \right) \sim \mathcal{O}(\alpha^{6-1}) = \mathcal{O}(\alpha^5)
\end{split}
\end{equation}
and hence we can ignore this contribution at the desired order. In $T^{0i}$ also, there are non-linear terms e.g. $\partial_l \Phi \partial^i \Phi^l$ where $\Phi^l \sim G \sum_a m_a \dot{X}_a / \vert \vec{x} - \vec{X}_a \vert \sim \mathcal{O}(\alpha^3)$. As a result $T^{0i} \sim \mathcal{O}(\alpha^9)$ and hence the contribution to the waveform will be at $\mathcal{O}(\alpha^5)$.

%%%%%%%%%%%%%%%%%%%%%%%%%%%%%%%%%%%%%%%%%%%%%%%%%%%%%%%%%%%%%%%%%%%%%%%%%%%%%%%%%%%%%%%%%%%%%%%%%%%%%%%%%%%%%%%%%%%
\subsection{Analysis of the error term $\mathcal{E}_3$} \label{Error3}
%%%%%%%%%%%%%%%%%%%%%%%%%%%%%%%%%%%%%%%%%%%%%%%%%%%%%%%%%%%%%%%%%%%%%%%%%%%%%%%%%%%%%%%%%%%%%%%%%%%%%%%%%%%%%%%%%%%
In this appendix we perform the analysis of the error $\mathcal{E}_3$ in eq.\eqref{r2:E3} that captures the correction to the Newtonian gravitational force due to the effect of additional forces. If there is an additional force at $\mathcal{O}(\alpha^5)$, integrating twice it can be shown that such a force induces a correction to the trajectory at $\mathcal{O}(\alpha^{-1})$. And as a result the mass quadrupole term will contribute to the waveform at 
\begin{equation}\label{A.21}
    \delta e^{ij}_{(1)}(w) \sim \frac{G \alpha^6}{R} \frac{d^2}{dw^2} \sum_a m_a \left( X_{a,N}^i \delta X_{a}^j + \delta X_a^i X_{a,N}^j \right) \sim \mathcal{O}(\alpha^{6-3}) = \mathcal{O}(\alpha^3)
\end{equation}
where $X_{a,N}$ is the Newtonian trajectory of the $a$-th particle and $\delta X_a$ denotes the correction to the Newtonian trajectory due to the additional force. Eq.\eqref{A.21} shows that due to such an additional force, the quadrupole moment will contribute to $\mathcal{O}(\alpha^3)$ waveform. In the following we will show that there is no $\mathcal{O}(\alpha^5)$ correction to the force.\\
We establish the required order directly from the geodesic equation.
A dot denotes differentiation with respect to the source time,
evaluated at $u$ in the multipole moments. In coordinate time the
geodesic equation reads
\begin{equation}
\begin{split}
 \ddot X_a^i={}&-\Gamma^i_{00}
 -2\Gamma^i_{0j}\dot X_a^j
 -\Gamma^i_{jk}\dot X_a^j\dot X_a^k +\dot X_a^i\left(
 \Gamma^0_{00}+2\Gamma^0_{0j}\dot X_a^j
 +\Gamma^0_{jk}\dot X_a^j\dot X_a^k\right).
 \label{eq:trajectory-geodesic}
\end{split}
\end{equation}
The scaling $w=\alpha^3u$, $\vec X_a (u)= \mathcal{O}(\alpha^{-2})$ gives $\dot{\vec X}_a(u)= \mathcal{O}(\alpha)$, $\partial_i\sim\alpha^2$ and $\partial_u\sim\alpha^3$.
In the weak-field approximation the metric components become
\begin{equation}
 g_{00}=-1+2\Phi+\mathcal{O}(\alpha^4),~~~ g_{0i}= \mathcal{O}(\alpha^3), ~~~ g_{ij} =\delta_{ij}+ \mathcal{O} (\alpha^2), \Phi=G\sum_b\frac{m_b}{|\vec x-\vec X_b|}
      =\mathcal{O}(\alpha^2).
 \label{eq:trajectory-metric}
\end{equation}
Here $\Phi$ is evaluated on the full trajectories. Expanding the Christoffel connections therefore gives
\begin{equation}
\begin{aligned}
 \Gamma^i_{00}&=-\partial_i\Phi+\mathcal{O}(\alpha^6),&
 \Gamma^i_{0j}&=\mathcal{O}(\alpha^5),&
 \Gamma^i_{jk}&=\mathcal{O}(\alpha^4),\\
 \Gamma^0_{00}&=\mathcal{O}(\alpha^5),&
 \Gamma^0_{0j}&=\mathcal{O}(\alpha^4),&
 \Gamma^0_{jk}&=\mathcal{O}(\alpha^5).
 \label{eq:trajectory-connection}
\end{aligned}
\end{equation}
Including the velocity factors in eq.\eqref{eq:trajectory-geodesic}, every contribution beyond
$\partial_i\Phi$ is thus $O(\alpha^6)$ or higher. Evaluating the regular external field on the trajectory yields
\begin{equation}
 \ddot X_a^i
 =-G\sum_{b\ne a}m_b
 \frac{X_a^i-X_b^i}{|\vec X_a-\vec X_b|^3}
 +\mathcal{O}(\alpha^6).
 \label{eq:trajectory-force}
\end{equation}
There is no $O(\alpha^5)$ correction to the Newtonian force law and consequently no correction to the waveform at the order retained in this work. This also shows that the error $\mathcal{E}_3$ appears to the force equation only at order $\alpha^6$.

%%%%%%%%%%%%%%%%%%%%%%%%%%%%%%%%%%%%%%%%%%%%%%%%%%%%%%%%%%%%%%%%%%%%%%%%%%%%%%%%%%%%%%%%%%%%%%%%%%%%%%%%%%%%%%%%%%%
\section{Comparison with the conjectured $A^{\mu\nu}_{(3)}$} \label{app B}
%%%%%%%%%%%%%%%%%%%%%%%%%%%%%%%%%%%%%%%%%%%%%%%%%%%%%%%%%%%%%%%%%%%%%%%%%%%%%%%%%%%%%%%%%%%%%%%%%%%%%%%%%%%%%%%%%%%
In \cite{Karan:2026jzs}, the authors have proposed a conjecture for all-order gravitational waveform for relativistic particles. In this appendix we will show that following their conjecture we can recover the waveform at order $(\ln w)^2/w^3$ in both the Newtonian and the next-to-Newtonian limit. For demonstration we will show the agreement of our result with the conjecture only for the late time waveform. However, the same replacement strategy described in section \ref{s2} will follow to get the waveform at early time.\\
In particular, we want to determine the following waveform from the conjecture
\begin{equation}
    \tilde{e}^{ij}(\omega) \sim \frac{1}{6} A^{ij}_{(3)} \omega^2 \left( \ln (\omega + i \epsilon) \right)^3
\end{equation}
Using Fourier transformation, one can write it in terms of the retarded time $u$ and then due to the scaling $w = \alpha^3 u$ we eventually write the waveform in terms of the scaled retarded time
\begin{equation}\label{eq4.2}
    e^{ij}(w) = \alpha^9 A^{ij}_{(3)} \frac{(\ln w)^2}{w^3}
\end{equation}
Note that the waveform in the time domain behaves as $\mathcal{O}(\alpha^2)$ (as derived in \cite{Karan:2026jzs}) and $\mathcal{O}(\alpha^3)$(in section \ref{s2} of this paper) in the Newtonian and next-to-Newtonian limit respectively. Therefore, to reproduce the waveform at these orders with the nonrelativistic limit dictated by $\alpha \ll 1$, the required order of the coefficient $A^{\mu\nu}_{(3)}$ should be $\mathcal{O}(\alpha^{-7})$ and $\mathcal{O}(\alpha^{-6})$ for the Newtonian and the next-to-Newtonian limit respectively.\\
From appendix-D of \cite{Karan:2026jzs} we can write down the corresponding coefficient of the wave tail as follows
{\allowdisplaybreaks
\begin{align}
 A_{(3)}^{ij}=\frac1R\sum_{a<b} \left[ \mathcal{T}_{ab}^{ij}+n^{(i}\xi_{ab}^{j)} -\frac12\eta^{ij}n\cdot\xi_{ab}\right],
 \label{eq:D7}
\end{align}
where,
\begin{align}
 \mathcal{T}_{ab}^{ij}=\frac{G^2}{(\sigma_{ab}^2-1)^{7/2}}\Bigg\{&
 -\frac{w_{ab}^{i}w_{ab}^{j}}{m_am_b}
 \bigg[(2\sigma_{ab}^2+3)(\zeta_{ab}^2+\zeta_{ba}^2)
 +10\sigma_{ab}\zeta_{ab}\zeta_{ba} -\frac13(\sigma_{ab}^2-1)
 (4\sigma_{ab}^4-12\sigma_{ab}^2+11) \mathcal{Z}_{ab}^2\bigg]\nonumber\\
 &+\frac43(\sigma_{ab}^2-1)
 \big[(3\sigma_{ab}\delta_a-\kappa_{ab}\delta_b)\zeta_{ab}
 +(3\sigma_{ab}\delta_b-\kappa_{ab}\delta_a)\zeta_{ba}\big]
 w_{ab}^{(i} \mathcal{Z}_{ab}^{j)}\nonumber\\
 &-\frac23(\sigma_{ab}^2-1)^2(2\sigma_{ab}^2-1)m_am_b
 (\delta_a^2+\sigma_{ab}\delta_a\delta_b+\delta_b^2)
 \mathcal{Z}_{ab}^{i} \mathcal{Z}_{ab}^{j}\Bigg\}.
 \label{eq:D8}
\end{align}}
where all the symbols are defined by the following expressions
\begin{align}\label{eq4.5}
 & d_a=-n\cdot p_a,\quad \delta_a=d_a/m_a,\quad
 \sigma_{ab}=-\frac{p_a\cdot p_b}{m_am_b},\quad
 \kappa_{ab}=2\sigma_{ab}^4-6\sigma_{ab}^2+1,\notag \\
 & \widehat C_a^\mu=\frac{2G}{p_a\cdot n}
 \sum_{b\ne a}\tau_{ab}n_\rho p_{[a}^{\mu}p_{b]}^{\rho},\quad
 \mathcal{Z}_{ab}=\widehat C_a-\widehat C_b,\quad
 w_{ab}^{\mu}=d_bp_a^\mu-d_ap_b^\mu, \notag \\
 & \zeta_{ab}=\frac{p_a\cdot \mathcal{Z}_{ab}}{m_a},\qquad
 \zeta_{ba}=-\frac{p_b\cdot \mathcal{Z}_{ab}}{m_b}, \qquad \tau_{ab} = - \frac{\sigma_{ab} \left( 2 \sigma_{ab}^2 - 3 \right)}{2 \left( \sigma_{ab}^2 - 1 \right)^{3/2} }.
\end{align}
From the expression of $\widehat C_a$ we see that due to the antisymmetric factor $p^\mu_{[a}p^\rho_{b]}$\footnote{In our convention, the antisymmetrization will not carry any factor of $1/2$.}, the inner product $n\cdot\widehat C_a=0$. Consequently $n\cdot \mathcal{Z}_{ab}$ vanishes exactly.\\
For physical velocities $\alpha\vec{v}_a$, we compute the following terms
\begin{align}
 & p_a^0 =m_a\left(1+\tfrac12\alpha^2\vec{v}_a^{\,2}\right)+\mathcal{O}(\alpha^4), \qquad  p_a^i =\alpha m_av_a^i+ \mathcal{O}(\alpha^3), \qquad \delta_a=1-\alpha\vec{n}\cdot\vec{v}_a+ \frac{\alpha^2}{2} v_a^2 + \mathcal{O}(\alpha^3) \notag \\
 & \sigma_{ab} =1+\tfrac12\alpha^2v_{ab}^2+ \mathcal{O}(\alpha^4),\qquad  \sigma_{ab}^2 -1=\alpha^2v_{ab}^2+ \mathcal{O}(\alpha^4), \qquad \tau_{ab}=\frac1{2\alpha^3v_{ab}^3}+ \mathcal{O}(\alpha^{-1}).
 \label{eq:basic}
\end{align}
Following the definition of the trajectory coefficient $\widehat C_a$ in \eqref{eq4.5}, we determine 
{\allowdisplaybreaks
\begin{align}
 \widehat C_a^i =2G\sum_{b\ne a}\tau_{ab}
 \left(\frac{d_b}{d_a}p_a^i-p_b^i\right)\nonumber = & \frac{G}{\alpha^2}\sum_{b\ne a}\frac{m_bv_{ab}^i}{v_{ab}^3}
 +\frac{G}{\alpha}\sum_{b\ne a}
 \frac{m_bv_a^i(\vec{n}\cdot\vec{v}_{ab})}{v_{ab}^3}+ \mathcal{O}(1)\nonumber\\
 =& \frac{c_a^i}{\alpha^2}
 +\frac{v_a^i(\vec{n}\cdot\vec{c}_a)}{\alpha}+ \mathcal{O}(1).
 \label{eq:Chat}
\end{align}}
Substituting the expression of $\widehat C_a^i$ in the definition of $\mathcal{Z}_{ab}$ from eq.\eqref{eq4.5} we find 
\begin{align}
 \mathcal{Z}_{ab}^i&=\frac{c_{ab}^i}{\alpha^2}
 +\frac{v_a^i(\vec{n}\cdot\vec{c}_a)-v_b^i(\vec{n}\cdot\vec{c}_b)}
 {\alpha}+ \mathcal{O}(1), \qquad 
 \mathcal{Z}_{ab}^0=\vec{n}\cdot \vec{\mathcal{Z}}_{ab}
 \notag\\
 w_{ab}^i&=\alpha m_am_b\left[v_{ab}^i
 + \frac{\alpha}{2}\left\{
 (\vec{n}\cdot\vec{v}_{ab})(v_a+ v_b)^i
 -\big(\vec{n}\cdot(\vec{v}_a+\vec{v}_b)\big)v_{ab}^i
 \right\}\right]+ \mathcal{O}(\alpha^3).
 \label{eq:w}
\end{align}
To find the expression of $\mathcal{T}^{ij}_{ab}$ we also need to evaluate the following scalar expressions
{\allowdisplaybreaks
\begin{align}
 & \zeta_{ab} = - \frac{\vec{n} \cdot \vec{c}_{ab}}{\alpha^2} + \frac{\vec{v}_a \cdot \vec{c}_{ab} - (\vec{n} \cdot \vec{v}_a) (\vec{n}\cdot \vec{c}_a) + ( \vec{n} \cdot \vec{v}_b) ( \vec{n} \cdot \vec{c}_b) }{\alpha} + \mathcal{O}(1)\\
 & \zeta_{ab}+\zeta_{ba} =\frac{\vec{v}_{ab}\cdot\vec{c}_{ab}}{\alpha} +\frac{v_{ab}^2}{2}\vec{n}\cdot(\vec{c}_a+\vec{c}_b)+ \mathcal{O}(\alpha), \notag \\
 & \zeta_{ab}\zeta_{ba} =-\frac{(\vec{n}\cdot\vec{c}_{ab})^2}{\alpha^4} +\frac{\vec{n}\cdot\vec{c}_{ab}}{\alpha^3} \big[ (\vec{v}_a+\vec{v}_b)\cdot\vec{c}_{ab} -2(\vec{n}\cdot\vec{v}_a )(\vec{n}\cdot\vec{c}_a) +2(\vec{n}\cdot\vec{v}_b)(\vec{n}\cdot\vec{c}_b) \big]+ \mathcal{O}(\alpha^{-2}), \notag \\
 & \mathcal{Z}_{ab}^2 =\frac{\vec{c}_{ab}^{\,2}-(\vec{n}\cdot\vec{c}_{ab})^2}{\alpha^4} +\frac{2}{\alpha^3}\Big\{ \vec{c}_{ab}\cdot \big[ \vec{v}_a(\vec{n}\cdot\vec{c}_a) -\vec{v}_b(\vec{n}\cdot\vec{c}_b)]-(\vec{n}\cdot\vec{c}_{ab}) \big[ (\vec{n}\cdot\vec{v}_a)(\vec{n}\cdot\vec{c}_a)-(\vec{n}\cdot\vec{v}_b)(\vec{n}\cdot\vec{c}_b) \big] \Big\}\nonumber \\
 & ~~~~~~~~~~~~~~~~~~~~~~~~~~~~~~~~~~~~~~~~~~~~~~~~~~~~~~~~~~~~~~~~~~~~~~~~~~~~~~~~ + \mathcal{O}(\alpha^{-2}).
\label{eq:Zsquare}
\end{align}}
Note that to obtain $\mathcal{T}_{ab}^{ij}$ it will be useful to evaluate all the scalar functions in the three lines of eq.\eqref{eq:D8}, namely the coefficients of $w_{ab}^i w_{ab}^j, w_{ab}^{(i}\mathcal{Z}_{ab}^{j)}$ and $\mathcal{Z}_{ab}^i \mathcal{Z}_{ab}^j$. Therefore, the scalar function in the first line of eq.\eqref{eq:D8} becomes the following 
\begin{equation}\label{eq:bracket2}
\begin{split}
    \bullet ~~ & (2\sigma_{ab}^2+3)(\zeta_{ab}^2+\zeta_{ba}^2) +10\sigma_{ab}\zeta_{ab}\zeta_{ba} -\frac13(\sigma_{ab}^2-1) (4\sigma_{ab}^4-12\sigma_{ab}^2+11) \mathcal{Z}_{ab}^2
   \\
   = & \frac{1}{\alpha^2} \big[ 5 \left( \vec{v}_{ab} \cdot \vec{c}_{ab} \right)^2 - v_{ab}^2 c_{ab}^2 \big] + \frac{1}{\alpha} \big[ 4 v_{ab}^2 \left( \vec{v}_{ab} \cdot \vec{c}_{ab} \right) \vec{n}\cdot \left( \vec{c}_a + \vec{c}_b \right) \big] + \mathcal{O}(1)
\end{split}
\end{equation}
The other two scalar combinations needed in the second and the third line of \eqref{eq:D8} are the following
\begin{align}
 & \bullet ~~\frac{4}{3} \left( \sigma_{ab}^2 -1 \right) \big[ (3\sigma_{ab}\delta_a-\kappa_{ab}\delta_b)\zeta_{ab}
 +(3\sigma_{ab}\delta_b-\kappa_{ab}\delta_a)\zeta_{ba} \big]
 \nonumber \notag \\
 & ~~~~~ = 8 \alpha v_{ab}^2 \vec{v}_{ab}\cdot\vec{c}_{ab} + 4 \alpha^2 v_{ab}^2 \Big[ v_{ab}^2\vec{n}\cdot(\vec{c}_a+\vec{c}_b) - [\vec{n}\cdot(\vec{v}_a+\vec{v}_b)] (\vec{v}_{ab}\cdot\vec{c}_{ab}) \Big] + \mathcal{O}(\alpha^3),\notag \\
 &  \bullet ~~ \frac23(\sigma_{ab}^2-1)^2(2\sigma_{ab}^2-1) \left( \delta_a^2+\sigma_{ab}\delta_a\delta_b+\delta_b^2 \right) = \frac{2}{3} \left( \alpha^4 v_{ab}^4  + 2 \alpha^6 v_{ab}^6  \right)\big[ 3-3\alpha\vec{n}\cdot(\vec{v}_a+\vec{v}_b) +\mathcal{O}(\alpha^2) \big].
 \label{eq:bracket3}
\end{align}

Now recall that in eq.\eqref{eq:D7}, except the term $\mathcal{T}^{ij}_{ab}$ there is a pure gauge part. We have to analyze that as well. As given in \cite{Karan:2026jzs}
\begin{align}\label{gauge}
    \xi_{ab}^\mu = & \frac{G^2}{(\sigma_{ab}^2 -1)^{3/2}} \Bigg{\lbrace}  \frac{14 \sigma_{ab}^2 - 13}{3} \mathcal{Z}_{ab \perp}^2 \Big[ ( \delta_a - \sigma_{ab} \delta_b ) m_b p_a^\mu - ( \sigma_{ab} \delta_a - \delta_b) m_a p_b^\mu \Big] \\ \notag
    & + \frac{2}{3} (\sigma_{ab}^2 - 1) ( 2 \sigma_{ab}^2 - 1) m_a m_b \Big[ \left( n \cdot \mathcal{Z}_{ab \perp} \right) \mathcal{Z}^{\mu}_{ab\perp} + \frac{5}{2} \mathcal{Z}^2_{ab \perp} n^\mu  \Big] \Bigg{\rbrace}
\end{align}
where
\begin{align}\label{Zperp}
    \mathcal{Z}_{ab \perp}^\mu = \mathcal{Z}_{ab}^\mu - \frac{( \zeta_{ab} + \sigma_{ab} \zeta_{ba}) p_a^\mu / m_a - ( \zeta_{ba} + \sigma_{ab} \zeta_{ab} ) p_b^\mu/m_b}{\sigma_{ab}^2 - 1} 
\end{align}
We will argue that both the Newtonian and the next-to-Newtonian waveform at the required logarithmic order does not receive any contribution from the pure gauge part.\\
Using the definition of the various symbols, we get
\begin{equation}
    \zeta_{ab} + \sigma_{ab} \zeta_{ba} = \zeta_{ab} + \zeta_{ba} + (\sigma_{ab} -1) \zeta_{ba} = \frac{\vec{v}_{ab} \cdot \vec{c}_{ab}}{\alpha} + v_{ab}^2 \vec{n}\cdot \vec{c}_a + \mathcal{O}(\alpha).
\end{equation}
Similarly,
\begin{equation}
    \zeta_{ba} + \sigma_{ab} \zeta_{ab} = \frac{\vec{v}_{ab} \cdot \vec{c}_{ab}}{\alpha} + v_{ab}^2 \vec{n}\cdot \vec{c}_b + \mathcal{O}(\alpha).
\end{equation}
Combining these, we can estimate the order of the projection $\mathcal{Z}^i_{ab\perp}$ and $\mathcal{Z}^0_{ab\perp}$. From eq.\eqref{Zperp} we see that for $\mu = i$, the second term behaves as $\mathcal{O}(\alpha^{-2})$ because the numerator is $\mathcal{O}(1)$ while the denominator scales as $\mathcal{O}(\alpha^{2})$ (see eq.\eqref{eq:basic}). Also in eq.\eqref{eq:w}, we determined that $\mathcal{Z}_{ab}^i$ scales as $\mathcal{O}(\alpha^{-2})$ as well. Therefore, it determines that $\mathcal{Z}^i_{ab\perp} \sim \mathcal{O}(\alpha^{-2})$. Interestingly, by taking $\nu=0$, it would be easy to show that due to some cancellation of the leading order terms, we will eventually see that $\mathcal{Z}^{0}_{ab\perp} \sim \mathcal{O}(\alpha^{-1})$. This implies that $\mathcal{Z}_{ab\perp}^2$ scales as $\mathcal{O}(\alpha^{-4})$ at leading order in $\alpha$. \\
To find how $\xi^\mu_{ab}$ scales with $\alpha$, we need to find the expression of the tensor structure in the first line of eq.\eqref{gauge}
\begin{align}
    & ( \delta_a - \sigma_{ab} \delta_b ) m_b p_a^0 - ( \sigma_{ab} \delta_a - \delta_b) m_a p_b^0 = - \alpha^2 m_a m_b v_{ab}^2 + \mathcal{O}(\alpha^3), \notag \\
    & ( \delta_a - \sigma_{ab} \delta_b ) m_b p_a^i - ( \sigma_{ab} \delta_a - \delta_b) m_a p_b^i = - \alpha^2 m_a m_b ( \vec{n} \cdot v_{ab}) v_{ab}^i + \mathcal{O}(\alpha^3)
\end{align}
So from eq.\eqref{gauge}, the first line inside the bracket scales as $\mathcal{O}(\alpha^{-2})$ since the prefactor is $\mathcal{O}(1)$ and $\mathcal{Z}_{ab\perp}^2$ scales as $\mathcal{O}(\alpha^{-4})$. Similarly in the next line, $\sigma_{ab}^2 - 1 \sim \mathcal{O}(\alpha^2), 2 \sigma_{ab}^2 - 1 \sim \mathcal{O}(1)$ and as a result the entire term scales as $\mathcal{O}(\alpha^{-2})$. Now taking into account the outside denominator we see that $\xi_{ab}^\mu \sim \mathcal{O}(\alpha^{-5})$. However, to get the waveform in the Newtonian and the next-to-Newtonian limit we must need $A^{ij}_{(3)}$ to scale as $\mathcal{O}(\alpha^{-7})$ and $\mathcal{O}(\alpha^{-6})$ respectively. Therefore we can safely ignore the contribution due to the pure gauge term at these orders in $\alpha$.

\subsection*{Comparison with the derived waveform in the Newtonian limit}

We now compute the waveform coefficient $A^{ij}_{(3)}$ using the expressions we have obtained. Note that to match the expression with the Newtonian limit we need to keep the terms from all the expression leading order in $\alpha$. Using the leading order $\alpha$ term from eq.\eqref{eq:w}, \eqref{eq:bracket2} and \eqref{eq:bracket3} in eq.\eqref{eq:D8} and \eqref{eq:D7} we obtain
\begin{align}
 \left.A_{(3)}^{ij}\right|_{\alpha^{-7}}
 =\frac{G^2}{R\alpha^7}\sum_{a<b}\frac{m_am_b}{v_{ab}^7}\Big\{&
 \big[v_{ab}^2\vec c_{ab}^{\,2}
 -5(\vec v_{ab}\cdot\vec c_{ab})^2\big]v_{ab}^iv_{ab}^j + 8v_{ab}^2(\vec v_{ab}\cdot\vec c_{ab})
 v_{ab}^{(i}c_{ab}^{j)}
 -2v_{ab}^4c_{ab}^ic_{ab}^j\Big\}.
 \label{eq:Npair}
\end{align}
From \cite{Karan:2026jzs} the waveform at the required order is given as follows 
\begin{equation}
 \left.e^{ij}(w)\right|_{\alpha^2w^{-3}\ln^2w}
 =\frac{4G\alpha^2}{R}\frac{(\ln w)^2}{w^3}
 \sum_a m_a\big(v_a^{(i}d_{2,a}^{j)}+c_a^{(i}d_{1,a}^{j)}\big).
 \label{eq:Ntarget}
\end{equation}
where the trajectory coefficients obey the following relations
{\allowdisplaybreaks
\begin{align}
 \vec d_{1,a}&=\frac G2\sum_{b\ne a}\frac{m_b}{v_{ab}^5}
 \big[3\vec v_{ab}(\vec v_{ab}\cdot\vec c_{ab})
 -v_{ab}^2\vec c_{ab}\big],\label{d1} \\
 \vec d_{2,a}&=\frac G4\sum_{b\ne a}\frac{m_b}{v_{ab}^7}
 \big[\vec v_{ab}\{v_{ab}^2 c_{ab}^{\,2}
 -5(\vec v_{ab}\cdot\vec c_{ab})^2\}
 +2v_{ab}^2\vec c_{ab}(\vec v_{ab}\cdot\vec c_{ab})\big] \label{d2}.
\end{align}}
Then one needs to substitute the expressions of the trajectory coefficients in eq.\eqref{eq:Ntarget}. After substitution, note that there will be two summation, one is $\sum_a$ and the other is $\sum_{b \neq a}$. Since the double sum contains both the contribution due to the pairs $(a,b)$ and $(b,a)$ we need to include the contributions due to both the pairs. While doing that it is important to note that under $a \leftrightarrow b$, $\vec{v}_{ab}$ and $\vec{c}_{ab}$ are odd and hence they will introduce a minus sign that should be carefully taken into account. For example, consider the following sum
\begin{equation}\label{sum cd1}
\begin{split}
    \sum_a  m_a c_a^{(i}d_{1,a}^{j)} = & \frac G2 \sum_a \sum_{b\ne a}\frac{m_a m_b}{v_{ab}^5} \big[3 c_a^{(i} v_{ab}^{j)} (\vec v_{ab}\cdot\vec c_{ab}) -v_{ab}^2 c_a^{(i }c_{ab}^{j)} \big]
    \\
    = & \frac{G}{4} \sum_{a < b}  \frac{m_a m_b}{v_{ab}^7} \big[6  c_{ab}^{(i} v_{ab}^{j)} (\vec v_{ab}\cdot\vec c_{ab}) v_{ab}^2 - 2 v_{ab}^4 c_{ab}^{(i }c_{ab}^{j)} \big]
\end{split}
\end{equation}
To obtain the last equality one needs to use the fact that under $a \leftrightarrow b$, $\vec{v}_{ab}$ and $\vec{c}_{ab}$ are odd and hence they will introduce a minus sign which will eventually turn the $c_a^{(i} v_{ab}^{j)}$ of the first line to $c_{ab}^{(i} v_{ab}^{j)}$ in the second line. In the similar way one could obtain 
\begin{equation}\label{sum vd2}
    \sum_a  m_a v_a^{(i}d_{2,a}^{j)} = \frac G4\sum_{a<b}\frac{m_a m_b}{v_{ab}^7} \big[v_{ab}^i v_{ab}^j \{v_{ab}^2 c_{ab}^{\,2} -5(\vec v_{ab}\cdot\vec c_{ab})^2\} + 2 v_{ab}^{(i} c_{ab}^{j)} (\vec v_{ab}\cdot\vec c_{ab}) v_{ab}^2 \big]
\end{equation}
Substitution of eq.\eqref{sum cd1} and \eqref{sum vd2} in eq.\eqref{eq:Ntarget} reproduces the expression for $A^{ij}_{(3)}$ what we have obtained from the conjecture in \cite{Karan:2026jzs} and hence it shows that the conjecture indeed agrees in the Newtonian limit of the $(\ln w)^2 / w^3$ coefficient.  

\subsection*{Comparison with the derived waveform in the PN expansion}

We now show that the conjecture for the targeted waveform $A^{ij}_{(3)}$ in the next-to-Newtonian order also agrees with the derived waveform in section \ref{s2}. \\
As argued below eq.\eqref{eq4.2} that to obtain the waveform in the next-to-Newtonian limit, we have to extract $A^{ij}_{(3)}$ at $\mathcal{O}(\alpha^{-6})$. Therefore, unlike the Newtonian case, we have to take the numerator at $\mathcal{O}(\alpha)$. Keeping this in mind, we now substitute eq.\eqref{eq:w}, \eqref{eq:bracket2} and \eqref{eq:bracket3} in eq.\eqref{eq:D8} and \eqref{eq:D7} to obtain the following expression \footnote{To get this form one needs to use the following relation 
\begin{equation}
    v_a^i(\vec{n}\cdot\vec{c}_a)-v_b^i(\vec{n}\cdot\vec{c}_b) = \frac{1}{2} \Big[ (\vec{n} \cdot \vec{c}_{ab}) (v_a + v_b)^i + \left[ \vec{n} \cdot ( \vec{c}_a + \vec{c}_b )\right] v_{ab}^i \Big]
\end{equation}
in the $\mathcal{O}(1/\alpha)$ term of $\mathcal{Z}_{ab}^i$.}
\begin{align}
 \left.A_{(3)}^{ij}\right|_{\alpha^{-6}}
 =&\frac{G^2}{R\alpha^6}\sum_{a<b}\frac{m_am_b}{v_{ab}^7} \Bigg\{
 \big[5(\vec v_{ab}\cdot c_{ab})^2-v_{ab}^2\vec c_{ab}^{\,2}\big]
 \Big[[\vec n\cdot(\vec v_a+\vec v_b)]v_{ab}^iv_{ab}^j
 -(\vec n\cdot\vec v_{ab})v_{ab}^{(i}(v_a+v_b)^{j)}\Big]\nonumber\\
 &+4v_{ab}^2(\vec v_{ab}\cdot\vec c_{ab})\Big[
 (\vec n\cdot\vec c_{ab})v_{ab}^{(i}(v_a+v_b)^{j)}
 +(\vec n\cdot\vec v_{ab})(v_a+v_b)^{(i}c_{ab}^{j)} -2[\vec n\cdot(\vec v_a+\vec v_b)]v_{ab}^{(i}c_{ab}^{j)}\Big]\nonumber\\
 &+2v_{ab}^4\Big[ [\vec n\cdot(\vec c_a+\vec c_b)]v_{ab}^{(i}c_{ab}^{j)} +[\vec n\cdot(\vec v_a+\vec v_b)]c_{ab}^ic_{ab}^j -(\vec n\cdot\vec c_{ab})(v_a+v_b)^{(i}c_{ab}^{j)}\Big]\Bigg\}.
 \label{eq:pairanswer}
\end{align}
This is the expression of $A^{ij}_{(3)}$ from the conjecture in \cite{Karan:2026jzs} and contributes to the waveform in the next-to-Newtonian order. \\

We now write down the corresponding waveform at the required order following the derived expression (see eq.\eqref{eij 1 multiparticle w} and set $r=3$) in section \ref{s2}  
\begin{align}
 \left.e_{(1)}^{ij}(w)\right|_{\alpha^3 w^{-3}\ln^2w}
 =\frac{G\alpha^3}{R}\frac{(\ln w)^2}{w^3}
 \sum_a m_a\Big[&
 2(\vec n\cdot\vec c_a)c_a^ic_a^j
 +8(\vec n\cdot\vec d_{2,a})v_a^iv_a^j\nonumber\\
 &-8(\vec n\cdot\vec v_a)
 (v_a^{(i}d_{2,a}^{j)}+c_a^{(i}d_{1,a}^{j)})
 +8(\vec n\cdot\vec d_{1,a})v_a^{(i}c_a^{j)}\Big].
 \label{eq:notetarget}
\end{align}
Using the explicit expressions of the trajectory coefficients from eq.\eqref{d1} we first compute the expression involving $d_{1,a}$ as follows
{\allowdisplaybreaks
\begin{align}
    & \sum_a m_a \Big[ - 8 (\vec{n} \cdot \vec{v}_a) c_a^{(i}d_{1,a}^{j)} + 8 ( \vec{n} \cdot \vec{d}_{1,a} ) v_a^{(i}c_a^{j)} \Big]
    \\
   =  &  4G \sum_a \sum_{b \neq a} \frac{m_a m_b}{v_{ab}^5} \Big[ - 3 (\vec{v}_{ab} \cdot \vec{c}_{ab}) ( \vec{n} \cdot \vec{v}_a ) v_{ab}^{(i} c_a^{j)} + v_{ab}^2 (\vec{n} \cdot \vec{v}_a ) c_{ab}^{(i} c_a^{j)}  
   \\
   & ~~~~~~~~~~~~~~~~~~~~~~~~~~~~~~~~~~ + \left[ 3 ( \vec{v}_{ab} \cdot \vec{c}_{ab}) (\vec{n} \cdot \vec{v}_{ab} ) -  v_{ab}^2 ( \vec{n} \cdot \vec{c}_{ab} ) \right] v_a^{(i}c_a^{j)} \Big]
\end{align}}
Likewise the Newtonian limit, here also we need to take the contribution of the double sum for both the pairs $(a,b)$ and $(b,a)$. Then we obtain the following expression 
\begin{equation}
\begin{split}
   4G \sum_{a < b} \frac{m_a m_b}{v_{ab}^5} & \Big{\lbrace} - 3 ( \vec{v}_{ab} \cdot \vec{c}_{ab} ) v_{ab}^{(i} \left[ (\vec{n} \cdot \vec{v}_a) c_a^{j)} - (\vec{n} \cdot \vec{v}_b ) c_b^{j)} \right] + v_{ab}^2 c_{ab}^{(i} \left[ (\vec{n} \cdot \vec{v}_a) c_a^{j)} - (\vec{n} \cdot \vec{v}_b ) c_b^{j)} \right] 
   \\
   & + \left[ 3 ( \vec{v}_{ab} \cdot \vec{c}_{ab} ) (\vec{n} \cdot \vec{v}_{ab}) - v_{ab}^2 (\vec{n} \cdot \vec{c}_{ab})\right] [ v_a^{(i} c_a^{j)} - v_b^{(i} c_b^{j)} ]\Big{\rbrace}.
\end{split}
\end{equation}
Now we use the following two identities 
\begin{equation}
\begin{split}
   & (\vec{n} \cdot \vec{v}_a) c_a^i - (\vec{n} \cdot \vec{v}_b ) c_b^i = \frac{1}{2} [ \vec{n} \cdot ( \vec{v}_a +\vec{v}_b )] c_{ab}^i + \frac{1}{2} (\vec{n} \cdot \vec{v}_{ab}) (c_a + c_b)^i
   \\
   & v_a^{(i} c_a^{j)} - v_b^{(i} c_b^{j)} = \frac{1}{2} (v_a + v_b)^{(i} c_{ab}^{j)} + \frac{1}{2} v_{ab}^{(i} ( c_a + c_b)^{j)}
\end{split}
\end{equation}
to rewrite the above expression in the following way 
\begin{align}
& \sum_a m_a \Big[ - 8 (\vec{n} \cdot \vec{v}_a) c_a^{(i}d_{1,a}^{j)} + 8 ( \vec{n} \cdot \vec{d}_{1,a} ) v_a^{(i}c_a^{j)} \Big] \notag \\
 =& G \sum_{a<b} \frac{m_a m_b}{v_{ab}^7} \Big[ v_{ab}^2(\vec v_{ab}\cdot\vec c_{ab}) \Big{\lbrace} -6 [\vec n\cdot(\vec v_a+\vec v_b)] v_{ab}^{(i}c_{ab}^{j)} + 6 (\vec n\cdot\vec v_{ab})
 (v_a+v_b)^{(i}c_{ab}^{j)} \Big{\rbrace} \nonumber\\
 + & 2v_{ab}^4\Big\{
 [\vec n\cdot(\vec v_a+\vec v_b)]c_{ab}^ic_{ab}^j
 -(\vec n\cdot\vec c_{ab})(v_a+v_b)^{(i}c_{ab}^{j)} 
 +(\vec n\cdot\vec v_{ab})c_{ab}^{(i}(c_a+c_b)^{j)}
 -(\vec n\cdot\vec c_{ab})v_{ab}^{(i}(c_a+c_b)^{j)}
 \Big\} \Big] .
 \label{eq:noteD1}
\end{align}
By following the exactly same strategy one can also find that the terms involving $d_{2,a}$ in eq.\eqref{eq:notetarget} gives the following expression
\begin{align} \label{eq:noteD2}
    & \sum_a m_a \Big[ - 8 (\vec{n}\cdot \vec{v}_a) v_a^{(i} d_{2,a}^{j)} + 8 (\vec{n} \cdot \vec{d}_{2,a}) v_a^i v_a^j \Big] \notag
    \\
    = & G \sum_{a < b} \frac{m_a m_b}{v_{ab}^7} \Big[ [5(\vec v_{ab}\cdot\vec c_{ab})^2 -v_{ab}^2\vec c_{ab}^{\,2}\big] \Big\{[\vec n\cdot(\vec v_a+\vec v_b)]v_{ab}^iv_{ab}^j -(\vec n\cdot\vec v_{ab})v_{ab}^{(i}(v_a+v_b)^{j)}\Big\}\nonumber\\
   +& v_{ab}^2(\vec v_{ab}\cdot\vec c_{ab}) \Big\{ 4 (\vec n\cdot\vec c_{ab}) v_{ab}^{(i}(v_a+v_b)^{j)} - 2 [\vec n\cdot(\vec v_a+\vec v_b)]v_{ab}^{(i}c_{ab}^{j)}
    - 2 (\vec n\cdot\vec v_{ab})(v_a+v_b)^{(i}c_{ab}^{j)} \Big\} \Big].
\end{align}
And finally, the terms involving only $c_a$ in eq.\eqref{eq:notetarget} can be written in the following form 
\begin{align}
 & \sum_a m_a \left[  2 (\vec{n} \cdot \vec{c}_a) c_a^i c_a^j \right] \notag
 \\
 = & G \sum_{a < b} \frac{m_a m_b}{v_{ab}^7} \Big[ 2v_{ab}^4\Big\{[\vec n\cdot(\vec c_a+\vec c_b)]v_{ab}^{(i}c_{ab}^{j)}+(\vec n\cdot\vec c_{ab})v_{ab}^{(i}(c_a+c_b)^{j)}-(\vec n\cdot\vec v_{ab})c_{ab}^{(i}(c_a+c_b)^{j)} \Big\} \Big].
 \label{eq:notecubic}
\end{align}
We now add up eq.\eqref{eq:noteD1}-\eqref{eq:notecubic} and substitute the resulting expression in eq.\eqref{eq:notetarget}, after various cancellation we indeed recover the same expression as given in eq.\eqref{eq:pairanswer} obtained from the conjecture.

\section{Quadratic gravitational vertex} \label{app C}
\label{app:vertex}

In this appendix we derive the spatial quadratic tensor entering the
vertex. We write $g_{\mu\nu}=\eta_{\mu\nu}+H_{\mu\nu}$, with subscripts
$(1)$ and $(2)$ denoting terms linear and quadratic in $H_{\mu\nu}$. All indices are raised and lowered with $\eta_{\mu\nu}$, and
$\operatorname{tr}X\equiv\eta^{\mu\nu}X_{\mu\nu}$.
The connection and Ricci tensors at first and second order in
$H_{\mu\nu}$ are
\begin{equation}
\begin{aligned}
\Gamma_{(1)\sigma\mu\nu}
&=\tfrac12\bigl(
  \partial_\mu H_{\sigma\nu}
 +\partial_\nu H_{\sigma\mu}
 -\partial_\sigma H_{\mu\nu}\bigr), \qquad
\Gamma_{(1)\mu\nu}^{\rho}
=\eta^{\rho\sigma}\Gamma_{(1)\sigma\mu\nu},
\qquad
\Gamma_{(2)\mu\nu}^{\rho}
=-H^{\rho\sigma}\Gamma_{(1)\sigma\mu\nu},\\
R_{(1)\mu\nu}
&=\partial_\rho\Gamma_{(1)\mu\nu}^{\rho}
 -\partial_\nu\Gamma_{(1)\mu\rho}^{\rho}, \qquad R_{(2)\mu\nu}
=\partial_\rho\Gamma_{(2)\mu\nu}^{\rho}
 -\partial_\nu\Gamma_{(2)\mu\rho}^{\rho}
 +\Gamma_{(1)\rho\lambda}^{\rho}\Gamma_{(1)\mu\nu}^{\lambda}
 -\Gamma_{(1)\nu\lambda}^{\rho}\Gamma_{(1)\mu\rho}^{\lambda}.
\end{aligned}
\label{khv:definitions}
\end{equation}
The quadratic operator entering the gravitational vertex is
\begin{equation}
\begin{aligned}
\mathcal{G}_{(2)}^{\mu\nu}[H]
={}&R_{(2)}^{\mu\nu}
 -H^{\mu\rho}R_{(1)\rho}{}^\nu
 -H^{\nu\rho}R_{(1)\rho}{}^\mu-\tfrac12\eta^{\mu\nu}
 \bigl(\operatorname{tr}R_{(2)}
       -H^{\rho\sigma}R_{(1)\rho\sigma}\bigr)
 +\tfrac12H^{\mu\nu}\operatorname{tr}R_{(1)}\\
& ~~~~~~~~~~~~~~~~~~~~+\tfrac12\operatorname{tr}H
 \bigl(R_{(1)}^{\mu\nu}
       -\tfrac12\eta^{\mu\nu}\operatorname{tr}R_{(1)}\bigr).
\end{aligned}
\label{khv:G2-definition}
\end{equation}

\paragraph{Non-relativistic reduction.}
At leading non-relativistic order, we set $H_{0i}=0$ and
$H_{ij}=\delta_{ij}H_{00}$. Since $\partial_u\sim O(\alpha^3)$
and $\partial_i\sim O(\alpha^2)$, time derivatives carry an
additional power of $\alpha$ relative to spatial derivatives
and are neglected at this order.\\
The nonzero components of the linear connection and its contracted
trace are
\begin{equation}
\begin{aligned}
\Gamma_{(1)0i}^{0}=\Gamma_{(1)i0}^{0}
&=-\tfrac12\partial_iH_{00},
\qquad
\Gamma_{(1)00}^{i}=-\tfrac12\partial_iH_{00}, \qquad \Gamma_{(1)\rho i}^{\rho}
=\partial_iH_{00}.\\
\Gamma_{(1)jk}^{i}
&=\tfrac12\bigl(
 \delta_{ik}\partial_jH_{00}
 +\delta_{ij}\partial_kH_{00}
 -\delta_{jk}\partial_iH_{00}\bigr).
\end{aligned}
\label{khv:Gamma1}
\end{equation}
It follows that
\begin{equation}
\begin{aligned}
R_{(1)00}
&=-\tfrac12\nabla^2H_{00},
\qquad R_{(1)0i}=0, \qquad R_{(1)ij}
=-\tfrac12\delta_{ij}\nabla^2H_{00},\\
\operatorname{tr}R_{(1)}
&=-\nabla^2H_{00},
\qquad
H^{\rho\sigma}R_{(1)\rho\sigma}
=-2H_{00}\nabla^2H_{00}.
\end{aligned}
\label{khv:R1}
\end{equation}
The quadratic connection is obtained by contracting the linear
connection with $-H^{\rho\sigma}$. Together with the connection
product entering $R_{(2)ij}$, the required identities are
\begin{equation}
\begin{aligned}
\Gamma_{(2)\mu\nu}^{0}
&=H_{00}\Gamma_{(1)\mu\nu}^{0},
\qquad
\Gamma_{(2)\mu\nu}^{i}
=-H_{00}\Gamma_{(1)\mu\nu}^{i}, \qquad \Gamma_{(2)i\rho}^{\rho}
=-2H_{00}\partial_iH_{00},\\
& \Gamma_{(1)j\lambda}^{\rho}\Gamma_{(1)i\rho}^{\lambda}
=\tfrac32(\partial_iH_{00})(\partial_jH_{00})
 -\tfrac12\delta_{ij}(\partial_kH_{00})(\partial_kH_{00}).
\end{aligned}
\label{khv:Gamma2}
\end{equation}
Substitution into eq.~\eqref{khv:definitions} gives
\begin{equation}
\begin{aligned}
R_{(2)ij}
={}&H_{00}\partial_i\partial_jH_{00}
 +\tfrac12(\partial_iH_{00})(\partial_jH_{00})+\tfrac12\delta_{ij}
 \bigl[(\partial_kH_{00})(\partial_kH_{00})
       +H_{00}\nabla^2H_{00}\bigr],\\
R_{(2)00}
={}&\tfrac12H_{00}\nabla^2H_{00}
 -\tfrac12(\partial_kH_{00})(\partial_kH_{00}), \qquad
 \operatorname{tr}R_{(2)} 
 ={}\tfrac52(\partial_kH_{00})(\partial_kH_{00})
 +2H_{00}\nabla^2H_{00}.
\end{aligned}
\label{khv:R2}
\end{equation}
Since $\operatorname{tr}H=2H_{00}$, the remaining terms in
eq.~\eqref{khv:G2-definition} reduce to
\begin{equation}
\begin{aligned}
& -H^{i\rho}R_{(1)\rho}{}^j
-H^{j\rho}R_{(1)\rho}{}^i
+\tfrac12H^{ij}\operatorname{tr}R_{(1)}
=\tfrac12\delta^{ij}H_{00}\nabla^2H_{00},\\
& -\tfrac12\delta^{ij}
 \bigl(\operatorname{tr}R_{(2)}
       -H^{\rho\sigma}R_{(1)\rho\sigma}\bigr)=-\delta^{ij}
 \bigl[\tfrac54(\partial_kH_{00})(\partial_kH_{00})
       +2H_{00}\nabla^2H_{00}\bigr],\\
& ~~~ \tfrac12\operatorname{tr}H
 \bigl(R_{(1)}^{ij}-\tfrac12\delta^{ij}\operatorname{tr}R_{(1)}\bigr)=0.
\end{aligned}
\label{khv:G2-contractions}
\end{equation}
Combining eqs.~\eqref{khv:R2} and \eqref{khv:G2-contractions}, we obtain
\begin{equation}
\begin{aligned}
\mathcal{G}_{(2)}^{ij}
={}&H_{00}\partial_i\partial_jH_{00}
 +\tfrac12(\partial_iH_{00})(\partial_jH_{00}) -\delta^{ij}H_{00}\nabla^2H_{00}
 -\tfrac34\delta^{ij}(\partial_kH_{00})(\partial_kH_{00}).
\end{aligned}
\label{khv:G2-spatial}
\end{equation}

\section{Angular average}\label{App D}

In this appendix, we show the detailed computation of the following angular average
\begin{equation}
\begin{aligned}
&\left\langle
\frac{\Phi_{ab}^{ij}(\chi_\phi;\phi)}
     {1+\alpha^3z\beta_{ab}}
\right\rangle = \frac{m_a^2m_b^2}{2(1+\alpha^3z\beta_{ab})}
\frac{1}{2\pi}\int_0^{2\pi}
d\phi
\left[ \frac{\delta^{ij}}{|\vec k_a-\vec k_b|^2}
-\frac{2 (k_a-k_b)^i(k_a-k_b)^j} {|\vec k_a-\vec k_b|^4} \right] + \mathcal{O}(\alpha^2).
\end{aligned}
\label{khavg:integration-task}
\end{equation}
Recall the expressions for $\vec{k}_a$ and $\vec{k}_b$ which will be essential to compute the integrand in the above angular average
\begin{equation}
\begin{aligned}
\vec k_a
&=\vec e_\phi+\chi_\phi\vec{\varpi}_{ab},
&\vec k_b
&=-\vec e_\phi+\chi_\phi(\vec n-\vec{\varpi}_{ab}),\\
\chi_\phi
&=-\frac{\alpha^3z\,e_\phi\cdot(C_a-C_b)}
        {1+\alpha^3z\beta_{ab}},
&e_\phi
&=e_1\cos\phi+e_2\sin\phi.
\end{aligned}
\label{khavg:substituted-definitions}
\end{equation}
Subtracting the two wave vectors yields
\begin{equation}
\vec k_a-\vec k_b
=2\vec e_\phi
-\frac{2\alpha^3z\,e_\phi\cdot(C_a-C_b)}
       {1+\alpha^3z\beta_{ab}}
 \left(\vec{\varpi}_{ab}-\frac{\vec n}{2}\right).
\label{khavg:exact-difference}
\end{equation}
Since $\beta_{ab}$ is independent of $\phi$ (see eq.\eqref{scalars and vectors of KH}), all angular dependence
is linear in $\cos\phi$ and $\sin\phi$. Consequently,
\begin{equation}
\vec k_a-\vec k_b
=\left.(\vec k_a-\vec k_b)\right|_0\cos\phi
+\left.(\vec k_a-\vec k_b)\right|_{\pi/2}\sin\phi.
\label{khavg:linear-angle}
\end{equation}
The subscripts $0$ and $\pi/2$ denote evaluation at $\phi=0$
and $\phi=\pi/2$, including the angular dependence of $\chi_\phi$. The integrands have period $\pi$. Setting $t=\tan\phi$ on
$(-\pi/2,\pi/2)$ converts the scalar average into a rational integral.
Completing the square in its quadratic denominator gives \footnote{To perform the integral over $t$ we have used the following formula
\begin{align}
    \int_{-\infty}^{+\infty} \frac{dt}{A + 2 B t+ C t^2} = \frac{\pi}{\sqrt{AC - B^2}} \notag
\end{align}}
\allowdisplaybreaks{
\begin{align}
\left\langle\frac{1}{|\vec k_a-\vec k_b|^2}\right\rangle =& \frac{1}{\pi}\int_{-\infty}^{\infty}
\frac{dt}{
 \left.|\vec k_a-\vec k_b|^2\right|_0
 +2t\left.(\vec k_a-\vec k_b)\right|_0\cdot
     \left.(\vec k_a-\vec k_b)\right|_{\pi/2}
 +t^2\left.|\vec k_a-\vec k_b|^2\right|_{\pi/2}}\\
= & \frac{1}{
 \left|
 \left.(\vec k_a-\vec k_b)\right|_0\times
 \left.(\vec k_a-\vec k_b)\right|_{\pi/2}
 \right|}.
\label{khdirect:scalar-result}
\end{align}}
The three moments required by the quadratic numerator in eq.\eqref{khavg:integration-task} follow by comparing coefficients in the master identity as follows
\begin{equation}
\begin{aligned}
\bullet ~ \left\langle
\frac{(\lambda\cos\phi+\mu\sin\phi)^2}
     {|\vec k_a-\vec k_b|^4}
\right\rangle =
\frac{
 \left|
 \lambda\left.(\vec k_a-\vec k_b)\right|_{\pi/2}
 -\mu\left.(\vec k_a-\vec k_b)\right|_0
 \right|^2}
 {2\left|
 \left.(\vec k_a-\vec k_b)\right|_0
 \times\left.(\vec k_a-\vec k_b)\right|_{\pi/2}
 \right|^3}
\end{aligned}
\label{khdirect:master}
\end{equation}
which leads to the following results
{\allowdisplaybreaks
\begin{align}\label{khdirect:moments-result}
 & \left\langle
\frac{\cos^2\phi}{|\vec k_a-\vec k_b|^4}
\right\rangle =\frac{\left.|\vec k_a-\vec k_b|^2\right|_{\pi/2}}
 {2\left|
 \left.(\vec k_a-\vec k_b)\right|_0\times
 \left.(\vec k_a-\vec k_b)\right|_{\pi/2}
 \right|^3}, \notag \\
 & \left\langle
\frac{\sin^2\phi}{|\vec k_a-\vec k_b|^4}
\right\rangle =\frac{\left.|\vec k_a-\vec k_b|^2\right|_0}
 {2\left|
 \left.(\vec k_a-\vec k_b)\right|_0\times
 \left.(\vec k_a-\vec k_b)\right|_{\pi/2}
 \right|^3} ,\notag \\
& \left\langle
\frac{\cos\phi\sin\phi}{|\vec k_a-\vec k_b|^4}
\right\rangle
=-\frac{
 \left.(\vec k_a-\vec k_b)\right|_0\cdot
 \left.(\vec k_a-\vec k_b)\right|_{\pi/2}}
 {2\left|
 \left.(\vec k_a-\vec k_b)\right|_0\times
 \left.(\vec k_a-\vec k_b)\right|_{\pi/2}
 \right|^3}.
\end{align}}
The three terms are evaluated using the corresponding lines of eq.~\eqref{khdirect:moments-result}. Combining them with the scalar average gives
\allowdisplaybreaks{
\begin{align}
&\left\langle
\frac{\delta^{ij}}{|\vec k_a-\vec k_b|^2}
-\frac{2(k_a-k_b)^i(k_a-k_b)^j}
      {|\vec k_a-\vec k_b|^4}
\right\rangle\\
&=\frac{1}{
 \left|
 \left.(\vec k_a-\vec k_b)\right|_0\times
 \left.(\vec k_a-\vec k_b)\right|_{\pi/2}
 \right|^3}
\Biggl\{
 \left|
 \left.(\vec k_a-\vec k_b)\right|_0\times
 \left.(\vec k_a-\vec k_b)\right|_{\pi/2}
 \right|^2\delta^{ij}\\
&\qquad
 -\left.|\vec k_a-\vec k_b|^2\right|_{\pi/2}
  \left.(k_a-k_b)^i\right|_0
  \left.(k_a-k_b)^j\right|_0 -\left.|\vec k_a-\vec k_b|^2\right|_0
  \left.(k_a-k_b)^i\right|_{\pi/2}
  \left.(k_a-k_b)^j\right|_{\pi/2}\\
&\qquad
 +\Bigl[
 \left.(\vec k_a-\vec k_b)\right|_0\cdot
 \left.(\vec k_a-\vec k_b)\right|_{\pi/2}
 \Bigr]
 \Bigl[
 \left.(k_a-k_b)^i\right|_0
 \left.(k_a-k_b)^j\right|_{\pi/2} +\left.(k_a-k_b)^i\right|_{\pi/2}
  \left.(k_a-k_b)^j\right|_0
 \Bigr]
\Biggr\}\\
&=\frac{
 \Bigl[
 \left.(\vec k_a-\vec k_b)\right|_0\times
 \left.(\vec k_a-\vec k_b)\right|_{\pi/2}
 \Bigr]^i
 \Bigl[
 \left.(\vec k_a-\vec k_b)\right|_0\times
 \left.(\vec k_a-\vec k_b)\right|_{\pi/2}
 \Bigr]^j}
 {\left|
 \left.(\vec k_a-\vec k_b)\right|_0\times
 \left.(\vec k_a-\vec k_b)\right|_{\pi/2}
 \right|^3}.
\label{khavg:identity}
\end{align}}
The last equality follows from the three-dimensional cross-product identity. To simplify the above expression further we use the following expressions 
\begin{equation}
\begin{aligned}
 & \left.\frac{\vec k_a-\vec k_b}{2}\right|_0 =\vec e_1-\frac{\alpha z(\vec e_1\cdot\vec c_{ab})}{1+\alpha^3z\beta_{ab}}\left(\vec{\varpi}_{ab}-\frac{\vec n}{2}\right)
 + \mathcal{O}(\alpha^2),\\
 & \left.\frac{\vec k_a-\vec k_b}{2}\right|_{\pi/2} =\vec e_2 -\frac{\alpha z(\vec e_2\cdot\vec c_{ab})}{1+\alpha^3z\beta_{ab}} \left(\vec{\varpi}_{ab}-\frac{\vec n}{2}\right) + \mathcal{O}(\alpha^2).
\end{aligned}
\label{khavg:wavevector}
\end{equation}
 The orthogonality conditions $p_a\cdot e_A=p_b\cdot e_A=0$,
together with $\vec p_a=\alpha p_a^0\vec v_a$, imply $e_A^0=\alpha\,\vec v_a\cdot\vec e_A =\alpha\,\vec v_b\cdot\vec e_A,~\text{and}~\vec v_{ab}\cdot\vec e_A=0.$
Thus, $\vec e_1$ and $\vec e_2$ span the spatial plane
perpendicular to $\vec v_{ab}$. Since $e_A\cdot e_B=\delta_{AB}$,
their spatial inner products satisfy
\[
\vec e_A\cdot\vec e_B
=\delta_{AB}+e_A^0e_B^0
=\delta_{AB}+O(\alpha^2).
\]
Choosing the orientation of the basis, we therefore have
\[
\vec e_1\times\vec e_2
=\frac{\vec v_{ab}}{|\vec v_{ab}|}+O(\alpha^2).
\]
Using Eq.~\eqref{khavg:wavevector} and the vector triple-product identity,
we obtain
\begin{equation}
\begin{aligned}
 &\left.(\vec k_a-\vec k_b)\right|_0\times
       \left.(\vec k_a-\vec k_b)\right|_{\pi/2}\\
 &=\frac{4}{|\vec v_{ab}|}
 \Bigg\{
 \left[1-\frac{\alpha z}{1+\alpha^3z\beta_{ab}}\,
       \vec c_{ab}\cdot
       \left(\vec{\varpi}_{ab}-\frac{\vec n}{2}\right)\right]
       \vec v_{ab}  +\frac{\alpha z}{1+\alpha^3z\beta_{ab}}
       \left[\vec v_{ab}\cdot
       \left(\vec{\varpi}_{ab}-\frac{\vec n}{2}\right)\right]
       \vec c_{ab}
 \Bigg\}+ \mathcal{O}(\alpha^2).
\end{aligned}
\label{khavg:cross-before}
\end{equation}
To combine the mixed terms appearing in the cross product, we use
\[
 \bullet ~\Big( (\vec e_2\cdot\vec c_{ab})\vec e_1
 -(\vec e_1\cdot\vec c_{ab})\vec e_2 \Big) \times \left(\vec{\varpi}_{ab}-\frac{\vec n}{2}\right) = \left[ \vec c_{ab}\times(\vec e_1\times\vec e_2) \right] \times \left(\vec{\varpi}_{ab}-\frac{\vec n}{2}\right)
\]
and the vector triple-product identity. \\
To simplify eq.\eqref{khavg:cross-before}, we expand \(\varpi_{ab}^{\mu}\) and \(\beta_{ab}\), defined in eq.\eqref{scalars and vectors of KH}, in the nonrelativistic limit.
\begin{equation}
\begin{aligned}
\vec{\varpi}_{ab}
&=\frac{\vec v_{ab}}{\alpha|\vec v_{ab}|^2}
 -\frac{(\vec n\cdot\vec v_b)\vec v_{ab}}{|\vec v_{ab}|^2}
 + \mathcal{O}(\alpha),\\
\varpi_{ab}^0
&=\frac{\vec v_a\cdot\vec v_{ab}}{|\vec v_{ab}|^2}
  \bigl[1-\alpha\vec n\cdot\vec v_b\bigr]
 + \mathcal{O}(\alpha^2).
\end{aligned}
\label{kh:varpi-expansion}
\end{equation}
Writing the temporal contractions explicitly gives
\begin{equation}
\begin{aligned}
\beta_{ab}
&=-C_b^0+\vec n\cdot\vec C_b
  -\varpi_{ab}^0(C_a^0-C_b^0)
  +\vec{\varpi}_{ab}\cdot(\vec C_a-\vec C_b)\\
&=\frac{\vec v_{ab}\cdot\vec c_{ab}}
        {\alpha^3|\vec v_{ab}|^2}
 +\frac{1}{\alpha^2}\left[
   \vec n\cdot\vec c_b
   -(\vec n\cdot\vec v_b)
    \frac{\vec v_{ab}\cdot\vec c_{ab}}{|\vec v_{ab}|^2}
 \right]
 + \mathcal{O}(\alpha^{-1}).
\end{aligned}
\label{kh:beta}
\end{equation}
The expressions in eq.\eqref{kh:varpi-expansion} and \eqref{kh:beta} follow from
\begin{equation}
\begin{aligned}
(p_a\cdot p_b)\vec p_a-p_a^2\vec p_b
&=-\alpha m_a^2m_b\,\vec v_{ab}+O(\alpha^3),\\
(p_a\cdot p_b)p_a^0-p_a^2p_b^0
&=\frac{\alpha^2m_a^2m_b}{2}
  \bigl[|\vec v_b|^2-|\vec v_{ab}|^2-|\vec v_a|^2\bigr]
  +O(\alpha^4)\\
&=-\alpha^2m_a^2m_b\,\vec v_a\cdot\vec v_{ab}
  +O(\alpha^4),\\
\frac{p_b\cdot n}{D_{ab}^2}
&=-\frac{1-\alpha\vec n\cdot\vec v_b+O(\alpha^2)}
        {\alpha^2m_a^2m_b|\vec v_{ab}|^2}.
\end{aligned}
\label{kh:varpi-substitution}
\end{equation}
Now to analyze eq.\eqref{khavg:cross-before} one needs to consider the following scalar combinations which can be simplified by using eqs.\eqref{kh:beta} and \eqref{kh:varpi-expansion} 
\[
\begin{aligned}
& 1+\alpha^3z\beta_{ab}
-\alpha z\,\vec c_{ab}\cdot
 \left(\vec{\varpi}_{ab}-\frac{\vec n}{2}\right) = 1+\frac{\alpha z}{2}
 \vec n\cdot(\vec c_a+\vec c_b)+ \mathcal{O}(\alpha^2),\\
& \alpha z\,\vec v_{ab}\cdot
 \left(\vec{\varpi}_{ab}-\frac{\vec n}{2}\right) =z-\frac{\alpha z}{2}
 \vec n\cdot(\vec v_a+\vec v_b)+ \mathcal{O}(\alpha^2).
\end{aligned}
\]
Consequently, the cross product in eq.\eqref{khavg:cross-before} takes the following form 
\allowdisplaybreaks{
\begin{align}
 & \left.(\vec k_a-\vec k_b)\right|_0\times \left.(\vec k_a-\vec k_b)\right|_{\pi/2}
 \\
 = & \frac{4}{|\vec v_{ab}|(1+\alpha^3z\beta_{ab})} \Bigg\{\vec v_{ab}+z\vec c_{ab} +\frac{\alpha z}{2} \Big[ [\vec n\cdot(\vec c_a+\vec c_b)]\vec v_{ab} -[\vec n\cdot(\vec v_a+\vec v_b)]\vec c_{ab}
 \Big]\Bigg\}+ \mathcal{O}(\alpha^2)\\
 = & \frac{4}{|\vec v_{ab}|(1+\alpha^3z\beta_{ab})}
 \Bigg\{
 \left[1+\frac{\alpha z}{2}\vec n\cdot(\vec c_a+\vec c_b)\right]
 (\vec v_{ab}+z\vec c_{ab}) 
 -\frac{\alpha z}{2}
 [\vec n\cdot(\vec v_a+\vec v_b+z(\vec c_a+\vec c_b))]
 \vec c_{ab}
 \Bigg\}+ \mathcal{O}(\alpha^2).
\label{khavg:cross-final}
\end{align}}
Using this expression the numerator of eq.\eqref{khavg:identity} can be evaluated as
\begin{equation}
\begin{aligned}
 &\left[
 \left.(\vec k_a-\vec k_b)\right|_0\times
 \left.(\vec k_a-\vec k_b)\right|_{\pi/2}
 \right]^i
 \left[
 \left.(\vec k_a-\vec k_b)\right|_0\times
 \left.(\vec k_a-\vec k_b)\right|_{\pi/2}
 \right]^j\\
 =& \left[\frac{4}
 {|\vec v_{ab}|(1+\alpha^3z\beta_{ab})}\right]^2
 \Bigg\{
 [1+\alpha z\,\vec n\cdot(\vec c_a+\vec c_b)]
 (v_{ab}^i+zc_{ab}^i)(v_{ab}^j+zc_{ab}^j)\\
 &\qquad~~~~~~~~~~~~~~~~~~~~~~ -\frac{\alpha z}{2}
 [\vec n\cdot(\vec v_a+\vec v_b+z(\vec c_a+\vec c_b))]
 \Big[
 c_{ab}^i(v_{ab}^j+zc_{ab}^j)
 +(v_{ab}^i+zc_{ab}^i)c_{ab}^j
 \Big]\Bigg\}
 + \mathcal{O}(\alpha^2).
\end{aligned}
\label{khavg:numerator-expanded}
\end{equation}
whereas the denominator can be written as follows  
\begin{equation}
\begin{aligned}
 &\frac{1}{
 \left|
 \left.(\vec k_a-\vec k_b)\right|_0\times
 \left.(\vec k_a-\vec k_b)\right|_{\pi/2}
 \right|^3}\\
 =& \left[\frac{4} {|\vec v_{ab}|(1+\alpha^3z\beta_{ab})}\right]^{-3}
 \frac{1}{|\vec v_{ab}+z\vec c_{ab}|^3}
 \Bigg[
 1-\frac{3\alpha z}{2}\vec n\cdot(\vec c_a+\vec c_b) 
 \\
 & ~~~~~~~~~~~~~~~~~~~~~~~~~~~~~~~~~~~~~~~~~~~~~~~~~~~ +\frac{3\alpha z}{2}
 [\vec n\cdot(\vec v_a+\vec v_b+z(\vec c_a+\vec c_b))]
 \frac{(\vec v_{ab}+z\vec c_{ab})\cdot\vec c_{ab}}
 {|\vec v_{ab}+z\vec c_{ab}|^2}
 \Bigg] + \mathcal{O}(\alpha^2).
\end{aligned}
\label{khavg:inverse-norm}
\end{equation}
Substituting eq.\eqref{khavg:numerator-expanded} and eq.\eqref{khavg:inverse-norm} in \eqref{khavg:identity} we get 
\allowdisplaybreaks{
\begin{align}
 &\left\langle
 \frac{\delta^{ij}}{|\vec k_a-\vec k_b|^2}
 -\frac{2(k_a-k_b)^i(k_a-k_b)^j}
       {|\vec k_a-\vec k_b|^4}
 \right\rangle\\
 &=\frac{|\vec v_{ab}|(1+\alpha^3z\beta_{ab})}{4}
 \Bigg\{
 \frac{(v_{ab}^i+zc_{ab}^i)(v_{ab}^j+zc_{ab}^j)}
      {|\vec v_{ab}+z\vec c_{ab}|^3} -\frac{\alpha z}{2}
 [\vec n\cdot(\vec c_a+\vec c_b)]
 \frac{(v_{ab}^i+zc_{ab}^i)(v_{ab}^j+zc_{ab}^j)}
      {|\vec v_{ab}+z\vec c_{ab}|^3}\\
 & ~~~~~~~~~~~~~~~~~~~~~~ \qquad-\frac{\alpha z}{2}
 [\vec n\cdot(\vec v_a+\vec v_b+z(\vec c_a+\vec c_b))]
 \frac{c_{ab}^i(v_{ab}^j+zc_{ab}^j)
       +(v_{ab}^i+zc_{ab}^i)c_{ab}^j}
      {|\vec v_{ab}+z\vec c_{ab}|^3}\\
 & ~~~~~~~~~ \qquad+\frac{3\alpha z}{2}
 [\vec n\cdot(\vec v_a+\vec v_b+z(\vec c_a+\vec c_b))]
 \frac{[(\vec v_{ab}+z\vec c_{ab})\cdot\vec c_{ab}]
       (v_{ab}^i+zc_{ab}^i)(v_{ab}^j+zc_{ab}^j)}
      {|\vec v_{ab}+z\vec c_{ab}|^5}
 \Bigg\}+ \mathcal{O}(\alpha^2).
 \label{khavg:angular-expanded}
\end{align}}

To put the above expression in a compact form one may use the following identities  
\begin{equation}
\begin{aligned}
 & \bullet \frac{d}{dz}\left\{
 [\vec n\cdot(\vec v_a+\vec v_b+z(\vec c_a+\vec c_b))]
 \frac{(v_{ab}^i+zc_{ab}^i)(v_{ab}^j+zc_{ab}^j)}
      {|\vec v_{ab}+z\vec c_{ab}|^3}
 \right\}\\
 =& [\vec n\cdot(\vec c_a+\vec c_b)]
 \frac{(v_{ab}^i+zc_{ab}^i)(v_{ab}^j+zc_{ab}^j)}
      {|\vec v_{ab}+z\vec c_{ab}|^3} +
 [\vec n\cdot(\vec v_a+\vec v_b+z(\vec c_a+\vec c_b))]
 \frac{d}{dz} \left( 
 \frac{(v_{ab}^i+zc_{ab}^i)(v_{ab}^j+zc_{ab}^j)}
      {|\vec v_{ab}+z\vec c_{ab}|^3} \right).
\end{aligned}
\label{khavg:product-rule}
\end{equation}
\begin{equation}
\begin{aligned}
 & \bullet \frac{d}{dz} \left\{ \frac{(v_{ab}^i+zc_{ab}^i)(v_{ab}^j+zc_{ab}^j)}
      {|\vec v_{ab}+z\vec c_{ab}|^3} \right\} \\
 = & \frac{c_{ab}^i(v_{ab}^j+zc_{ab}^j)
       +(v_{ab}^i+zc_{ab}^i)c_{ab}^j}
      {|\vec v_{ab}+z\vec c_{ab}|^3} -
 \frac{3[(\vec v_{ab}+z\vec c_{ab})\cdot\vec c_{ab}]
       (v_{ab}^i+zc_{ab}^i)(v_{ab}^j+zc_{ab}^j)}
      {|\vec v_{ab}+z\vec c_{ab}|^5}.
\end{aligned}
\label{khavg:rational-derivative}
\end{equation}
Using these we get 
\begin{equation}
\begin{aligned}
 &\left\langle
 \frac{\Phi_{ab}^{ij}(\chi_\phi;\phi)}
      {1+\alpha^3z\beta_{ab}}
 \right\rangle
 =\frac{m_a^2m_b^2|\vec v_{ab}|}{8}
 \Bigg[
 \frac{(v_{ab}^i+zc_{ab}^i)(v_{ab}^j+zc_{ab}^j)}
      {|\vec v_{ab}+z\vec c_{ab}|^3}\\
 &\qquad~~~~~~~~~~~~~~~~~~~~~~~~ -\frac{\alpha z}{2}\frac{d}{dz}\left\{
 [\vec n\cdot(\vec v_a+\vec v_b+z(\vec c_a+\vec c_b))]
 \frac{(v_{ab}^i+zc_{ab}^i)(v_{ab}^j+zc_{ab}^j)}
      {|\vec v_{ab}+z\vec c_{ab}|^3}
 \right\}
 \Bigg]+ \mathcal{O}(\alpha^2).
\end{aligned}
\label{khavg:final-average}
\end{equation}

\bibliographystyle{JHEP}
\bibliography{newtonian_references}
\end{document}